\documentclass[acmtog,authorversion]{acmart}

\usepackage{booktabs} 

\setcitestyle{square}

\usepackage[ruled]{algorithm2e} 

\SetAlFnt{\small}
\SetAlCapFnt{\small}
\SetAlCapNameFnt{\small}
\SetAlCapHSkip{0pt}
\IncMargin{-\parindent}

\acmJournal{TOG}

\acmSubmissionID{xxx}

\setcopyright{authorversion}

\definecolor{amber}{rgb}{1.0, 0.49, 0.0}
\definecolor{darkgreen}{rgb}{0.0, 0.5, 0.0}
\definecolor{aquamarine}{rgb}{0.1,0.55,0.65}

\newcommand{\customComment}[3]{\unskip}

\def\equationautorefname~#1\null{%
  Equation~#1\null
}

\newcommand{\figurePath}{none}

\newlength{\beautyHeight}

\newcommand{\Trans}{T}
\newcommand{\Transctrl}{\Trans_{\mathrm{c}}}
\newcommand{\Transres}{\Trans_{\mathrm{r}}}
\newcommand{\optd}{\tau}

\newcommand{\optdmaj}{\bar{\optd}}
\newcommand{\optdmajres}{\bar{\optd}_{\mathrm{r}}}
\newcommand{\optdctrl}{\optd_{\mathrm{c}}}
\newcommand{\optdres}{\optd_{\mathrm{r}}}

\newcommand{\estoptd}{X}
\newcommand{\estoptdres}{Y}
\newcommand{\X}{X}
\newcommand{\g}{c}
\newcommand{\M}{M}

\newcommand{\Prob}{\text{Pr}}
\newcommand{\Variance}{\text{Var}}
\newcommand{\Eff}{\text{Eff}}
\newcommand{\Cost}{\text{Cost}}

\newcommand{\coef}{\mu}

\newcommand{\nul}{\coef_{\mathrm{n}}}
\newcommand{\ext}{\coef}
\newcommand{\extctrl}{\coef_{\mathrm{c}}}
\newcommand{\extres}{\coef_{\mathrm{r}}}
\newcommand{\majres}{\maj_{\mathrm{r}}}

\newcommand{\maj}{\bar{\coef}}
\newcommand{\control}{\extctrl}

\newcommand{\len}{\ell}

\DeclareMathOperator*{\E}{\mathbb{E}}
\DeclareMathOperator*{\Var}{\mathrm{Var}}

\providecommand{\myfloor}[1]{\left \lfloor #1 \right \rfloor }

\usepackage[export]{adjustbox}
\usepackage{paralist}
\usepackage{accents} 
\usepackage{commath} 
\usepackage{amsmath} 
\usepackage{breqn}
\usepackage{mathtools} 
\usepackage{pgfplots} 
\pgfplotsset{compat=1.17}
\usepackage[normalem]{ulem} 
\usepackage{subfigure}
\usepackage{wrapfig}
\usepackage[makeroom]{cancel}

\begin{document}

\title{An unbiased ray-marching transmittance estimator}

\author{Markus Kettunen} \affiliation{\institution{NVIDIA}}
\author{Eugene d'Eon} \affiliation{\institution{NVIDIA}}
\author{Jacopo Pantaleoni} \affiliation{\institution{NVIDIA}}
\author{Jan Nov\'ak} \affiliation{\institution{NVIDIA}}

\renewcommand{\shortauthors}{Kettunen et al.}

\begin{abstract}
  We present an in-depth analysis of the sources of variance in state-of-the-art unbiased volumetric transmittance estimators, and propose several new methods for improving their efficiency. These combine to produce a single estimator that is universally optimal relative to prior work, with up to several orders of magnitude lower variance at the same cost, and has zero variance for any ray with non-varying extinction.  We first reduce the variance of truncated power-series estimators using a novel efficient application of U-statistics.  We then greatly reduce the average expansion order of the power series and redistribute density evaluations to filter the optical depth estimates with an equidistant sampling comb. Combined with the use of an online control variate built from a sampled mean density estimate, the resulting estimator effectively performs ray marching most of the time while using rarely-sampled higher order terms to correct the bias.
\end{abstract}



\newcommand{\glassFigureCaption}{\label{fig:teaser}We propose a new unbiased Monte Carlo estimator for volumetric transmittance based on a power-series expansion.  The zeroth-order term in our estimator corresponds to a novel variant of ray-marching.  Higher order terms ensure a bias-free estimate and are evaluated infrequently.  The result can have multiple orders of magnitude less variance than previous work with similar number of density evaluations.}


  \renewcommand*{\figurePath}{figures/glass_figure}%
\newlength{\teaserLengthA}
\newlength{\teaserLengthB}
\begin{teaserfigure}
\centering
\setlength{\fboxrule}{5pt}%
\setlength{\tabcolsep}{.3pt}%
\setlength{\teaserLengthA}{2.4pt}
\setlength{\teaserLengthB}{1.2pt}
\renewcommand{\arraystretch}{0.0}%
\footnotesize%
\setlength{\beautyHeight}{5cm}
\begin{tabular}{crrcccc}
    &
    &
    &Ratio tracking
    &P-series CMF
    &Unbiased ray marching
    &Biased ray marching
    \\%
    &
    &
    &\cite{cramer1978application}
    &\cite{georgiev2019integral}
    &(ours)
    &(ours)
    \\[2pt]%
    \hspace{-2mm}%
    \begin{adjustbox}{valign=t}%
        \rotatebox{90}{%
            \makebox[\beautyHeight]{\textsc{Glass}}%
        }%
    \end{adjustbox}\,\,%
    &%
    \begin{adjustbox}{valign=t}%
        \includegraphics[height=\beautyHeight+\teaserLengthB]{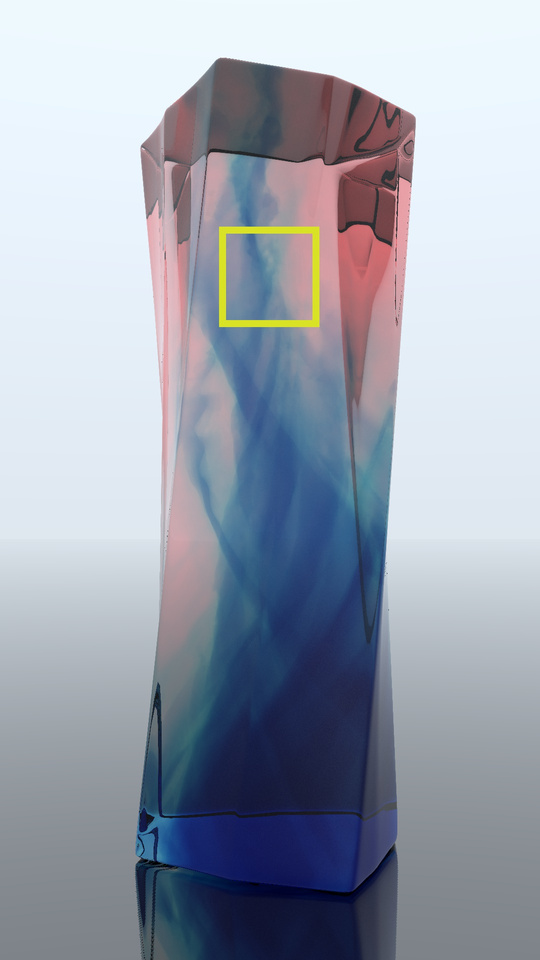}%
    \end{adjustbox}%
    \hspace{\teaserLengthA}\strut%
    &%
    \begin{adjustbox}{valign=t}%
        \rotatebox{90}{%
        	\makebox[\beautyHeight/2]{Variance}\hspace{\teaserLengthB}%
        	\makebox[\beautyHeight/2]{Inset}
        }%
    \end{adjustbox}%
        &%
        \begin{adjustbox}{valign=t}%
            \begin{adjustbox}{totalheight=1.0\beautyHeight+\teaserLengthB,tabular={c}}%
                \includegraphics[height=0.5\beautyHeight]{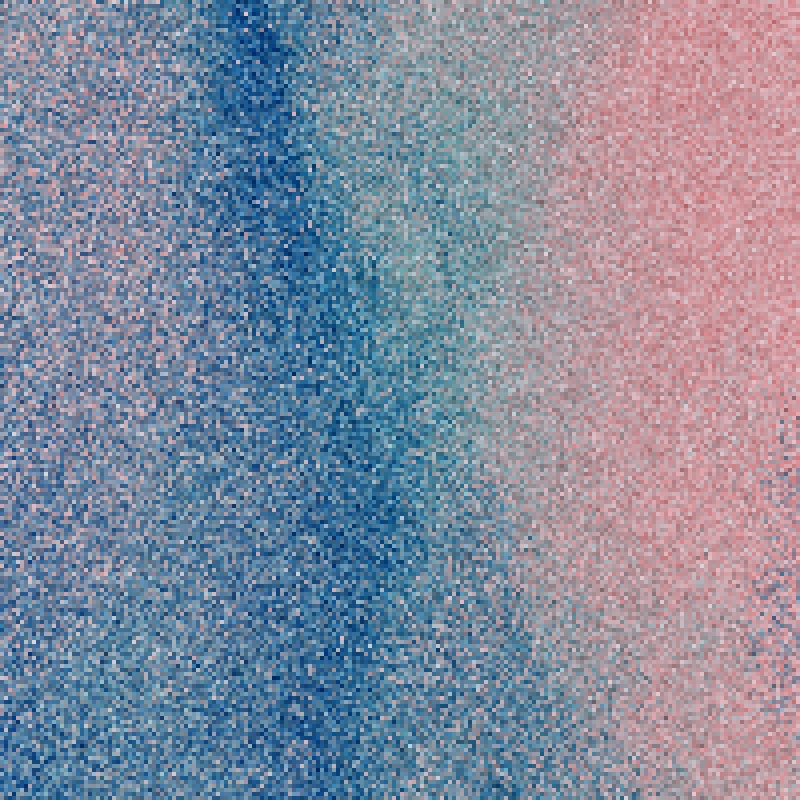}\vspace{\teaserLengthB}\\%
                \includegraphics[height=0.5\beautyHeight]{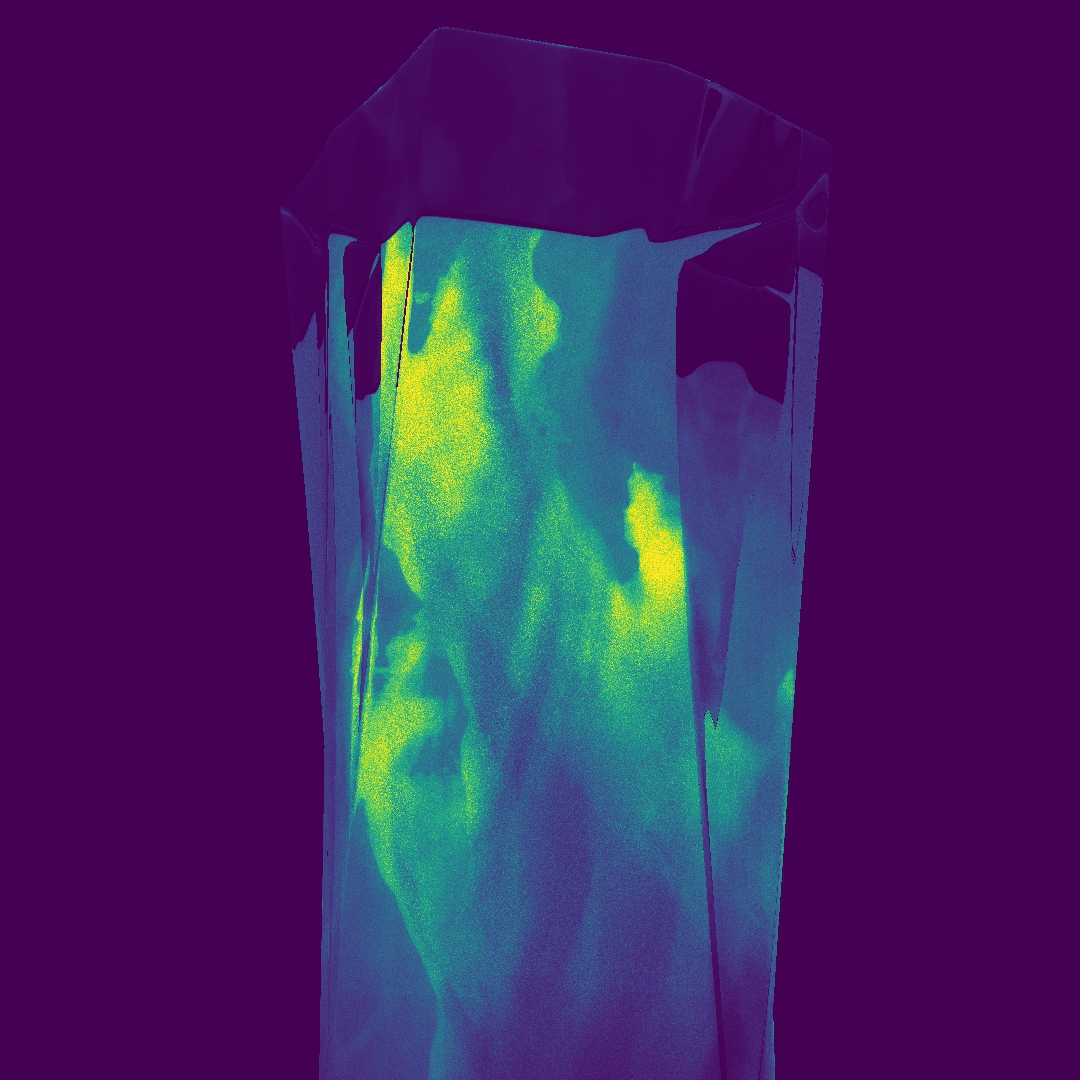}
            \end{adjustbox}%
        \end{adjustbox}%
        &%
        \begin{adjustbox}{valign=t}%
            \begin{adjustbox}{totalheight=1.0\beautyHeight+\teaserLengthB,tabular={c}}%
                \includegraphics[height=0.5\beautyHeight]{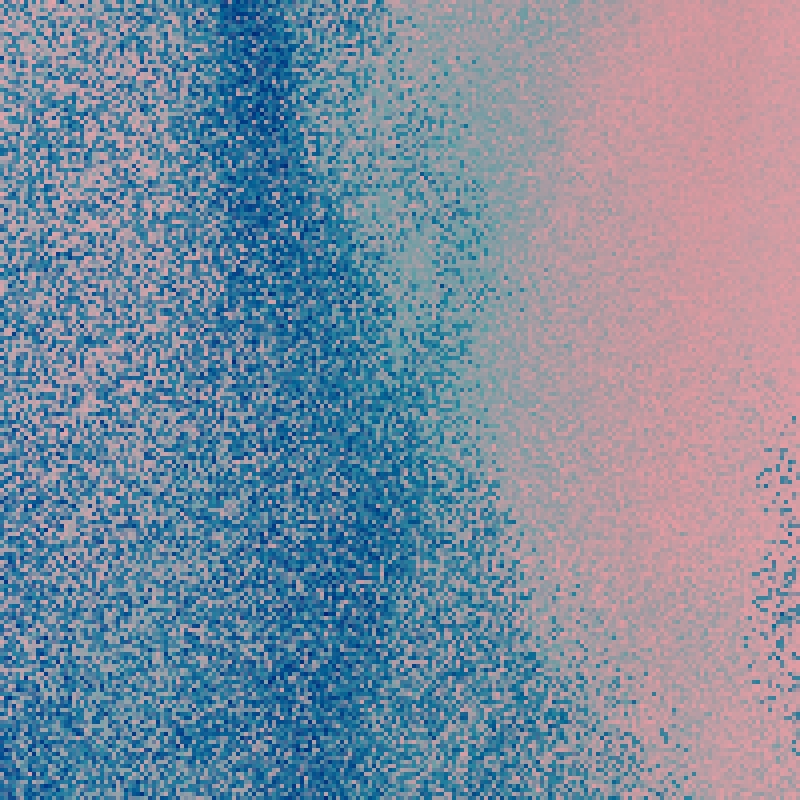}\vspace{\teaserLengthB}\\%
                \includegraphics[height=0.5\beautyHeight]{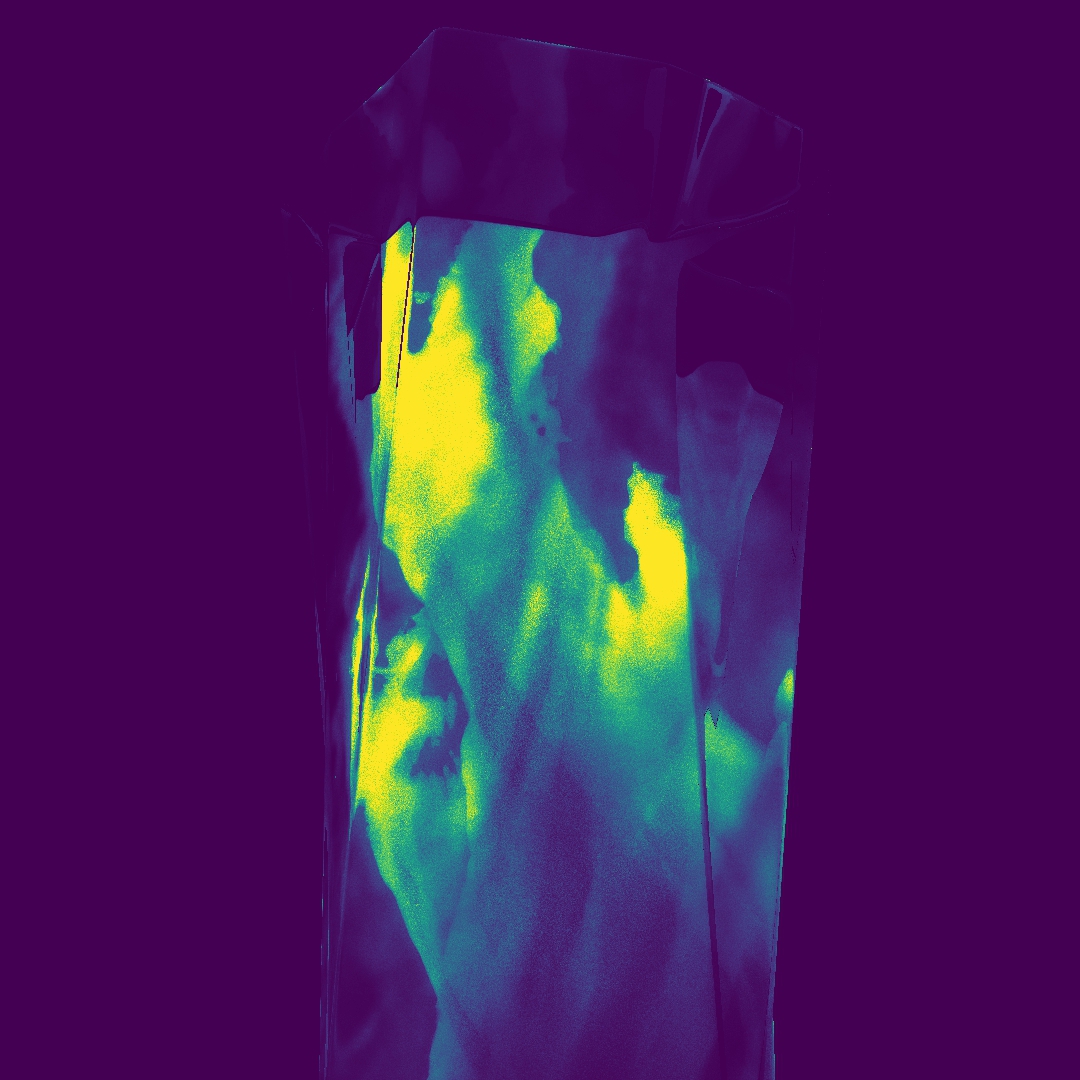}
            \end{adjustbox}%
        \end{adjustbox}%
        &%
        \begin{adjustbox}{valign=t}%
            \begin{adjustbox}{totalheight=1.0\beautyHeight+\teaserLengthB,tabular={c}}%
                \includegraphics[height=0.5\beautyHeight]{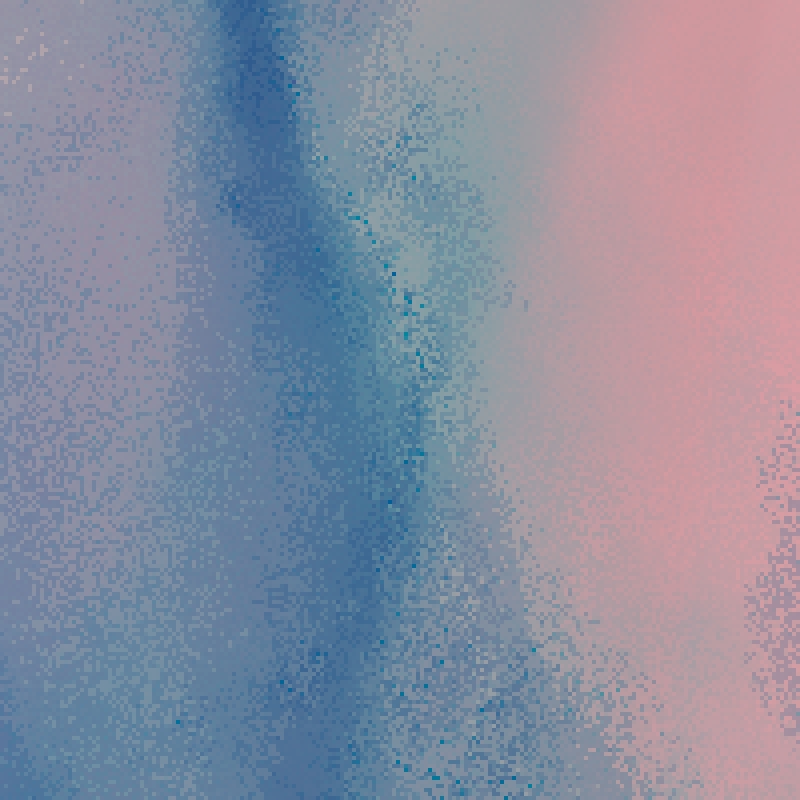}\vspace{\teaserLengthB}\\%
                \includegraphics[height=0.5\beautyHeight]{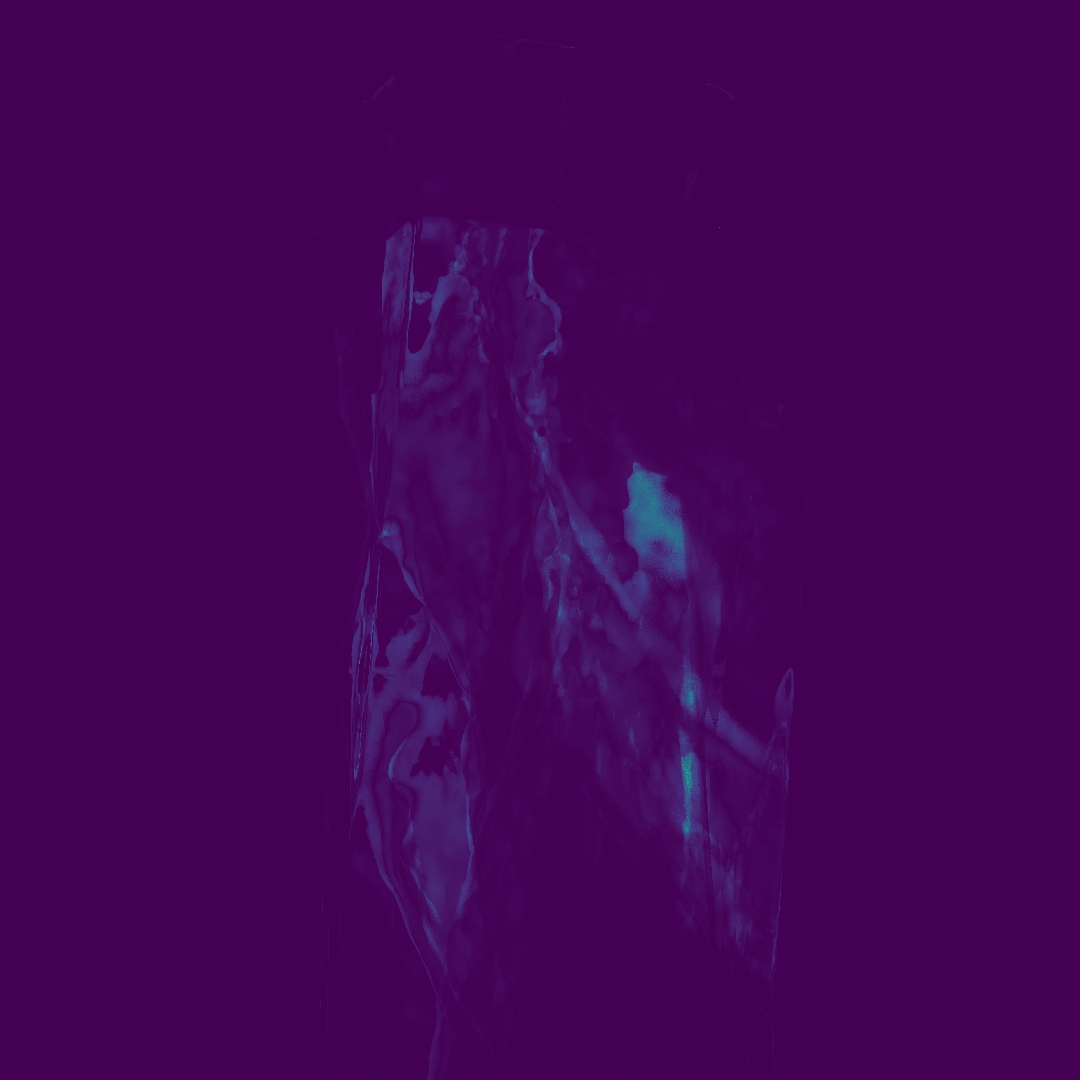}
            \end{adjustbox}%
        \end{adjustbox}%
        &%
        \begin{adjustbox}{valign=t}%
            \begin{adjustbox}{totalheight=1.0\beautyHeight+\teaserLengthB,tabular={c}}%
                \includegraphics[height=0.5\beautyHeight]{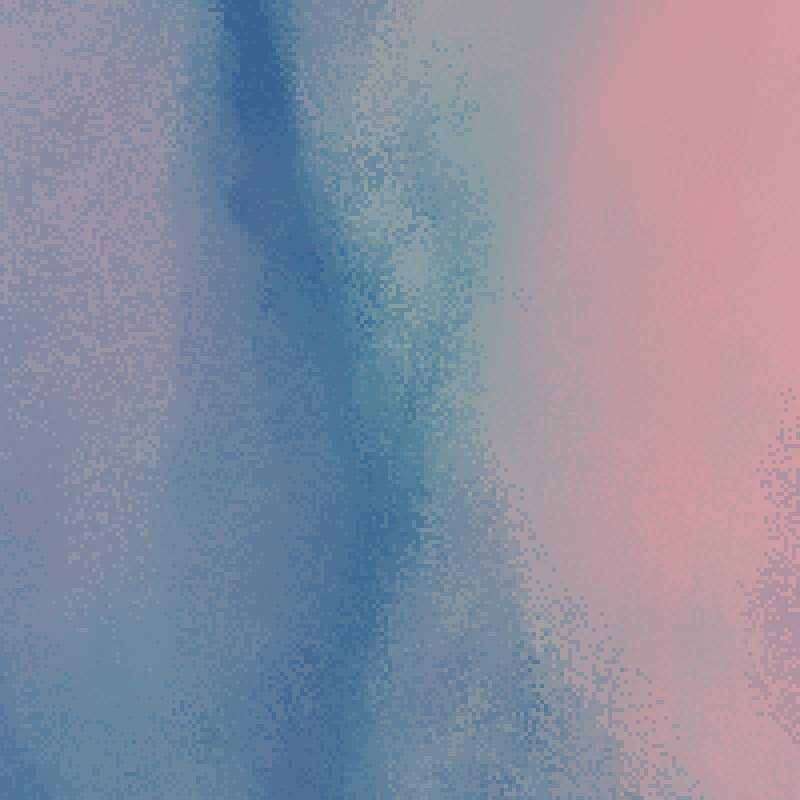}\vspace{\teaserLengthB}\\%
                \includegraphics[height=0.5\beautyHeight]{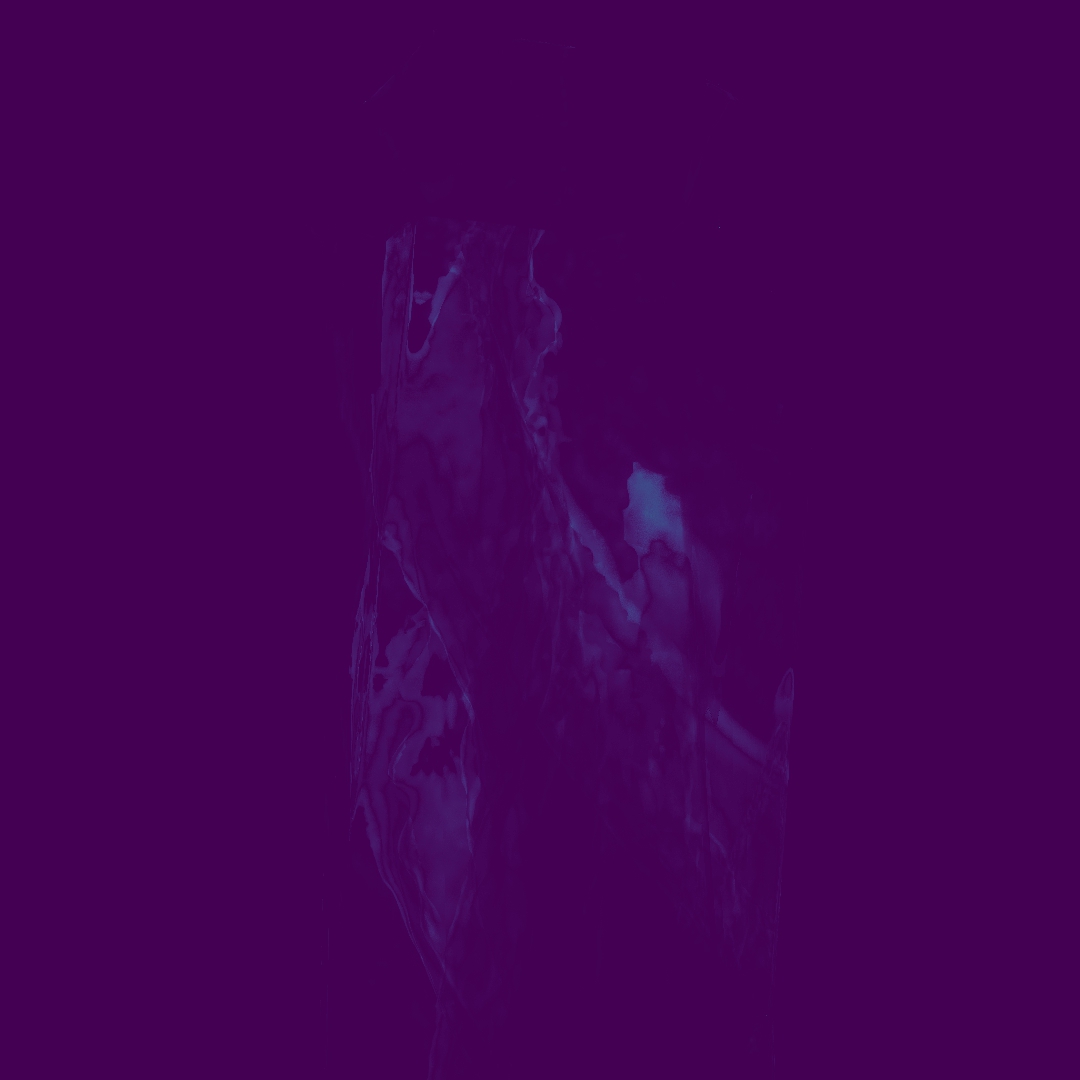}
            \end{adjustbox}%
        \end{adjustbox}%
        \vspace{1mm}
    \\%
    &Variance (MSE) at equal cost:\hspace{\teaserLengthB}
    &
            &1.52e-3
            &1.24e-3
            &9.49e-5
            &4.87e-5
        \\[1pt]
\end{tabular}
\caption{\glassFigureCaption}
\end{teaserfigure}


\maketitle

\section{Introduction}

    The visibility between two points in a scene is a fundamental quantity in light transport simulation.  In a vacuum, it takes on a binary value.  In a participating medium, however, scalar radiative transfer~\cite{chandrasekhar60} is used to statistically account for the presence of scattering and absorbing particles.  The number of particles intersecting a given ray is a random variable and visibility becomes a fractional quantity: the probability of traversing uncollided from $a$ to $b$,
    \begin{equation}\label{eq:T}
        \Trans(a,b) = \exp \left( -\int_a^b \ext(x) dx \right),  
    \end{equation}
    where $\ext(x)$ is a known deterministic non-negative function (the \emph{extinction coefficient} at position $x$).  The probability $T(a,b)$ is sometimes called \emph{transmittance}, and efficiently computing this value is essential for rendering scenes with haze, fog, and clouds.
    
    The integral in \autoref{eq:T} is rarely known in closed form.  Exceptions include piecewise-homogeneous volumes, and simple atmospheric models~\cite{novak18}.  The general-purpose approach, therefore, is to use Monte Carlo to estimate transmittance by point-sampling $\ext(x)$ at a number of locations $x$ along the ray.  A number of estimators have been proposed for this purpose, but no one estimator is optimal in all cases, and their efficiency depends on several parameters that are difficult to determine automatically.
    
    In this paper, we present new methods for unbiased estimation of \autoref{eq:T}.  After reviewing previous work in \autoref{sec:related}, we present a new parametric variance-analysis in \autoref{sec:variance} that reveals several key factors that limit the performance of these estimators.  This inspires a number of novel variance-reduction methods, which we detail in Section~\ref{sec:power-series}.  Combining all of these methods together, we propose a new estimator in Section~\ref{sec:power-series:assembly} that seems universally more efficient than prior work, and can in many cases yield transmittance estimates with orders of magnitude less variance at the same cost. 

     Our proposed estimator is based on a low-order Taylor series expansion of the exponential function near a relatively accurate estimate of the real optical depth obtained by multiple density lookups. The use of a low-order expansion frees up sampling budget for 
     more accurate evaluation of both the expansion point and the Taylor series terms, which further allows lowering the evaluation order and improving the samples. This self-reinforcing loop leads to an unbiased low-variance estimator that most of the time only evaluates the already quite accurate zeroth order term. The proposed evaluation of this term can be identified with the classical jittered ray-marching solution, whereas the remaining terms can be seen as probabilistically-sampled correction terms that make it unbiased, so we refer to our technique as \emph{unbiased ray marching}.
    
     \begin{table}
        \footnotesize
        \renewcommand{\arraystretch}{1.6}
        \begin{tabular}{ll}
        $x \in [a,b]$ & coordinate along a ray \\
        $\ext(x) $ & extinction coefficient \\
        $\maj(x)$ & majorant extinction \\
        $\extctrl(x) $ & control extinction coefficient \\
        $\extres(x) = \ext(x) - \extctrl(x) $ & residual extinction coefficient \\
        $\majres(x) = \maj(x) - \extctrl(x) $ & majorant residual extinction coefficient \\
        $\optd = \int_a^b \ext(x)\, dx$ & optical depth \\
        $\optdmaj = \int_a^b \maj(x)\,dx$ & majorant optical depth\\
        $\optdmajres = \int_a^b \majres(x)\, dx$ & majorant residual optical depth \\
        $\len = b - a$ & length of interval \\
        \end{tabular}
        \caption{Symbols.}
        \label{tab:symbols}
    \end{table}

\section{Background and Related Work}\label{sec:related}
    In this section we review the main approaches for transmittance estimation in light and particle transport literature and also identify new connections to work outside of transport theory.  For brevity, we will sometimes abuse the term ``density'' to mean the extinction coefficient $\ext(x)$ (which is the product of the number density of particles at $x$ with the total cross section), and the interval will sometimes be omitted (e.g.\ $\Trans$ refers to $\Trans(a,b)$).
    
    \subsection{Ray-marching and bias}
    Transmittance is the exponential of the negative optical depth,
    \begin{equation}\label{eq:optd}
        \Trans(a,b) = \exp \left(-\optd(a,b)\right) = \exp \left(-\int_a^b \ext(x) dx\right).
    \end{equation}
    The \emph{optical depth} $\optd$ can be easily approximated using \emph{ray marching} (uniformly-spaced samples along the interval) or by jittered and unbiased Monte Carlo approaches, but the exponential of these estimates will result in a biased estimator of $\exp(-\tau)$ ~\cite{raab2006unbiased}. The \emph{jackknife} method and its generalizations~\cite{miller1974jackknife} can be used to reduce the bias in some cases, but the error may still not be acceptable for certain applications. The key challenge of transmittance estimation, then, is to form unbiased estimates of $\Trans$ given only point samples of $\ext(x)$.  This relates more broadly to estimating a functional $\exp(-\lambda)$ when $\lambda$ is easily estimated in an unbiased fashion (see \citet{jacob2015nonnegative} for an extensive analysis of the challenges posed by general unbiased functional integration).
    
    \subsection{Poisson point processes}
    
    Many unbiased methods have been devised to estimate exponentiated integrals like \autoref{eq:T}, and these methods are closely related to the zero-order estimation problem for point processes. A \emph{point process} $N(\len)$ is a stochastic counting process of the number of events (such as particles) occurring in some time (or along a ray of length) $\len$.  For a \emph{Poisson} point process (PPP), the events are independent and $N(\len) \sim \text{Po}(\lambda_\len)$ is Poisson-distributed with \emph{rate} $\lambda_\len$ \cite{cox1966statistical}.  This rate is the integral of the \emph{intensity function} $\lambda(x)$ of the process over the interval
    \begin{equation}
        \lambda_\len = \int_0^\len \lambda(x) dx
    \end{equation}
    and allows the mean density of points to vary over the domain.
    It is well known that this PPP is exactly the process governing the scattering and absorption events in classical radiative transfer~\cite{cox1966statistical,mikhailov1992optimization}, due to the assumption of independent scattering centers.  The correspondence between the two is established by equating the rate of the point process $\lambda(x)$ to the extinction coefficient of the medium $\ext(x)$ as the particle moves across the interval when starting from $a$, $\lambda(x) = \ext(a + x)$.  Transmittance is then the probability of finding no points/particles along the interval
    \begin{equation}
        T(a,b) = \Prob\left[ N(\len) = 0 \right], \quad \len = b-a.
    \end{equation}
    Since the mean of a PPP is the rate $\lambda_\len = \E[N(\len)] = \optd(a,b)$, the exponential free paths of classical radiative transfer follow from the zero-order probability of the Poisson distribution (the probability mass function of a Poisson distribution with rate $\optd$ is $e^{-\tau } \tau ^k / k!$, 
    which is an exponential for $k = 0$).

    \subsection{Tracking estimators}
    The most well-known unbiased transmittance estimators are called \emph{tracking estimators} due to the fact that they track a particle moving from $a$ to $b$ by sampling a PPP to determine an ordered sequence of collisions with the medium.  For a constant-density medium, the exponentially-distributed free-path lengths between collisions are easily sampled~\cite{novak18}.  For a nonhomogeneous medium, the PPP can be sampled using the method of \emph{delta-tracking} \cite{Butcher58,Zerby61,Bertini63,woodcock1965techniques,Skullerud68, coleman1968mathematical,mikhailov1970method,galtier:2013}. Using a \emph{majorant} $\maj(x) \ge \ext(x)$, a denser process is sampled whose rate/optical-depth is easily computable
    \begin{equation}
        \optdmaj(a,b) = \int_a^b \maj(x) dx.
    \end{equation}
    A rejection process is then used to thin the denser process down to the desired result whereby each sampled point $x_i$ is kept with probability $\ext(x_i) / \maj(x_i)$.  This rejection embodies the fictitious/null collision concept of the transport literature.  We note that this is equivalent to a method in the point process literature known as \emph{thinning} \cite{pasupathy2010generating} (identification of this correspondence appears to be new).  The earliest use of either method would appear to be attributed to von Neumann shortly after the war, according to \citet{carter72}.
    
    While the majorant $\maj$ is often a constant, efficiency of delta-tracking is improved with a majorant that more tightly bounds the target density.  Piecewise-linear \cite{klein1984time} or piecewise-polynomial \cite{szirmay2011free} majorants can be efficiently sampled.  For a general survey of methods for sampling nonhomogeneous PPPs, see \cite{pasupathy2010generating}.
    
    Somewhat remarkably, without knowing $\optd$, delta-tracking can sample the number of collisions $N$ from the Poisson distribution $N(\len) \sim \text{Po}(\optd)$.  Given $n$ samples of $N(\len)$ with mean $\bar{v}$, the minimum-variance unbiased estimator for transmittance (given only $\bar{v}$) is \cite{johnson1951estimators}
    \begin{equation}\label{eq:T:johnson}
        \widehat{\Trans}_{\mathrm{J}} = \left(1-\frac{1}{n}\right)^{n \bar{v}}. 
    \end{equation}
    The single-sample ($n = 1$) form of this estimator produces (assuming $0^0 = 1$) the delta-tracking\footnote{Also known as the track-length transmittance estimator~\cite{georgiev2019integral}} transmittance estimator \cite{cramer1978application,novak18}, which returns a binary estimate depending upon whether or not $N = 0$.  The case $n > 1$ provides an interesting generalization of this estimator (which we call \emph{Johnson's estimator}) and, to the best of our knowledge, it has not been applied to light transport.  While this estimator is optimal (given only $\bar{v}$), in practice it can be improved upon by using the sampled densities $\ext(x_i)$ directly.
    Another related estimator had been proposed by \cite{raab2006unbiased}, obtained by averaging together $n$ partially stratified delta-tracking estimates.
    
    \subsubsection{Ratio tracking}
    
    Weighted tracking on a line \cite{cramer1978application} (also known as \emph{ratio tracking} in graphics~\cite{novak2014residual}) applies an expected-value optimization to the $n=1$ delta-tracking estimator to form a product of ratios of densities (\emph{null density} $\nul(x) = \maj(x) - \ext(x)$ to total density $\maj(x)$).  This is closely related to a distance-sampling scheme known as weighted delta tracking (see e.g.\ \citet{galtier:2013} or \citet{legrady2017woodcock}).
    Like delta tracking, a majorant PPP samples $N \sim \text{Po}(\optdmaj)$ points $x_i$ in the interval $(a,b)$.  Instead of returning $0$ as soon as a real particle is sampled, the ratio-tracking estimator imparts a fractional opacity to each sampled particle based on its probability of being fictitious,
    \begin{equation}\label{eq:RT}
        \widehat{\Trans}_{\mathrm{rt}} = \prod_{i=1}^N \left(1 - \frac{\ext(x_i)}{\maj(x_i)}\right) = \prod_{i=1}^N \frac{\nul(x_i)}{\maj(x_i)}.
    \end{equation}
    Ratio tracking outperforms delta tracking in most cases.  However, delta tracking can use early termination after the first real particle is sampled and avoid many unnecessary density evaluations.  Therefore it can be beneficial to switch to delta tracking after the running product in \autoref{eq:RT} goes below some threshold \cite{novak2014residual}.
    
    \subsection{Control variates}
    \label{section:bg-control-variate}
    A common theme in transmittance estimation is the utilization of auxiliary density functions (null-collision density, control density, etc.). While these auxiliary functions can serve different purposes, for example to facilitate sampling of collisions and/or to reduce variance, they (or the combination of them) can be interpreted as a control variate (CV) \cite{georgiev2019integral}.
    Given an analytically integrable control variate $\extctrl(x)$ with $\optdctrl=\int_a^b \extctrl(x)\,\dif x$,
    the optical depth integral can be rewritten as\footnote{In standard literature the CV and its integral are typically weighted by a coefficient that controls the strength of applying the CV. 
    Since we design our CVs heuristically with the goal of maximizing positive correlations, we simply absorb the scaling factor into the CV for brevity.}
    \begin{align}
        \optd(a,b) = \optdctrl(a,b) + \int_a^b \ext(x) - \extctrl(x) \dif x.
    \end{align}
    We will refer to $\extctrl(x)$ and $\extres(x) = \ext(x) - \extctrl(x)$ as the \emph{control} and \emph{residual} extinction coefficients, and to their respective integrals $\optdctrl$ and $\optdres$ as the control and residual optical depth.  
    The \emph{majorant residual} coefficient $\majres(x) = \maj(x) - \extctrl(x)$ and optical depth $\optdmajres = \optdmaj - \optdctrl$ follow.
    The transmittance is then
    \begin{equation}\label{eq:CV:xform}
        \Trans(a,b) = \Transctrl(a,b) \, \Transres(a,b) = \exp\left({-\optdctrl}\right) \exp \left( -\int_a^b \extres(x) dx \right).
    \end{equation}
    This transformation can dramatically reduce variance, particularly when the control closely matches the true density.
    
    \subsubsection{Residual ratio tracking / Poisson estimator}
    
    The first application of control variates in transmittance estimation was with ratio tracking to produce the \emph{residual ratio tracking} (RRT) estimator~\cite{novak2014residual}.  Applying the transformation in \autoref{eq:CV:xform}, the estimator reads
    \begin{equation}\label{eq:RRT}
        \widehat{\Trans}_{\mathrm{rrt}} = \exp\left({-\optdctrl}\right) \prod_{i=1}^N \left(1 - \frac{\extres(x_i)}{\majres(x_i)}\right).
    \end{equation}
    where $N \sim \text{Po}(\optdmajres)$ points $x_i$ in the interval $(a,b)$ are generated by sampling a PPP with residual
    intensity $\lambda(x) = \maj(x) - \extctrl(x) \ge 0$. 
    While different articles may propose different approaches for setting the residual intensity $\lambda(x)$ of the PPP, this estimator is conceptually equivalent to the one known as the \emph{Poisson estimator}~\cite{beskos2006exact, fearnhead2008particle, papaspiliopoulos2011monte,chen2012brownian,jacob2015nonnegative,jonsson2020direct}, first presented by Wagner~\shortcite{wagner1987unbiased}.
    
    Concurrently, \citet{jonsson2020direct} have also connected the ratio tracking and Poisson estimator literature and proposed several new variants of RRT that use online estimation of a constant control.  We also propose online control estimation, but include additional variance-reduction techniques such as comb-filtering.  We then apply this idea to a power-series formulation, which improves performance and naturally includes ray marching as a biased member of the general formalism.

    \subsection{Power-series formulation}
    
    Another family of unbiased transmittance estimators follows from a power-series (Taylor) expansion of the exponential in \autoref{eq:T}.
    This approach has been suggested as early as~\cite{cameron1954generalized} and has been used in estimation problems involving transformed observations~\cite{neyman1960correction}.  One such form, which is used to estimate the exponential of the Hamiltonian in particle physics Markov-chain Monte Carlo simulations, due to Bhanot and Kennedy \cite{bhanot1985bosonic, wagner1987unbiased,wagner1988monte,lin2000anoisy}, can be applied directly to transmittance estimation.  Related applications of this idea to transmittance estimation were independently presented by several authors \cite{longo2002direct,elhafi2018three,georgiev2019integral,jonsson2020direct}.  Similar work has been proposed by \cite{lyne2015russian} in the context of Bayesian inference.
    
    Georgiev et al. \shortcite{georgiev2019integral} first introduced this formulation to computer graphics, showing how it can in fact be seen as a very general framework for expressing and analysing all transmittance estimators.
    Following their derivation ~\citep[Equations (15) and (16)]{georgiev2019integral}, transmittance (\ref{eq:T}) can be expressed as:
    \begin{align}\label{eq:T:pseries}
        \Trans(a,b)
        &= \sum_{k=0}^\infty \frac{\left(-\optd\right)^k}{k!} 
        = \sum_{k=0}^{\infty}\frac{1}{k!} \prod_{i=1}^k \left(-\int_a^b \ext(x)\,\dif x \right) \nonumber\\
        &= 1 - \frac{\optd}{1!} + \frac{\optd^2}{2!} - \frac{\optd^3}{3!} + \cdots \, .
    \end{align}
    Monte Carlo estimation is then applied to each integer power of the optical depth $\optd^k$ in the expansion.  This is typically\footnote{See the appendix of \citet{glasser1962minimum} for an interesting alternative.} 
    achieved by using $k$ numerical estimates $\{\estoptd_1, \dots ,\estoptd_k\}$ of the negative optical depth.
    As long as $\{\estoptd_1, \dots ,\estoptd_k\}$ are \emph{independent} and \emph{unbiased}, i.e.\:
    \begin{align}
        \E[\estoptd_i] = -\optd(a,b) = -\int_a^b \ext(x) \,\dif x\, ,
    \end{align}
    it follows that their product provides an unbiased estimate of the $k$-th power of $-\optd$ (we drop the index since all $X_i$ have the same expectation):
    \begin{align}
        \E{\left[\prod_{i=1}^k \estoptd_i\right]} = \prod_{i=1}^k \E{\left[\estoptd_i\right]} = \E{\left[\estoptd\right]}^k  = \left(-\optd\right)^k.
    \end{align}
    This observation allows formulating the transmittance function as the series of products of unbiased, independent estimates of (negative) optical depth:
    \begin{align}
        \label{eq:T:pseries-estimator}
        \Trans(a,b) = e^{\E[\estoptd]} = \sum_{k=0}^\infty \frac{1}{k!} \E{\left[\prod_{i=1}^k \estoptd_i\right]}.
    \end{align}
    Various estimators then follow from estimating random finite portions of this expansion with the appropriate weight corrections (explained below).
    
    The above derivation highlights the importance of using independent and unbiased estimates of $\optd$ within a single term $\optd^k$ of the power series. Correlations across the terms of the sum, however, are perfectly acceptable. In fact, high-order terms are typically computed from the low-order ones using the \emph{recursive formulation} \cite{bhanot1985bosonic,georgiev2019integral}
    \begin{equation}\label{eq:pseries:recursive}
        T(a,b) = 1 - \frac{\optd}{1} \left(1 - \frac{\optd}{2} \left( 1 - \frac{\optd}{3} \left( \dots \right. \right. \right. ,
    \end{equation}
    which has been used in \autoref{eq:T:pseries-estimator}.
    
    A control variate $\extctrl(x)$ is often applied to \autoref{eq:T:pseries-estimator} to yield
    \begin{align}
        \label{eq:T:pseries-estimator:pivot}
        \Trans(a,b) = \Transctrl(a,b) \, \Transres(a,b)
        &= e^{-\optdctrl} e^{-\optdres} \nonumber\\
        &= e^{-\optdctrl} \sum_{k=0}^\infty \frac{1}{k!} \prod_{i=1}^k  \left(-\int_a^b  \extres(x) \,\dif x \right) \nonumber\\
        &= e^{-\optdctrl} \sum_{k=0}^\infty \frac{1}{k!} \E{\left[\prod_{i=1}^k \estoptdres_i\right]},
    \end{align}
    where $\estoptdres_i$ are unbiased, independent estimates of (negative) residual optical depth.  \citet{galtier:2013}, \citet{elhafi2018three} and \citet{georgiev2019integral} proposed to set the control variate to a strict majorant, $\extctrl(x) = \maj(x)$, to avoid sign oscillations in power-series estimates of $\Transres(a,b)$. Based on our analysis from Section 3, we will revisit this decision and propose a new way of setting the control variate in Section 4.
	Notice how $-\optdctrl$ effectively acts as a pivot for the Taylor series expansion. This interpretation is central to our investigations and we will refer to $-\optdctrl$ as the \emph{pivot} in the rest of the text.

    \subsubsection{Numerical evaluation}
    In practice, the evaluation of the infinite power series needs to be limited to sampling a finite number of terms; 
    \citet{georgiev2019integral} have shown that virtually all unbiased transmittance estimators can be ultimately related to sampling this power series expansion.
    In this respect, existing unbiased estimators can be classified into two broad categories.

    \paragraph{Single-term estimation}
    \citet{georgiev2019integral} showed that the delta-tracking and ratio-tracking estimators can be described in the power-series formulation by noting that these estimators estimate a single term in \autoref{eq:T:pseries} at a time; when $N$ points are sampled by the $\maj$-driven PPP, these estimators estimate $(-\tau)^N/N!$.  The general form of the single-term power-series estimator is called the \emph{generalized Poisson estimator} \cite{fearnhead2008particle},
    \begin{align}\label{tracking:general-estimator}
        \widehat{\Trans}_{\mathrm{single}} = e^{-\optdctrl}  \frac{1}{N! P(N)} \prod_{i=1}^N \estoptdres_i\, ,
    \end{align}
    where $P(N)$ is the probability mass function of $N$.  Using delta-tracking results in a Poisson distribution, $N \sim \text{Po}(\optdmajres)$, but other distributions can be used~\cite{fearnhead2008particle,jonsson2020direct}.  For the standard delta-tracking estimator, the random variable $\estoptdres$ is replaced by a $\extctrl$-weighted Bernoulli random variable and the power-series derivation of this estimator is a special case of a more general derivation \cite[appendix]{glasser1962minimum}.
    
    \paragraph{Truncated-series estimators}
    The recursive power-series relation in \autoref{eq:pseries:recursive} directly produces a \emph{truncated-series} estimator that estimates all terms in the Taylor expansion up to and including $\tau^N$.  If $N$ is a discrete random variable and $Q(k) = \Prob[N \geq k]$, then the truncated estimator for \autoref{eq:T:pseries-estimator:pivot} is:
    %
    \begin{align}\label{pseries:general-estimator}
        \widehat{\Trans}_{\mathrm{trunc}}
        = e^{-\optdctrl} \sum_{k=0}^N \frac{1}{k! Q(k)} \prod_{i=1}^k \estoptdres_i\,.
    \end{align}
    Instead of selecting $N$ from a Poisson process, Russian roulette is commonly employed and $Q(k)$ becomes the product of the continuation probabilities.
    
    \subsubsection{Bhanot \& Kennedy roulette}
    \label{sec:BK-roulette} 
        A useful scheme where the expansion is \emph{always} evaluated up to order $K$ and then terminated using term-wise roulette decisions follows from the equivalence (which we generalize here)  \cite{bhanot1985bosonic}
        \begin{equation}\label{eq:bhanot}
            e^x = \sum_{k=0}^K \frac{x^k}{k!} + \frac{\g}{K+1}\left( \frac{x^{K+1}}{\g K!} + \frac{\g}{K+2} \left( \frac{x^{K+2}}{\g^2 K!} + \dots \right. \right. \, .
        \end{equation}
        Here $\g$ is a roulette control parameter restricted to $0 < \g < K + 1$ and $\g / (K+k)$ is the probability of expanding from order $K+k-1$ to order $K+k$.  Bhanot and Kennedy proposed using a continuous \emph{expansion parameter} $c > 0$, setting $K = \myfloor{c}$, which we will refer to as the \emph{Bhanot \& Kennedy} (BK) estimator.
        The BK roulette, specifically the parameter $c$, provides a very explicit control over the cost of the estimator. 
        
        Independently, \citet{georgiev2019integral} proposed the \emph{p-series CMF} estimator that sets $\g = \optdmaj$ to the majorant and selects $K$ such that $99\%$ of the majorant cumulative mass function (CMF) is accumulated (assuming $\optdmaj$ is a safe and reasonable guess for the true optical depth when selecting $K$).

    \subsection{Additional related work}
    
    Delta tracking and ratio tracking each have variations known as the \emph{next-flight} estimators \cite{cramer1978application, novak18} that fall somewhat in between the tracking and truncated series forms.  Also, \citet{georgiev2019integral} introduced a number of additional estimators, including the p-series cumulative estimator that employs a different roulette strategy than described above, but concluded the p-series CMF was best overall.  
	We refer the reader to these works for further details.  
    
    In this work we use multiple correlated density evaluations per estimate of optical depth, which was mentioned by \citet{georgiev2019integral} but, to the best of our knowledge, has not been applied before.

\section{Efficiency Analysis}\label{sec:variance}

In this section, we investigate the efficiency of single-term and truncated power-series estimators.  We measure the sensitivity of each estimator's variance to various factors, which leads to key insights that inform the design of new estimators.  

\subsection{Efficiency and cost}
\begin{figure*}
        \centering
        \includegraphics[scale=0.3]{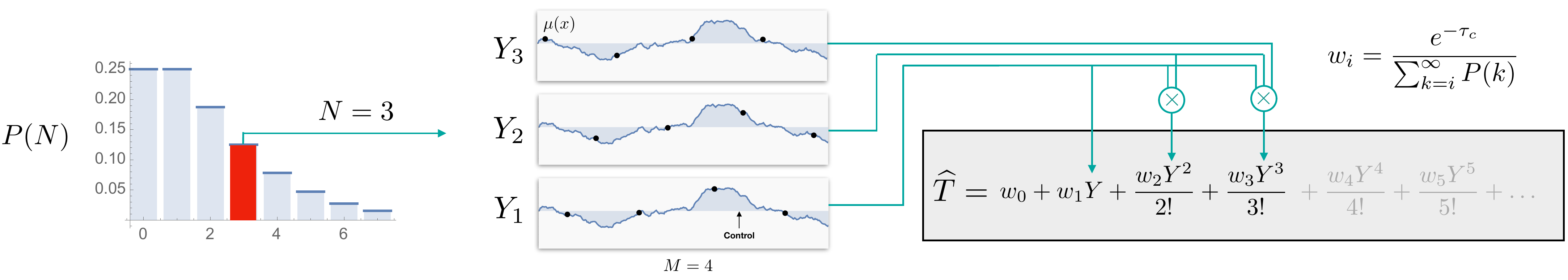}
        \vspace{-.1in}
        \caption{Transmittance estimation via truncated power series using a constant control variate and multiple ($\M = 4$) density evaluations per estimate $Y_i$ of the negative residual optical depth.}
        \label{fig:estimator:diagram}
    \end{figure*}

Following prior work we define the \emph{efficiency} of an estimator to be the reciprocal of variance times cost,
\begin{equation}
    \Eff[\widehat{T}] = \frac{1}{\Var[\widehat{T}] \Cost[\widehat{T}]}
\end{equation}
where $\Cost[\widehat{T}]$ is the mean number of density $\ext(x)$ evaluations. For both single-term and truncated power-series estimators, this will depend on $N$: the number of unbiased estimates of negative residual optical depth ($\estoptdres_i$) needed to estimate a subset of the power series in \autoref{eq:T:pseries-estimator:pivot}.  By abandoning the physical picture of tracking estimators, the power series formulation permits a new parameter $\M$ that we call the \emph{query size}, which is the number of density evaluations per estimate $\estoptdres_i$ (see \autoref{fig:estimator:diagram}).  The estimator for $Y$ is then
\begin{equation}\label{eq:Y:estimator}
    \widehat{Y} = -\frac{1}{\M} \sum_{i=1}^{\M} \frac{\extres(x_i)}{p(x_i)} = -\frac{1}{\M} \sum_{i=1}^{\M} \frac{\ext(x_i)-\control(x_i)}{p(x_i)}
\end{equation}
where $p(x)$ is the density for sampling $x \in (a,b)$ and total cost is $\Cost[\widehat{T}] = \E[N] \cdot \M$.

The efficiency of a given estimator will depend on a number of parameters that can be adjusted: the control variates, the query size $\M$, and (in the case of BK roulette) power-series expansion parameter $c$.  Ideally, an automatic procedure would optimally configure these parameters given only limited knowledge of the density statistics in the scene.  To discern more about how this could be achieved, we need a detailed picture of how these parameters influence variance.  While it is known that the efficiency of residual ratio tracking improves with increasing majorant~\cite{georgiev2019integral} and basic heuristics for setting control variates have been discussed~\cite{novak2014residual,jonsson2020direct}, to the best of our knowledge, little to no detailed investigation of query size $\M$ or expansion parameter $c$ has been presented for any truncated estimator. 

\subsection{Roulette variance in the uniform medium}
Ultimately, variance in a transmittance estimator will arise due to two factors, which we will call $Y$-\emph{variance} and \emph{roulette variance}, and we will show that they are in fact weakly coupled.  Transmittance estimators are random functions $f_{\widehat{T}}(\estoptdres_1,\dots,\estoptdres_N)$ of $N$ random variables $Y_i$.  By $Y$-\emph{variance}, we mean the variance in the optical depth estimates $Y_i$ themselves, which leads to variance in $\widehat{T}$ upon insertion into $f_{\widehat{T}}$. To better understand the influence of $Y$-variance, we can turn it off by considering a uniform medium and uniform sampling $p(x_i) = 1 / \len$. 
The only variance that remains is then due to $N$ being a random variable causing $f_{\widehat{T}}$ to evaluate different portions of the power series.  This variance arises due to the roulette scheme of a truncated estimator (or PPP sampling for a single-term estimator), and we call it \emph{roulette variance}.

\begin{figure}
        \centering
        \includegraphics[width=0.99\linewidth]{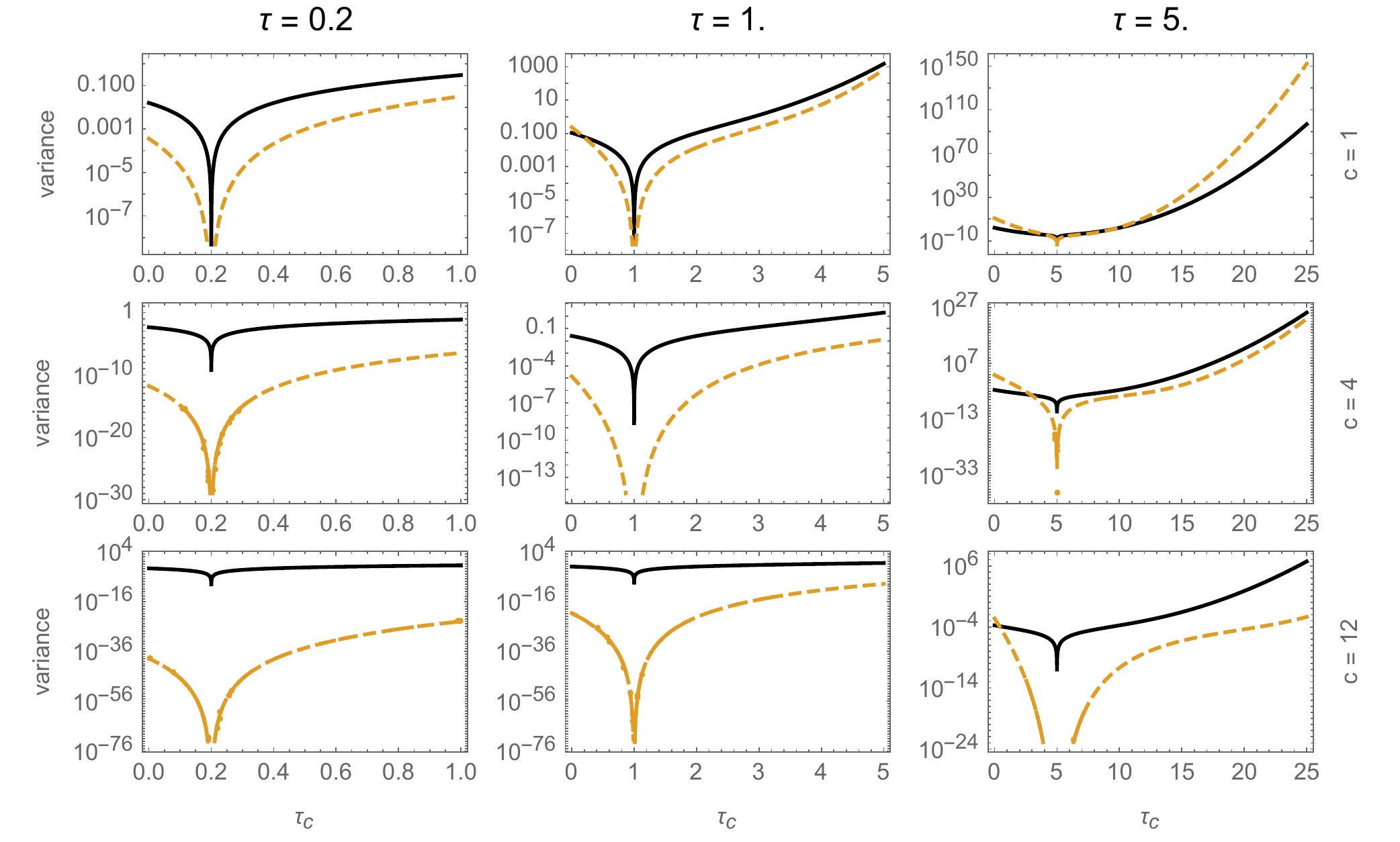}
        \vspace{-.12in}
        \caption{Variance of cost-matched RRT (\emph{black})  and BK (\emph{dashed}) estimators for uniform media as a function of negative pivot $\optdctrl$ in three configurations of optical depth $\optd$ and expansion parameter $c$.
        \label{fig:BK:RRT:var:compare}}
    \end{figure}

In \autoref{fig:BK:RRT:var:compare} we compare the roulette variance of single-term and truncated power-series estimators.
We use residual ratio tracking, with known variance ($\autoref{eq:var:RRT}$), as the single-term estimator and the Bhannot \& Kennedy estimator to represent truncated power-series; we derive the variance of the BK estimator in Appendix~\ref{sec:var:BK:constdens}.
In each plot, the RRT rate is cost-matched to the BK estimator by adjusting the majorant $\maj$ such that $\optdmajres = \optdmaj - \optdctrl = \len (\maj - \control) = \E[N_{BK}]$, where $\len$ is the length of the estimation interval (see also \autoref{eq:cost:BK}). The most efficient estimator is the one with the lowest variance.  We observe two important trends as the pivot $-\optdctrl$ is varied;  First, we can achieve arbitrarily low variance by moving the negative pivot close to the true optical depth of the medium ($\optdctrl = \optd$).  Second, when the pivot is near this optimal value, the truncated (BK) estimator is universally better than the single-term estimator (RRT).

This analysis hints at the possibility of finding a single transmittance estimator that performs best in all cases (truncated), but also highlights the need for an accurate pivot estimator in order to achieve this.  It is known that ratio tracking can outperform truncated estimators in some cases~\cite{georgiev2019integral} and we see this again here for the uniform medium: when the pivot is far from its optimal value, the truncated estimator sees a significant explosion of variance, while the single-term estimator sees far less. This issue is lessened by estimating more of the series via the expansion parameter $c$, but at a cost of more density evaluations. Increasing $c$ also widens the performance gap between the two estimators (\autoref{fig:BK:RRT:var:compare}, bottom row) and extends the range of pivots where the truncated estimator is better than the single-term.

\begin{figure}
        \centering
        \includegraphics[width=0.99\linewidth]{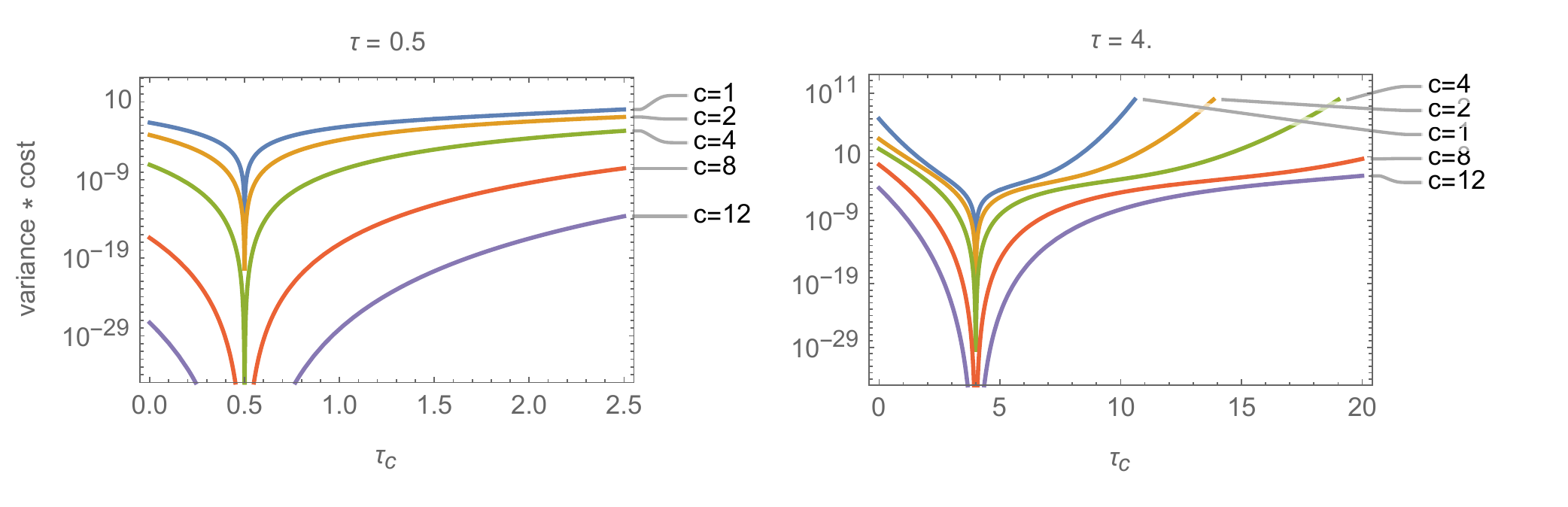}
        \vspace{-.12in}
        \caption{Inverse efficiency of the BK estimator in a uniform medium as a function of negative pivot $\optdctrl$ in two configurations of optical depth $\optd$ and five configurations of expansion parameter $c$.}
        \label{fig:roulette-variance}
    \end{figure}

\autoref{fig:roulette-variance} further demonstrates how the expansion parameter $c$ influences the efficiency of power-series estimators by plotting inverse efficiency.  We see that increasing $c$ widens the range of pivots where high efficiency can be obtained.  It also shows a universal trend shared with RRT: increasing $\E[N]$ monotonically improves the overall efficiency, regardless of the pivot.

    \subsection{Varying media}
    
    While rays passing through uniform density are common in practice (especially empty space with $T = 1$), the analysis above is not representative of the full picture.  We now introduce increasing amounts of $Y$-variance to observe how and when the total variance changes.
    
\begin{figure}
        \centering
        \includegraphics[width=0.99\linewidth]{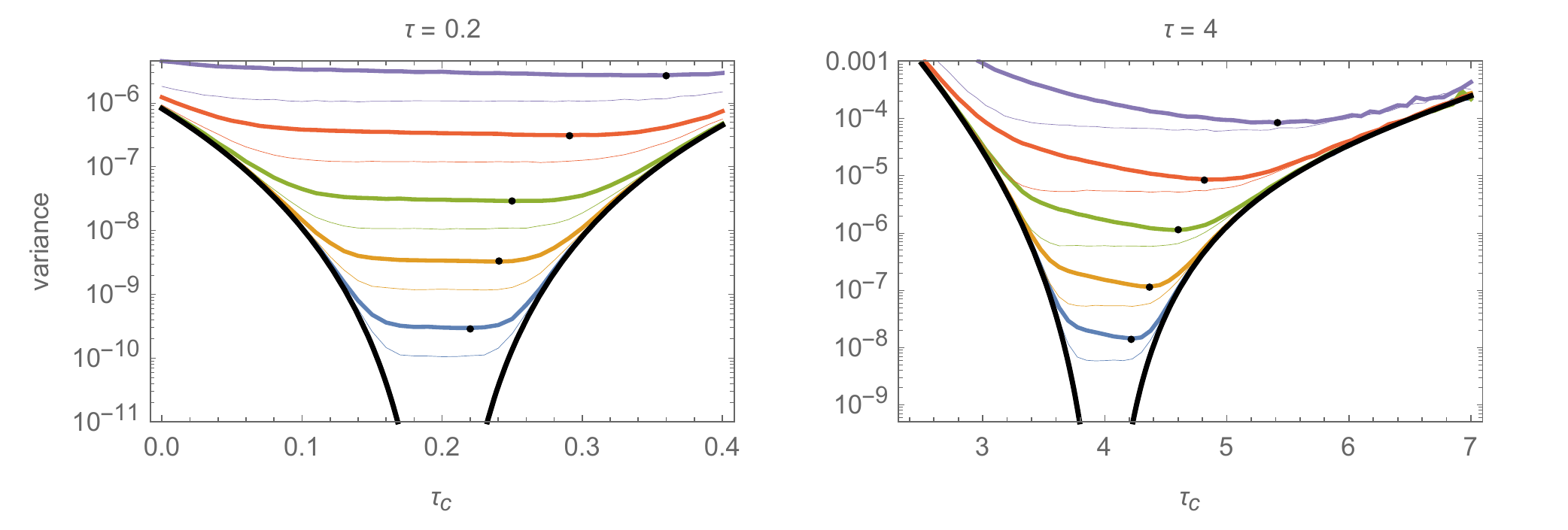}
        \vspace{-.12in}
        \caption{Variance of the BK (thick) and U-BK (thin) estimators ($K = c = 2$) for two different optical thicknesses.  The uniform medium (black) is compared to five different levels of density fluctuations (colored). The pivot with the lowest variance for the BK estimator is indicated by black dots and shifts to the right with increasing density fluctuations, making the optimal pivot difficult to predict.  Application of U-statistics always reduces the variance and also widens the range of pivots where the minimum variance is achieved.}
        \label{fig:BK:general-variance}
    \end{figure}

    In \autoref{fig:BK:general-variance} we compare the variance of the truncated estimator as different amounts of fluctuation in the density $\ext(x)$ are introduced while preserving the mean of $\ext(x)$.  Each thick colored line corresponds to a different amount of fluctuation in $\ext(x)$ (thereby introducing $Y$-variance).  The uniform medium (pure roulette variance) is shown in black for reference.  This comparison is comprehensive in that,
    like with ratio tracking, it follows from the power-series formulation that the variance of the BK estimator is purely a function of the mean and variance of the $Y$ estimates (together with the pivot value)---the exact profile of the density fluctuations is irrelevant (this is because the BK roulette is independent of $Y_i$).  
    
    We find that the variance of the BK estimator is dominated by either $Y$-variance or roulette variance: \emph{they are weakly coupled}. Far from the optimal pivot (where the black curves merge with the rest) the variance is essentially the same as that of a medium with constant density, so increasing $\M$ will have little impact. 
	Conversely, no matter how good the pivot, variance in the samples $Y_i$ limits the minimum-achievable variance. Further, as $Y$-variance decreases (by increasing $\M$, say), the pivot needs to be closer to the optimal value to avoid roulette variance limiting the gains (to stay inside the black curves in \autoref{fig:BK:general-variance}), suggesting that the sample budget of any online pivot estimation should be positively correlated to $\M$.
   
   \subsection{Summary} From the analysis in this section we take away several key insights:
       \begin{compactitem}
        \item Regardless of the optical depth or variation of density along a ray, the pivot is a critical parameter for achieving optimal efficiency with either ratio tracking or truncated estimators.
        \item Near the optimal pivot value, truncated estimators outperform ratio tracking when $Y$-variance is low.
        \item The lowest achievable variance of the estimator is ultimately limited by variance in the optical depth estimates.
    \end{compactitem}
    In the next section we introduce a novel truncated power-series estimator that builds on these insights with the goal of ``climbing down into the valley of zero variance`` in \autoref{fig:BK:general-variance}: the region between the black curves.  This is achieved by combining an accurate online pivot estimation with $Y$-variance reduction and symmetrizing the power-series estimator to further reduce variance and sensitivity to the estimated pivot.

\section{High-Efficiency Power-Series Estimators}
\label{sec:power-series}

    In this section, we propose new truncated power-series estimators inspired by the previous analysis.  These estimators builds on previous work through the introduction of several novel methods, largely under the theme of $Y$-variance reduction and pivot estimation.  We describe each of these methods separately,  with both experimental and theoretical motivation for each, before detailing their combination.
    For notational simplicity, and unless stated otherwise, we will discuss estimation of transmittance \emph{without} the application of the control variate.
    Extending the proposed improvements to residual transmittance $\Transres$ is trivial, necessitating mere substitution of the corresponding terms.

    \subsection{Symmetrization via U-Statistics}\label{section:combinations}
       
        In order to estimate transmittance with a power-series estimator that evaluates all terms up to order $N$ (\autoref{eq:T:pseries-estimator}), we need to obtain $N$ estimates of the negative optical thickness, $\X_1$ to $\X_N$, and evaluate the following sum: 
        \begin{equation}\label{pseries:fixed-order-expansion}
             \widehat{\Trans}_{\mathrm{trunc}} = 1 + \frac{\X_1}{1!\,Q(1)} + \frac{\X_1 \X_2}{2!\,Q(2)} + \cdots + \frac{\X_1 \cdots \X_N}{N!\,Q(N)}\,,
        \end{equation}
        where $Q(k)$ is the probability of evaluating \emph{at least} $k$ orders.  This specific estimator follows from the recursive formulation of the power series~\cite{bhanot1985bosonic,georgiev2019integral} but is not the only unbiased estimator with the correct expectation.  We show how to reduce the variance of this estimator with no additional density evaluations.
        
        The key insight in reducing the variance of \autoref{pseries:fixed-order-expansion} is noting that the first estimate $X_1$ appears in all of the terms, but the last estimate $X_N$ is used in only once, and so increasing $N$ has little impact on the variance of the 
        linear term, and so on.  Our goal is to ensure that all estimates are in a symmetric position with respect to impacting the sum, and that we utilize the estimates maximally for each term in the estimator.
        We can achieve this for the first-order term in \autoref{pseries:fixed-order-expansion} by replacing $\X_1$ by the mean of all estimates:
        \begin{equation}
            m_1 \coloneqq \frac{\X_1 + \cdots + \X_N}{N}
        \end{equation}
        Analogously, we replace the $\X_1\X_2$ product in the second-order term by the mean of all two-term products $\X_i \X_j$:
        \begin{equation}
             m_2 \coloneqq \frac{\X_1 \X_2 + \cdots + \X_1 \X_N + \X_2 \X_3 + \cdots + \X_{N-1} \X_{N}}{\binom{N}{2}}.
        \end{equation}

        In order to generalize this idea to the $k$-th order, we sum the products of all possible $k$-wide combinations---the $k$-th \emph{elementary symmetric sum}:
        \begin{equation}
            s_k \coloneqq \sum\limits_{1 \le i_1 < \cdots < i_k \le N} { \X_{i_1} \cdots \X_{i_k}}\,,
        \end{equation} 
        where $s_0 \coloneqq 1$, and divide it by the number of $k$-wide combinations; this yields a general formula for computing the $k$-th symmetric mean:
        \begin{equation}\label{Equation:mkAsEk}
            m_k \coloneqq \frac{s_k}{\binom{N}{k}}\,.
        \end{equation}
        
        This variance-reduction procedure is well-known in probability theory: \autoref{pseries:fixed-order-expansion} is a statistic $f(X_1,\cdots,X_N)$ of $N$ independent and identically-distributed random variables.  It is known~\cite{halmos1946theory} that the unique and minimum-variance estimator of such a statistic is the symmetric function $f^{[N]}(X_1,\cdots,X_N)$ that is invariant to the order of the $X_i$ inputs.  The generalized means $m_k$ are known as U-statistics~\cite{lee2019u}.
 
        Utilizing U-statistics as the numerators in \autoref{pseries:fixed-order-expansion} yields a novel \emph{U-statistics power-series estimator} of transmittance:
        \begin{equation}\label{pseries:combination-estimator}
             \widehat{\Trans}_{\mathrm{U}} = 1 + \frac{m_1}{1!\,Q(1)} + \frac{m_2}{2!\,Q(2)} + \cdots + \frac{m_N}{N!\,Q(N)}.
        \end{equation}
        This approach can lower the variance of any estimator that estimates more than one term of the power series at the same time (tracking estimators are already fully symmetric).  When combined with the generalized Bhanot \& Kennedy roulette scheme (\ref{eq:bhanot}) we refer to this estimator as the \emph{U-BK estimator}.  
        
\begin{figure}
        \centering
        \includegraphics[width=0.99\linewidth]{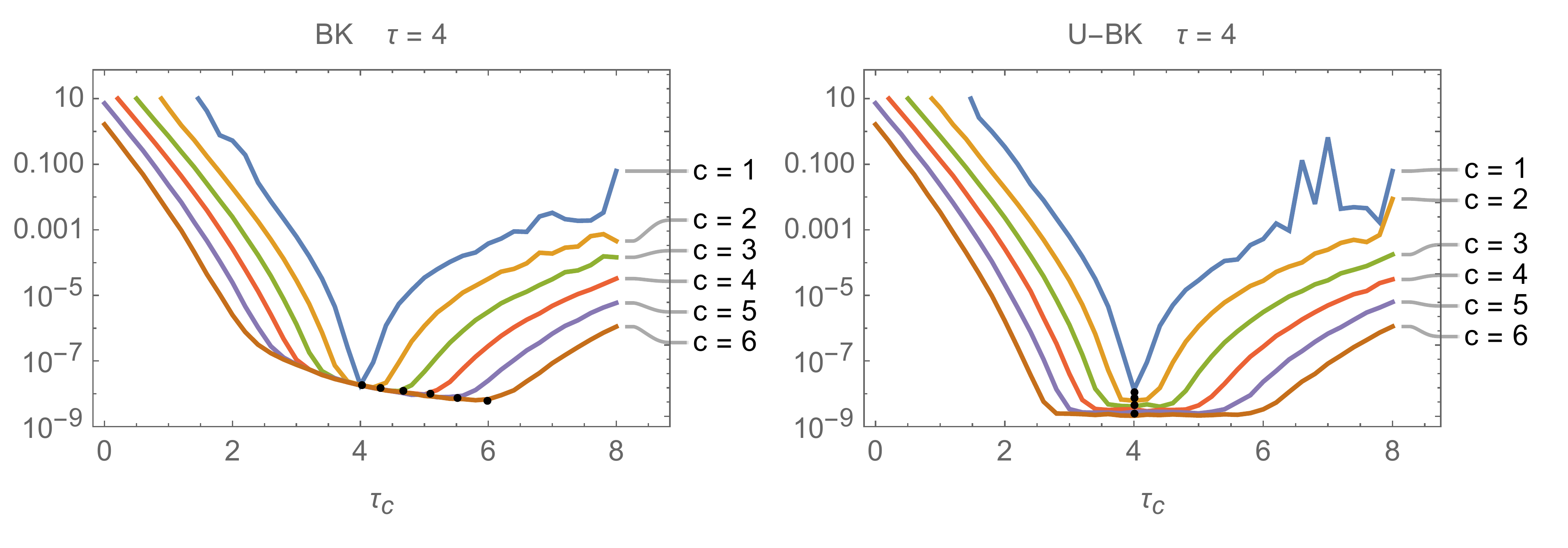}
        \vspace{-.12in}
        \caption{Variance of BK (left) and U-BK (right) estimators for a ray with optical depth $\tau = 4$ and small $Y$-variance.  The pivot that minimizes variance for each value of expansion parameter $c$ is indicated (approximately) by the black dots.  In addition to reducing variance, U-statistics flattens the variance profile with respect to pivot.  This helps to mitigate any increase in variance due to errors in online pivot estimation and provides a simple common goal for pivot estimation---the negative optical depth of the ray.}
        \label{fig:BK:vary_c}
    \end{figure}

        In addition to reducing variance, U-statistics makes it easier to find the optimal pivot.  This can be seen in the variance comparisons in \autoref{fig:BK:general-variance} (varying $Y$-variance with $c$ fixed) and \autoref{fig:BK:vary_c} (varying $c$ with $Y$-variance fixed).  For BK, the negative pivot $\optdctrl$ that achieves minimum variance shifts to the right of $\optd$ as $Y$-variance or $c$ increases, making this a difficult parameter to automatically determine.  In addition to univerisally lowering the variance, we see that U-statistics flattens the variance profile in regions not dominated by roulette variance.  Importantly, regardless of $Y$-variance or $c$, the true optical depth $\optd$ is a (near) optimal setting for $\optdctrl$ in all cases, making the goal of pivot estimation simple: to estimate the negative optical depth.

        The main caveat of naively evaluating a U-statistics estimator is the exponential computational complexity: the total number of combinations required for evaluating the series up to order N is $\binom{N}{0} + \binom{N}{1} + \cdots + \binom{N}{N} = 2^N$. 
        A simple approximation of the optimal estimator could be built in $O(N^2)$ time by averaging $N$ rotations of the $X_i$ estimates (taking the average of $\{ f(X_1,X_2,\dots,X_N),$ $f(X_2,X_3,\dots,X_1), \dots \}$).
        However, efficient full symmetrization is possible by using
        the Girard-Newton formulas \cite{mead:1992:newton}, independently found by Albert Girard and Isaac Newton in the 17th century, which relate numbers $x_1$ to $x_N$ to their elementary symmetric sums $s_k$.  By precomputing the power sums $P_k = \sum_{i=1}^N x_i^k$, we have
        \begin{equation} 
        \begin{split}
            s_k & = \frac{1}{k} \sum_{i=1}^k (-1)^{i-1} s_{k - i} P_i ,
        \end{split}
        \end{equation}
        a simple and efficient recurrence relation for the elementary symmetric sums.
        
        Although the Girard-Newton formulas provide a convenient way for directly calculating the elementary symmetric means, we found them to suffer from numerical precision problems.
        Algorithm~\ref{Algorithm:ElementarySymmetricMeansNew} provides pseudo-code for a novel incremental algorithm to compute the elementary symmetric means that we designed to address these robustness issues while potentially allowing to add new samples on-the-fly.
        An explanation of the algorithm is provided in Appendix~\ref{appendix:ElementarySymmetricMeans}.
        We recommend using this version in practical implementations.
 
        Our algorithm and the Girard-Newton formulas both run in time $O(NZ)$, where $N$ is the number of samples and $Z$ is the number of orders evaluated. Normally we evaluate all orders ($Z = N$) but if the sample count $N$ is high enough, the highest orders might not contribute and we might want to make $Z$ smaller than $N$ to save time. These algorithms reduce the time of evaluating the elementary symmetric means from $O(2^N)$ to $O(N^2)$, or $O(N Z)$, and make the combination estimator practical.
 
        \begin{algorithm}
            \SetAlgoLined
            \SetKwInOut{Input}{Input}
            \SetKwInOut{Output}{Output}
            \Input{Samples $x_1, \cdots, x_N$; Evaluation order $Z$}
            \Output{Elementary symmetric means $m_0, ..., m_Z$}
            $m_0 = 1$ \;
            $m_k = 0$ (for $k = 1$ to $Z$) \;
            \For{$n = 1$ to $N$} {
                \For{$k = \min(n, Z)$ to $1$} {
                    $m_k = m_k + \frac{k}{n} \left(m_{k-1} x_n - m_k\right)$ \;
                }
            }
            \caption{\label{Algorithm:ElementarySymmetricMeansNew}ElementaryMeans}
        \end{algorithm}

    \subsection{Selecting a Pivot}
    
    We presented empirical evidence in \autoref{sec:variance} that the pivot plays a key role in minimizing the variance of transmittance estimators and that, with U-statistics, the negative optical depth is a universally good choice.  We now present additional theoretical motivation for this observation before discussing online pivot estimation.
    
\begin{figure}
        \centering
        {\includegraphics[width=\linewidth]{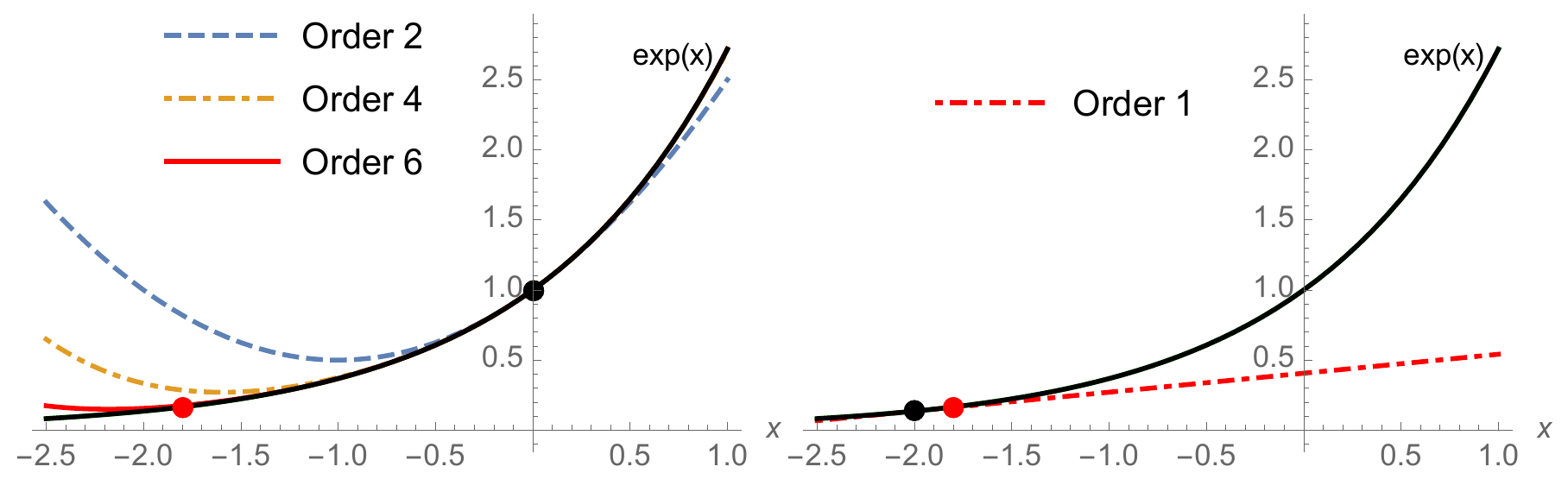}}
        \caption{Moving the pivot (black) of the Taylor series expansion closer to the point where we want to evaluate it (red) has a dramatic effect on its convergence.  Under the power-series formulation of transmittance estimation this means that a more accurate control variate permits more aggressive roulette on higher order terms in the series, lowering the cost.
        }
        \label{fig:pivot}
    \end{figure}

    Interpreting the negative control thickness $-\optdctrl$ as the expansion point, or pivot, of the Taylor series of the exponential provides an insightful new way to analyze power series estimation.
    Consider the general problem of evaluating $e^x$, for a given $x \in \mathbb{R}$, using its series expansion centered at point $p$:
    \begin{equation}
    e^x = e^p \sum_{k=0}^\infty \frac{(x - p)^k}{k!} \approx e^p \sum_{k=0}^N \frac{(x - p)^k}{k!} .
    \end{equation}
    Different values of the pivot $p$ correspond to different polynomial fits:
    the closer $p$ is to $x$, the faster the Taylor polynomial converges to the true value at $x$ (see \autoref{fig:pivot} for demonstration).  For transmittance estimation, this is another way of saying that the ``mass'' that each term in the series contributes to the final estimate of $T$ shifts as the pivot changes (see \cite[Figure (5) and (6)]{georgiev2019integral}).
    
    \begin{figure}
        \centering
        \includegraphics[width=1.0\linewidth]{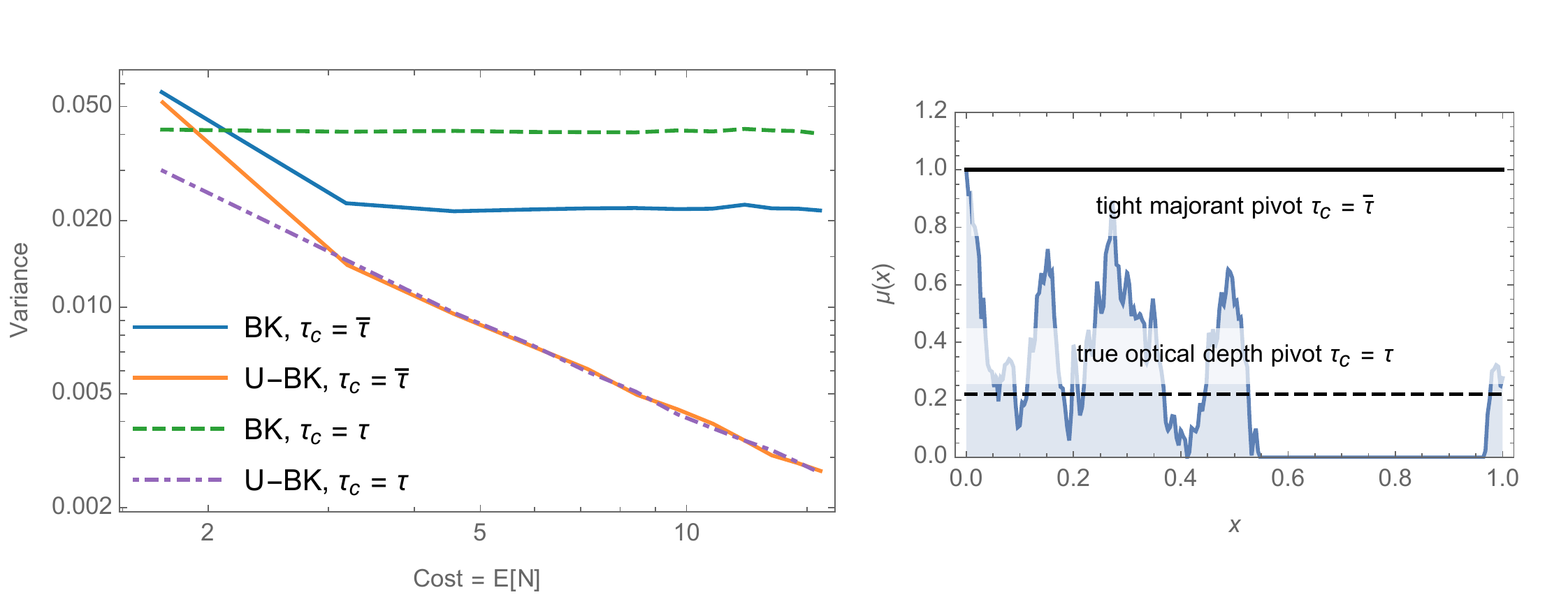}
        \caption{The variance of a non-symmetric p-series (BK) estimator quickly reaches a plateau as the expansion order is raised (increasing $c$) while maintaining a fixed pivot.  This is because the additional samples are used only in the higher-order terms, which have an insignificant contribution in the Taylor expansion of $T$.  Moving the pivot closer to the true optical depth worsens the efficiency because even fewer terms play a significant role in the expansion.
        Our symmetric estimator (UBK) continually improves with larger $c$ because additional samples improve all terms in the expansion.}
        \label{fig:mean-pivot-variance}
    \end{figure}
    
    For the uniform medium, where all optical depth estimates are deterministic (zero variance), we found that the optimal estimator is the one with the pivot that has the Taylor expansion converge at the zeroth order: $e^x = e^{-\tau} \left(1 + \optdres + \cdots\right)$ where $\optdres = 0$.  This may seem like a purely theoretical curiosity, because knowing $\optd$ immediately gives $T$, but it does ensure that any estimator using an estimate of $-\optd$ for $p$ will have zero variance for rays with uniform density, such as through empty portions of the volume.
    
    Whether or not $p = -\optd$ is a good choice in general, though, is more complicated. Without a symmetric estimator, once the optical depth estimates are random, a pivot derived from a majorant density to suppress alternating signs at consecutive orders tends to result in lower variance \cite{galtier:2013, elhafi2018three,georgiev2019integral}.  This results from a complicated interaction between the sampling probabilities for each term in the series and their expected contributions (masses).  We note that (lack of) symmetry can shed more light on why this happens.
    
    In \autoref{fig:BK:general-variance} and \autoref{fig:BK:vary_c} we compare the variance of the truncated BK estimator as the pivot changes. As the expansion parameter $c$ or the level of Y-variance change, the optimal pivot (black dots) moves.  Note how in \autoref{fig:BK:vary_c} (left) at each of the optimal pivot locations, an increase in $c$, which provides more samples to the estimator, has no effect on the variance---it plateaus.  We show this as well in \autoref{fig:mean-pivot-variance} with the medium and pivot held fixed as more samples are used by the estimator.  For either the tight majorant pivot (blue line) or using the true optical depth $\optd$ (yellow line), eventually the extra samples are wasteful.
    In the latter case, the plateau is reached instantly.

    To a first order, the explanation for this effect is the lack of estimator symmetry, resulting in bad sampling of the low order terms: the $k$-th term in the Taylor series is estimated by a \emph{single} product of noisy samples $\estoptdres_1 \cdots \estoptdres_k$, and this sampling error is never corrected by sampling higher orders: i.e. sampling the order $k+1$ with a new density sample $\estoptdres_{k+1}$ will not improve the estimates of the previous orders $\{1, ..., k\}$.
    
    This means that regardless of the pivot, the convergence will plateau as the expected contribution of the higher orders eventually tends to zero and additional samples stop improving the result. With a more accurate pivot, which results in even higher relative expected contribution of the low order terms, the problem is exacerbated, making the convergence plateau even sooner.
    
    The situation changes once we utilize U-statistics because each additional sample improves all of the terms in the Taylor expansion.
    Computing the pivot using the majorant density no longer improves performance, while the approximate density mean always yields better variance, especially in the low sample-count settings (\autoref{fig:mean-pivot-variance}).  As we noted above, in addition to lowering the variance relative to the non-symmetric estimator, U-statistics creates a range of pivots containing $p = -\optd$ where the variance is near optimal, and so we will refer to $p = -\optd$ as the \emph{optimal pivot}.  Appendix~\ref{appendix:OptimalityOfTheMeanPivot} contains a more thorough analysis of the effects of using the approximate mean pivot.

    \subsubsection{Variance of higher-order terms}\label{section:VarianceAnalysis}
    Assuming we have obtained a good pivot, a natural next question is to understand how the $Y$-variance relates to the required order of the Taylor expansion.
    Let us assume a relatively accurate pivot, $p \approx \E[X]$ such that our shifted samples $Y_i \coloneqq X_i - p$ have approximately zero expectation.
    For simplicity, let us also assume that our estimates for the terms $\E[Y]^k$ are given by simple products $Y_1 \cdots Y_k$. The variance of the product is
    \begin{equation}
    \begin{split}
        \Var[Y_1 \cdots Y_k] \approx \E[Y^2]^k \approx \Var[Y]^k ,
    \end{split}
    \label{eq:product-variance}
    \end{equation}
    which relies on the assumption that $\E[Y]^2 \approx 0$. Since shifting a random variable does not change its variance, the immediate follow-up is that if we decrease the variance of our samples $X_i$ to a factor $s$, the variance of our estimate for $\E[X - p]^k$ will fall geometrically to the factor of $s^k$, quickly making the higher-order terms insignificant. For instance, decreasing the variance of our samples by $50\%$ would result in a decrease of the variance of the 10th order term to around 1/1000 of its original value.
    
    This means that even a small reduction in $Y$-variance makes the Taylor series converge with fewer terms. This, in turn, allows us to save computation by more aggressive Russian roulette. We can use this freed sampling budget for bringing the variance of the samples down even more -- a potential self-amplifying feedback loop. However, all of this needs a good pivot which might not always be available.
    
    \subsubsection{Sampled pivots and additional symmetry}
    One way of obtaining accurate approximations of the optimal pivot is to subdivide the volume, precompute localized statistics, and query them along each ray.
    We take a lighter approach and propose to estimate the optimal pivot on-the-fly by taking an additional independent sample $X_{N+1}$ of the integral $-\tau$.

    A single sample might not seem enough to estimate a mean:
    however, we observe that we can apply a procedure analogous to the rotations briefly mentioned in Section 4.1 to effectively increase the total number of samples to $N+1$: If we indicate by $X$ the entire set of samples $\{X_1, \cdots, X_{N+1}\}$, we may consider all $N+1$ estimators resulting from taking each unbiased sample $X_i$ as the pivot in turn, and using the samples $X \setminus \{X_i\}$ to build our symmetrized estimator from section~\ref{section:combinations}, and averaging the result.
    Formally, such unbiased estimator of order $N$ reads:
    \begin{equation}\label{eq:sampled-mean-pivot-T}
         \widehat{T} = \frac{1}{N+1} \sum_{i=1}^{N+1} f_N \left( 
            X_i, \, X \setminus \{X_i\}
            \right)
    \end{equation}
    where the function $f_N$ is given by:
    \begin{equation}
         f_N(p, Y) = e^{p} \sum_{k=0}^N 
         \frac{
            m_k( Y - p ) 
         }{k! Q(k)} ,
    \end{equation}
    where $m_k$ is the $k$-th symmetric means, each $X_i$ is an independent unbiased estimator of $\int_a^b -\ext (x) \, \dif x$, and $Y - p$ subtracts $p$ from each of the remaining samples.  This ensures that all samples in the expanded set have a symmetric contribution in the new estimator.
    

    \subsection{Combed Estimators}
    \label{sec:combed-estimators}

   In this section, we focus on reducing the $Y$-\emph{variance}.
   \autoref{eq:product-variance} suggests that the variance of higher order terms $k$ is proportional to the $k$-th power of the variance of the estimators $X_i$: hence, even a small reduction in variance of each individual $X_i$ will transform into much larger reductions for the higher order terms of the series expansion.
    We propose to use an unbiased, multi-sample estimator that strikes better quality-cost tradeoff than single-sample estimators: $X_i = -\frac{ \ext(x_i) }{ p(x_i) }$, where $x_i \sim p$.
    The estimator is based on randomized Cranley-Patterson (CP) rotations of equidistant points; other sampling patterns are briefly discussed in \autoref{sec:lds-tuples}. 

    Without loss of generality, we assume the integration interval to be $[0,\len)$.
    We use an $M$-tuple of equidistant points
    $\{u_j: \ell j M^{-1}\}_{j=1..M}$
    that we randomly offset and wrap around the $[0,\len)$ interval using the CP rotation. 
    For each order $i$, we use a single random number $x_i \in [0,\len)$ to obtain the rotated set $\{x_{ij} : (x_i + u_j) \, mod\, \len\}_{j=1..M}$ and estimate the optical depth as:
    \begin{equation}
    X_i = -\frac{1}{M} \sum_{j=1}^M \frac{ \ext( x_{ij} ) }{ p( x_{ij} ) }.
    \end{equation}

    Notice that this estimator is equivalent to convolving the integrand with an $M$-point Dirac comb: 
    \begin{equation}\label{eq:dirac-convolution}
    \ext^{\otimes}(s) = \frac{1}{M} \sum_{j=1}^M \ext\left( (s + u_j) \,mod\, \len\right).
    \end{equation}
    Henceforth, we will refer to the resulting estimators as
    \emph{combed estimators}.
    
    \paragraph{Fast convergence rate of equidistant sampling}

    Using multiple density evaluations induces higher evaluation cost than single-sample optical-depth estimators.
    It is thus important to consider whether the U-statistics estimator, which utilizes all estimates maximally, yields lower variance with few high-quality estimates or with many low-quality ones.

    The reason why using combs dedicating $M$ evaluations to each $Y$ estimate is advantageous is to be found in the very fast convergence rate of
    integration by equidistant sampling: in fact, whereas with pure random sampling $M$ evaluations would yield an integration error reduction of only $O(1/\sqrt{M})$, if the integrand has bounded slope (which is common in practice, at least locally) integrating with equidistant combs features a convergence rate of $O(1/M)$, meaning that variance goes does down as fast as $O(1/M^2)$.
    Hence, even if at equal sample count we are reducing the number of combinations available for the U-statistics estimator, we have observed that this is more than compensated by the much lower variance of the individual estimates.

    Our proposed algorithms will take this idea to a logical maximum: we try to maximally benefit from the improved convergence rate by utilizing as dense sampling combs as possible (i.e. a large $M$), and compensate for the larger $M$ by a very aggressive Russian roulette to keep the truncation order $N$ low. 

    \subsubsection{Combing as density reshuffling}
\begin{figure}
        \centering
        {\includegraphics[clip, trim=0.0cm 0.0cm 5.6cm 0.0cm, width=0.95\linewidth]{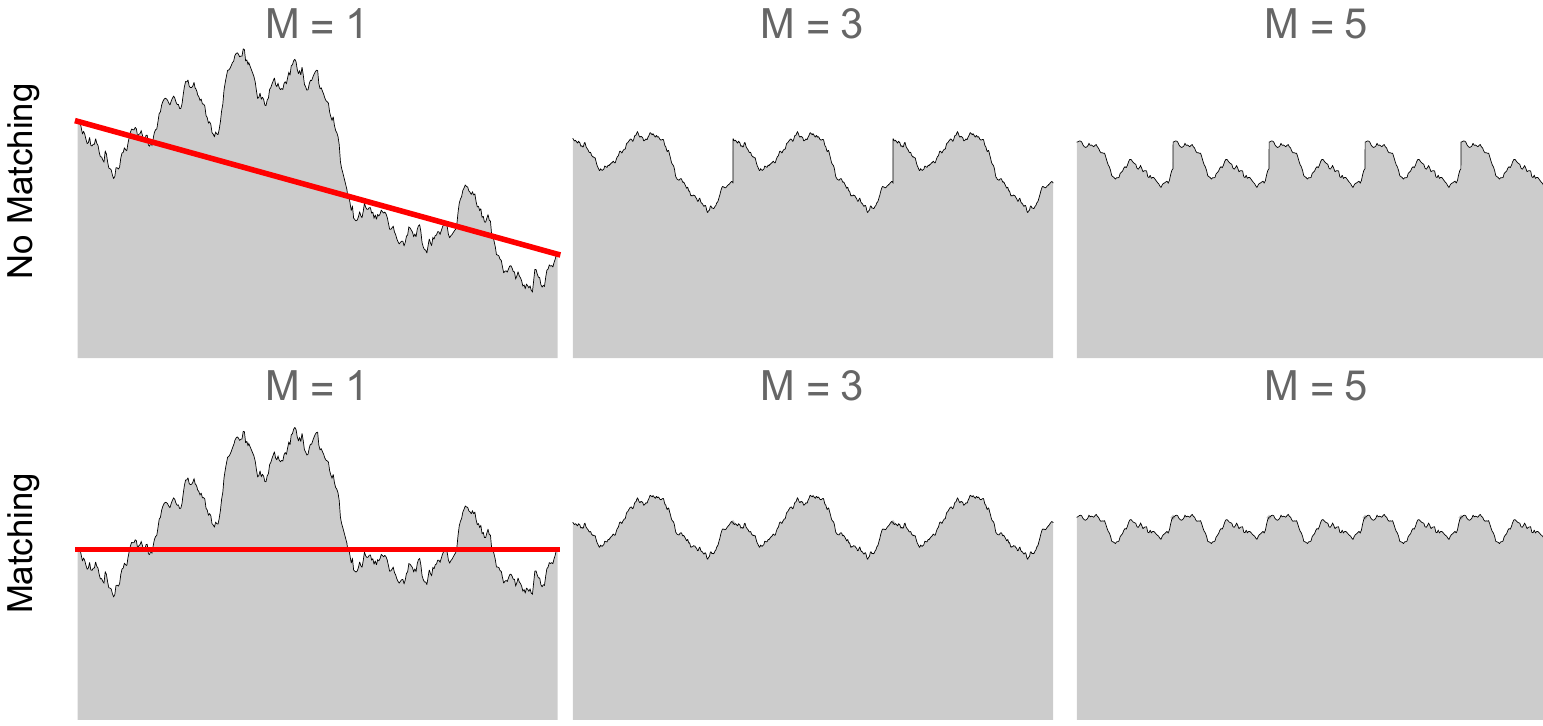}}
        \caption{Top row: $\M$-tap Dirac comb filtering is used to reshuffle the original density $\ext(x)$ (left) to reduce $Y$ variance.  Bottom row: an affine transformation (red) that preserves optical depth is applied to the density to match the endpoints and remove the discontinuities in the combed densities with only two extra density evaluations.}
        \label{fig:density-reshuffling}
    \end{figure}

    As shown by equation (\Ref{eq:dirac-convolution}), an M-point equidistant sample of the density function corresponds to a single evaluation of the convolution of the density with an M-point Dirac comb. This convolution does not change the value of the density integral: it merely reshuffles its density into a form that is more suitable for Monte Carlo estimation (see \autoref{fig:density-reshuffling}).  
    This inspires a new general \emph{invariance principle for transmittance estimation}: we can alter the density along the ray in any integral-preserving way that we like and not change the result.  With this principle, we can maintain the physical picture of a particle traversing the interval or use the Volterra integral formulation of transmittance~\cite{georgiev2019integral} and still benefit from $Y$-variance reduction using a query size $\M$.
    We can also design additional density-reshuffling transformations that further reduce $Y$-variance.

    \subsection{Endpoint Matching}
    \label{sec:endpoint-matching}
    
    The CP rotation utilized in the combed estimator may introduce an artificial discontinuity. This is easy to realize when noticing that rotating the set of samples  around the integration interval is equivalent to rotating the integrand (while keeping the set of samples fixed).
    The original interval endpoints $a$ and $b$ coincide at a new location $x_i$ where the rotated integrand $\mu^{\mathrm{cp}}$ features discontinuity:
    \begin{gather}
    \lim_{s \rightarrow x_i^{-}} \mu^{\mathrm{cp}}(s) = \mu(0) \nonumber \\
    \lim_{s \rightarrow x_i^{+}} \mu^{\mathrm{cp}}(s) = \mu(\len) . 
    \end{gather}
    \begin{wrapfigure}{r}{0.3\linewidth}
      \begin{center}
        \includegraphics[clip, trim=0.1cm 0.1cm 0.7cm 0.1cm, width=\linewidth]{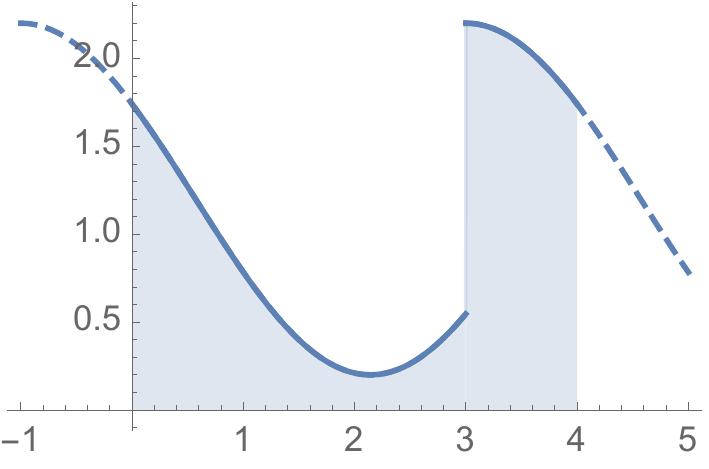}
      \end{center}
      \caption{\label{fig:cp-discontinuity}}
    \end{wrapfigure}
    In practice, if the original $\mu$ had a bounded maximum slope, which is often the case in practice, $\mu^{\mathrm{cp}}$ no longer does (see ~\autoref{fig:cp-discontinuity}). This breaks the assumption that guarantees the improved convergence rate of equidistant sampling.

    We can remedy the discontinuity by another \emph{density reshuffling} operation, namely subtracting a zero-mean affine control variate that interpolates the endpoints:
    \begin{equation}
        \mu^{\star}(s)  := \mu(s) + \left(\frac{1}{2} - \frac{s}{\len}\right)(\mu(\len) - \mu(0)) .
    \end{equation}
    This modification eliminates the discontinuity caused by the random offset ($\mu^{\star}(0) = \mu^{\star}(\len)$) while the integral remains unchanged,
    re-enabling the improved convergence rate from equidistant sampling.
    See the bottom row of Figure~\ref{fig:density-reshuffling} for an illustration.
    Appendix~\ref{appendix:epm} provides further formulas and simplifications.
        
    \subsection{Russian Roulette}
    \label{section:aggressive-roulette}
    
    Sampling the order of the series expansion $N$ is typically performed incrementally by Russian roulette, and several different methods have been
    discussed~\citep{bhanot1985bosonic,booth2007unbiased,papaspiliopoulos2011monte,girolami2013playing,moka2019unbiased,georgiev2019integral}.
    Our work builds on the \emph{Bhanot \& Kennedy} roulette described in \autoref{sec:BK-roulette}, which we modify to incorporate the following observations.
    
    Using a sampled pivot that is close to the true optical thickness reduces the expected
    contributions of the first and higher orders of the Taylor expansion.
    Moreover, with the combed and endpoint-matched $M$-sample U-BK estimator, both the bias and variance of the zeroth order term are often very small, while most of the variance comes form the higher order (correction) terms.
    
    Using all the samples for the pivot would improve the estimator (which has superlinear convergence in $M$), but it would also make the method biased; however, we can still take advantage of this observation by allocating a larger portion of the samples to the zeroth order, and sampling the higher order terms more infrequently.
    
    In order to do that, we terminate the series at the zeroth order with probability $p_{Z}$,
    and only sample the first and higher order terms with probability $1 - p_{Z}$ times the original BK roulette probabilities.
    This is equivalent to using the original BK roulette with the acceptance probability of the first-order term multiplied by $1 - p_{Z}$, leading to the following probabilities of evaluating the first $K$ terms:
    \begin{align}
        P_0 &= 1 \\
        P_1 &= \cdots = P_K = 1 - p_{Z} \,,
    \end{align}
    and the following \emph{conditional} probability for adding the subsequent terms:
    \begin{align}
        P_{k| k-1}= \frac{c}{k}, \quad \text{where } k > K \,.
    \end{align}
    
    This amortizes the cost of the correction terms, allowing us to use larger tuple sizes which in turn improve the pivot and exponentially reduce the expected contribution and variance of the higher-order terms. 

    In practice, we found that truncating the series at the zeroth term in $90\%$ of cases, i.e.\ $p_{Z} = 0.9$, provides a large increase in efficiency across all of our tests.
    Using the BK scheme with parameters $K = c = 2$, this lowers the expected evaluation order from $e-1 \approx 1.71828$ to about $0.31945$, decreasing our expected sample count from $e \approx 2.71828$ to $1.31945$, and allowing us around twice larger tuple sizes $M$. The superlinear convergence obtained by increased tuple sizes more than offsets the variance increase caused by the higher weights of the correction terms.
    
    Raising this probability to $99\%$ yields even lower variance, but at the cost of occasional outliers (manifesting as ``fireflies''). 
    Moreover, evaluating only the zeroth order term $90\%$ of the time already provides $90\%$ of the possible cost savings, so higher values are unlikely to strike much better efficiency.

        \begin{algorithm}
            \SetAlgoLined
            \SetKwInOut{Input}{Input}
            \SetKwInOut{Output}{Output}
            \Input{$p_{Z} = 0.9$ }
            \Output{maximum order $k$ and weights $w_0, ..., w_k$}
            
            \BlankLine
            \BlankLine

            $w_0 = 1$\;
            \BlankLine
            $P =  1 - p_{Z}$\;
            \BlankLine

            $u$ = rand()\;
            \BlankLine

            // Stop at the zeroth order term with probability $p_{Z}$ \\
            \If{$P \leq u$}{
                return $0$\;
            }
            \BlankLine

            // BK with $K$ = $c$ = $2$; \\
            $K$ = $c$ = $2$\;
            \BlankLine

            \For{$i = 1$ to $K$} {
                $w_i$ = $1/P$ \;
            }
            
            \For{$i = K+1$ to $\infty$} {
                // Compute the continuation probabilities  \\
                $c_i$ = $\min(\g / i, 1)$ \;

                // Update the probability of sampling at least order i \\ 
                $P$ = $P \cdot c_i$\;

                // Russian roulette termination \\ 
                \If{$P \leq u$}{ 
                    return $i - 1$\; 
                }

                // Final weight for order i \\ 
                $w_i = 1 / P$\;
            }

            \caption{\label{Algorithm:BKRouletteExpMean}AggressiveBKRoulette}
        \end{algorithm}

        \begin{algorithm}
            \SetAlgoLined
            \SetKwInOut{Input}{Input}
            \SetKwInOut{Output}{Output}
            \Input{$p_Z$, $\g$, $K = \myfloor{\g}$ assumed}
            \Output{expected evaluation order of our roulette}
            
            // Evaluate $\E[N_{BK}] = K + \left(K! / \g^K\right) \left(e^\g - \sum_{k=0}^K \g^k / k!\right)$ \\
            $K$ = $\myfloor{\g}$ \;
            $t$ = 1\;
            $sum$ = 1\;
            \For{$k = 1$ to $K$} {
                $t$ = $t * \g / k$\;
                $sum$ = $sum + t$\;
            }
            $E_N$ = $K + (\exp(\g) - sum) / t$\;
            // Non-zero orders are evaluated with probability $1 - p_Z$ \\
            return $(1 - p_{Z}) \cdot E_N$\;

            \caption{\label{Algorithm:BKExpectedSampleCount}BKExpectedEvalOrder}
        \end{algorithm}

        \begin{algorithm}
            \SetAlgoLined
            \SetKwInOut{Input}{Input}
            \SetKwInOut{Output}{Output}
            \Input{control optical thickness $\optdmaj$}
            \Output{tuple size M matching the p-series CMF cost}
            
            $N_{CMF} = \left\lceil\sqrt[\leftroot{2}\uproot{2}3]{(0.015 + \optdmaj)(0.65 + \optdmaj)(60.3 + \optdmaj)}\right\rceil$ \;

            $N_{BK}$ = BKExpectedEvalOrder(2) ; // $\approx 0.31945$.
            
            return $\max(1, \myfloor{ N_{CMF} / (N_{BK} + 1) + 0.5 })$ \;

            \caption{\label{Algorithm:DetermineTupleSize}DetermineTupleSize (for unbiased ray marching)}
        \end{algorithm}
        
\subsection{Tuple size deduction}
\label{sec:tuple-size-deduction}
    With all the above improvements, we obtain an estimator that has superlinear convergence properties in the tuple size.
    While in itself this is very powerful, designing a strategy to determine optimal tuple sizes may depend on all the sources of noise surrounding transmittance estimation (for example, in a rendering problem, all the sources of noise in path sampling), and we consider it outside the scope of this paper.
    Our objective is designing an estimator that works well even at relatively low sample counts.  We found that a sample count related to the one used in the p-series CMF estimator works well in practice, and we describe its evaluation and use here.

    As a first step, we employed a simple grid search to obtain a fit for the expected sample count used by the p-series CMF with 99\% mass (given the control / majorant optical thickness $\optdmaj$)
    \begin{equation}\label{eq:sample-count-fit}
       \E[N_{CMF}] \approx \left\lceil\sqrt[\leftroot{2}\uproot{2}3]{(0.015 + \optdmaj)(0.65 + \optdmaj)(60.3 + \optdmaj)}\right\rceil .
    \end{equation}
    This approximation is by empirical analysis asymptotically correct, has
    mean absolute error of $0.34$ samples
    for $\optdmaj < 10$ and a maximum relative error of $9\%$ for $\optdmaj \ge 10$.  We found this more efficient than the approach used by \citet{georgiev2019integral}.
    
    We then use the Algorithm \ref{Algorithm:BKExpectedSampleCount} to solve for $\M$ such that a generalized BK roulette with given $K$ produces the same mean number of density evaluations as p-series CMF.  This uses an exact formula, \autoref{eq:cost:BK}, for the expected evaluation order $\E[N_{BK}]$.
	We need one sample per order plus one for the pivot, and hence to achieve the same cost, the expected sample count $\E[N_{BK}] + 1$ times the tuple size $M$ must match $\E[N_{CMF}]$, or in other words, our desired tuple size is given by
    \begin{equation}
        M = \frac{
            \E[N_{CMF}]
        }{
            \E[N_{BK}] + 1
        } .
    \end{equation}
    Algorithm \ref{Algorithm:DetermineTupleSize} provides pseudo-code for the final algorithm.

    In practice, in order to not oversample high-density but low-variance volumes, we recommend using the difference between the majorant and minorant optical thicknesses as the control parameter $\optdmaj$, when a minorant is available. 
    
    \subsection{Assembling the Estimators}
    \label{sec:power-series:assembly}

    The following paragraphs summarize the construction of our final estimators.

    \subsubsection{The Unbiased Ray-marching Estimator}
    Our final unbiased estimator is summarized in Algorithm~\ref{Algorithm:FinalUnbiasedEstimator} and works as follows: We first determine the number of density evaluations, $\M$, for estimating each sample of negative optical thickness $X_i$ (as described in \autoref{sec:tuple-size-deduction}), and we determine the highest order of the power series, $N$, using the BK roulette (\autoref{section:aggressive-roulette}); these first two steps do not impact each other.
    
    Then we compute $N+1$ combed estimates $(X_1 \cdots X_{N+1})$ using equidistant, CP-rotated evaluations (\autoref{sec:combed-estimators}) and apply endpoint matching (\autoref{sec:endpoint-matching}).
    
    Finally, we use each $X_i$ as the pivot (\autoref{eq:sampled-mean-pivot-T}) and evaluate the Taylor series thereof using the symmetrized estimator (\autoref{section:combinations}). Specifically, we use our new elementary-means algorithm (\autoref{Algorithm:ElementarySymmetricMeansNew}) with the remaining $N$ estimates $X \setminus \{X_i\}$.

    Notice that at very low sample counts, e.g. $M < 6$, we disable endpoint matching, since we found that the additional overhead of its two additional evaluations $\mu(0)$ and $\mu(\len)$ was not worth the resulting variance reduction.
    
    Most of the time our estimator evaluates only the zeroth order term, when roulette samples $N = 0$.  In this case, the estimate is simply $e^{X_1}$, where $X_1$ is the single estimation of negative optical depth using equidistant evaluations of the density from $a$ to $b$.  On its own, this estimate is virtually the same as jittered ray marching \cite{pauly2000metropolis} potentially coupled with the endpoint-matching control variate.  The higher order terms correct the bias, so we call this estimator \emph{unbiased ray-marching}.

        \begin{algorithm}
            \SetAlgoLined
            \SetKwInOut{Input}{Input}
            \SetKwInOut{Output}{Output}
            \Input{Interval length $\ell$; control optical thickness $\optdmaj$}
            \Output{Transmittance $T$}
            $M = \text{DetermineTupleSize}(\optdmaj);$ \\
            $N = \text{AggressiveBKRoulette}(K=c=2, p_{Z}=90\%);$ \\
            $\mu_\ell, \mu_0 = \text{EvalDensity}(\ell), \text{EvalDensity}(0)$;  // optional \\
            \For{$i = 1$ to $N+1$} { 
                $u_i = \text{rand}();$ \\
                $X_i = -\frac{\ell}{M} \cdot \sum_{i=0}^{M-1} \text{EvalDensity}(\frac{\ell}{M}(u + i));$ \\
                $X_i = X_i - \frac{\ell}{M}\left(\frac{1}{2} - u_i\right)(\mu_\ell - \mu_0);$  // optional \\
            }
            T = 0; \\
            \For{$j = 1$ to $N+1$} {
                $m_0, \cdots, m_N = \text{ElementaryMeans}(\{X \setminus {X_j}\} - X_j);$ \\
                $T = T + \frac{1}{N+1}\, e^{X_j} \sum_{k=0}^N \frac{m_k}{k!\, p_k}$;
            }
            \caption{\label{Algorithm:FinalUnbiasedEstimator}Unbiased ray marching}
        \end{algorithm}

    \subsubsection{The Biased Ray-marching Estimator}
    
    One of the surprising conclusions from section~\Ref{section:aggressive-roulette} is that with all our optimizations in place and enough equidistant samples and a sampled pivot, we can make the Russian roulette most often truncate at the constant term -- and still obtain very little variance. This is possible because our sampled pivots become increasingly good estimates for the real integral with the addition of more equidistant samples, and hence even a zeroth order Taylor polynomial often results in a very good -- and a very cost-effective -- estimate for the real integral.
    
	This surprising behavior is partially explained by the following observation: When the pivot $X_1$ is an unbiased estimate for the integral, the zeroth order approximation $e^{X_1}$ is actually in a sense accurate to the first order, essentially gaining an order of accuracy for free: 
    \begin{equation}
    \begin{split}
        \E[e^{X_1} - e^{\E[X]}] & = e^{\E[X]} \E\left[e^{X_1 - \E[X]} - 1\right] \\
        & = e^{\E[X]} \E\left[\left(X_1 - \E[X]\right) + \frac{(X_1 - \E[X])^2}{2!} + \cdots\right] \\
        & = e^{\E[X]} \E\left[0 + \frac{(X_1 - \E[X])^2}{2!} + \cdots\right] .
    \end{split}
    \end{equation}
 
    This suggests that we can build an effective low-bias estimator by \emph{always} truncating the series at the zeroth order, that is to say evaluating only:
    \begin{equation}
    e^{-\optd} \approx e^{X_1} \cdot 1 = e^{X_1} ,
    \end{equation}
    where $X_1$ is an unbiased estimator for the integral of $\ext$ which we obtain with combing and by using all of the transmittance budget to increase the tuple size $M$. We couple this technique with the endpoint matching control variate (see Algorithm~\ref{Algorithm:FinalBiasedEstimator}).
    
    This is again the same as jittered ray marching applied to the endpoint-matching-reshuffled density: a surprisingly simple algorithm.

        \begin{algorithm}
            \SetAlgoLined
            \SetKwInOut{Input}{Input}
            \SetKwInOut{Output}{Output}
            \Input{Interval length $\ell$; control optical thickness $\optdmaj$}
            \Output{Transmittance $T$}
            $M = \left\lceil\sqrt[\leftroot{2}\uproot{2}3]{(0.015 + \optdmaj)(0.65 + \optdmaj)(60.3 + \optdmaj)}\right\rceil$ \;
            $u = \text{rand}();$ \\
            $X = -\frac{\ell}{M} \cdot \sum_{j=0}^{M-1} \text{EvalDensity}(\frac{\ell}{M} (u + i));$ \\
            $X = X - \frac{\ell}{M}\left(\frac{1}{2} - u\right)\left(\text{EvalDensity}(\ell) - \text{EvalDensity}(0)\right)$ ; // optional \\
            $T = e^{X};$ \\
            \caption{\label{Algorithm:FinalBiasedEstimator}Biased ray marching}
        \end{algorithm}

\section{Results}
    In this section we compare our proposed unbiased and biased transmittance estimators to ratio tracking (RT)~\cite{cramer1978application}, residual ratio tracking (RRT)~\cite{novak2014residual} and the p-series CMF~\cite{georgiev2019integral} estimators in a variety of scenes featuring participating media.
    For the unbiased methods we report variance, and for our biased ray marching we measure mean-square-error of one sample.
    
    In \autoref{fig:teaser} and \autoref{fig:image-comparison} we study the performance of the individual estimators in a path tracer. 
    All volumes in the figure are stored using the VDB data structure~\cite{Museth:2013:VDB} that additionally provides aggregate volumetric statistics (minimum, maximum, and mean density) over $8\times8\times8$ voxel regions---super voxels.
    We utilize the statistics for computing tighter (residual) majorants for tracking estimators. 
    For p-series CMF and our estimators, we use the mean densities in super voxels to ``warp'' the lookups: along each ray, we perform regular tracking~\cite{amnatides:1987:voxeltraversal} through the super grid and build a piecewise-constant probability density function (PDF) from the super-voxel means, and distribute the lookup points proportional to the PDF using the inversion method.
    For our estimators, specifically, we generate CP-rotated equidistant samples in the $[0,1]$ primary interval and then transform them into a warped comb along the ray.
    
    The insets in the figure show results for different estimators at one path sample per pixel. Since the efficiency of certain estimators improves with higher lookup counts, we normalize the comparison by adjusting them to yield approximately equal number of density lookups \emph{per transmittance estimate}.
    We use the p-series CMF estimator as the baseline and uniformly increase local (residual) majorants such that the tracking estimators utilize approximately the same number of lookups (predicted by \autoref{eq:sample-count-fit}).
    For our methods we employ the automatic tuple size mechanism discussed in \autoref{sec:tuple-size-deduction}.

    In the following we list the specifics of individual scenes:
\begin{itemize}
    \item \textsc{Plume} features absorptive smoke. Transmittance estimation is the only source of noise in this scene; this setting is thus the most representative one (out of the four scenes) of the relative performance between the estimators.
    \item \textsc{Box} features indirect illumination from an area light surrounded by an absorptive medium. Despite the extra noise from simulating up to four light bounces, the impact of the different transmittance estimators is still clearly visible.
    \item \textsc{Cloud} features single-scattering illumination due to two point lights. We use equiangular sampling~\cite{kulla:2011:importance} to sample collisions along primary rays. Transmittance along the primary ray and the shadow rays is estimated with the studied estimator. The improvement from the transmittance estimation is partially masked by other sources of noise.
    \item \textsc{Glass with Smoke} features frequency-dependent absorptive smoke in a reflective glass box and shows how improved transmittance estimation can affect the quality of volume rendering either directly or through reflections.
    \item \textsc{Glass} (\autoref{fig:teaser}) features another frequency-dependent absorptive medium in a glass embedding.
\end{itemize}

    \newcommand{\bigFigureCaption}{\label{fig:image-comparison} A comparison of our unbiased and biased estimators (two rightmost columns) to ratio tracking, residual ratio tracking and the p-series CMF estimators on a variety of rendered content featuring participating media. The \emph{Plume}, \emph{Box} and \emph{Glass with Smoke scenes} contain purely absorptive media, while the \emph{Cloud} scene shows single-scattering illumination by point lights, rendered using equiangular sampling to sample collisions along primary rays.}
    
  \renewcommand*{\figurePath}{figures/big_figure}%
\newlength{\bigFigureLengthA}
\newlength{\bigFigureLengthB}
\begin{figure*}
\setlength{\fboxrule}{5pt}%
\setlength{\tabcolsep}{.3pt}%
\setlength{\bigFigureLengthA}{2.4pt}
\setlength{\bigFigureLengthB}{1.2pt}
\renewcommand{\arraystretch}{0.0}%
\centering
\footnotesize%
\setlength{\beautyHeight}{3.73cm}
\begin{tabular}{crrccccc}
    &
    &
    &Ratio tracking (RT)
    &Residual RT
    &P-series CMF
    &Unbiased ray march.
    &Biased ray march.
    \\%
    &
    &
    &\cite{cramer1978application}
    &\cite{novak2014residual}
    &\cite{georgiev2019integral}
    &(ours)
    &(ours)
    \\[2pt]%
	\begin{adjustbox}{valign=t}%
        \rotatebox{90}{%
            \makebox[\beautyHeight]{\textsc{Plume}}%
        }%
    \end{adjustbox}\,\,%
    &%
    \begin{adjustbox}{valign=t}%
        \includegraphics[height=\beautyHeight+2\bigFigureLengthB]{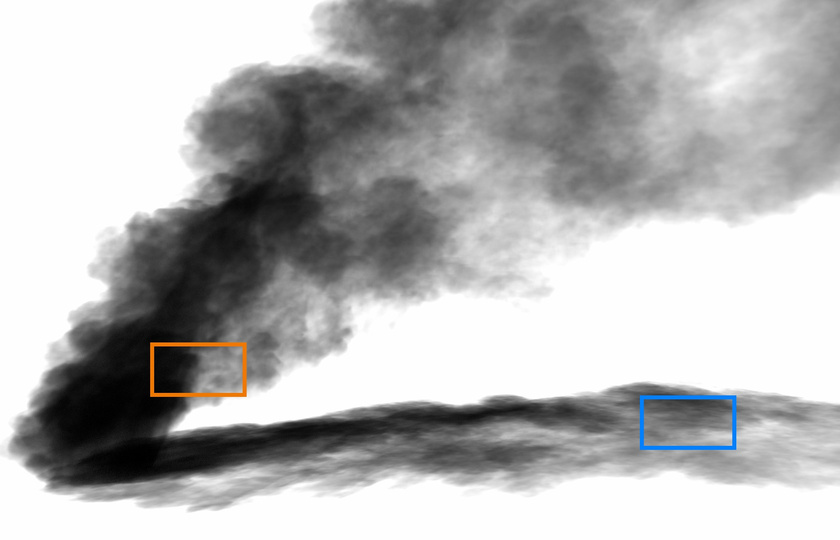}%
    \end{adjustbox}%
    \hspace{\bigFigureLengthA}\strut%
    &%
    \begin{adjustbox}{valign=t}%
        \rotatebox{90}{%
        	\makebox[\beautyHeight/3]{Inv.eff.}\hspace{\bigFigureLengthB}%
        	\makebox[\beautyHeight/3]{Inset 2}\hspace{\bigFigureLengthB}%
        	\makebox[\beautyHeight/3]{Inset 1}%
        }%
    \end{adjustbox}%
            &%
        \begin{adjustbox}{valign=t}%
            \begin{adjustbox}{totalheight=1.0\beautyHeight+2\bigFigureLengthB,tabular={c}}
                \includegraphics[height=\beautyHeight/3]{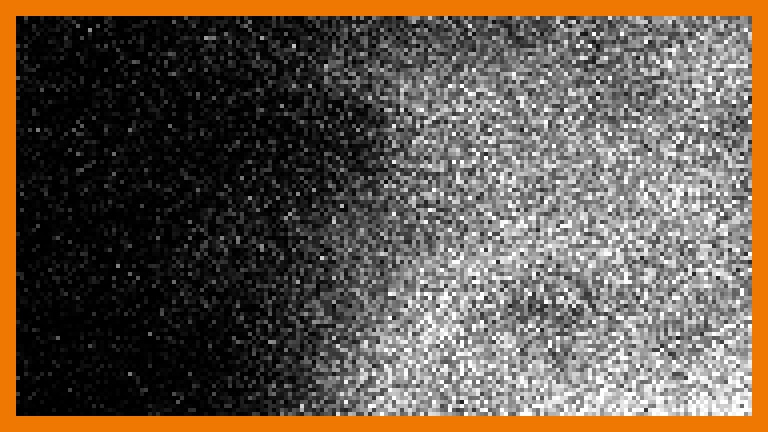}\vspace{\bigFigureLengthB}\\
                \includegraphics[height=\beautyHeight/3]{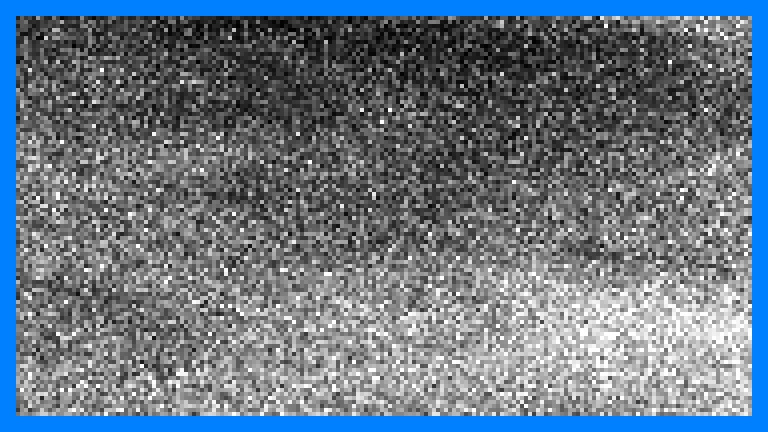}\vspace{\bigFigureLengthB}\\
                \includegraphics[height=\beautyHeight/3]{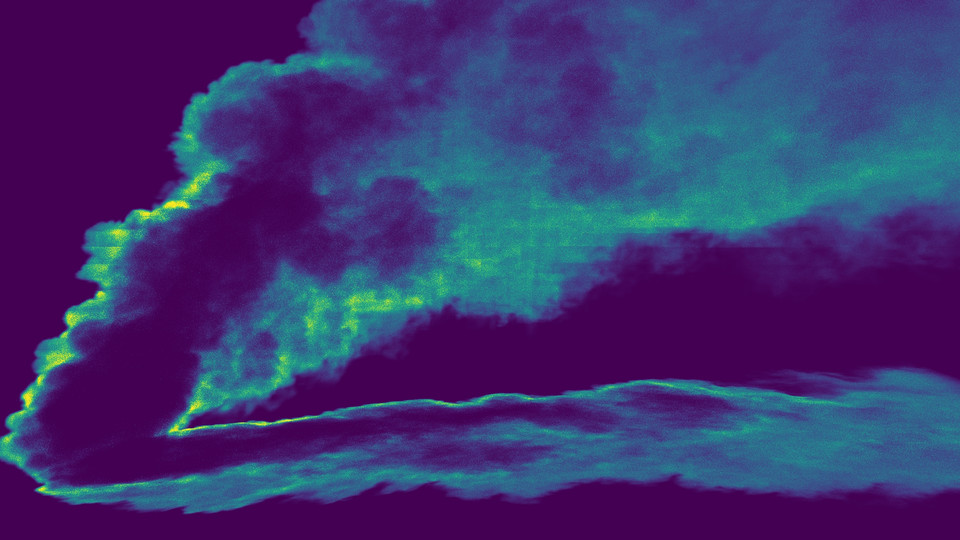}%
            \end{adjustbox}%
        \end{adjustbox}%
            &%
        \begin{adjustbox}{valign=t}%
            \begin{adjustbox}{totalheight=1.0\beautyHeight+2\bigFigureLengthB,tabular={c}}
                \includegraphics[height=\beautyHeight/3]{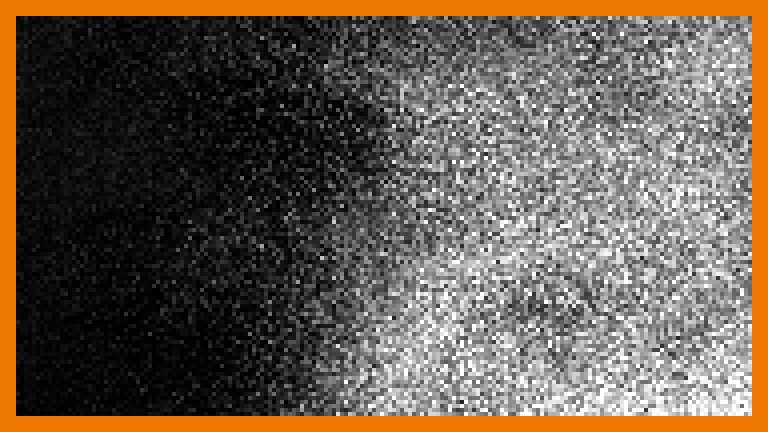}\vspace{\bigFigureLengthB}\\
                \includegraphics[height=\beautyHeight/3]{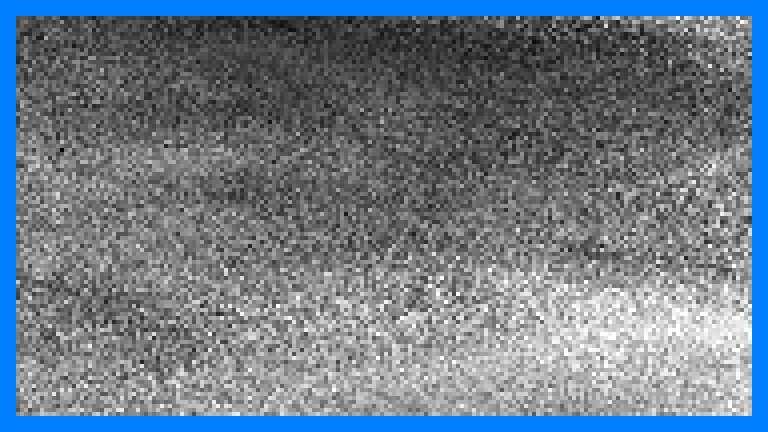}\vspace{\bigFigureLengthB}\\
                \includegraphics[height=\beautyHeight/3]{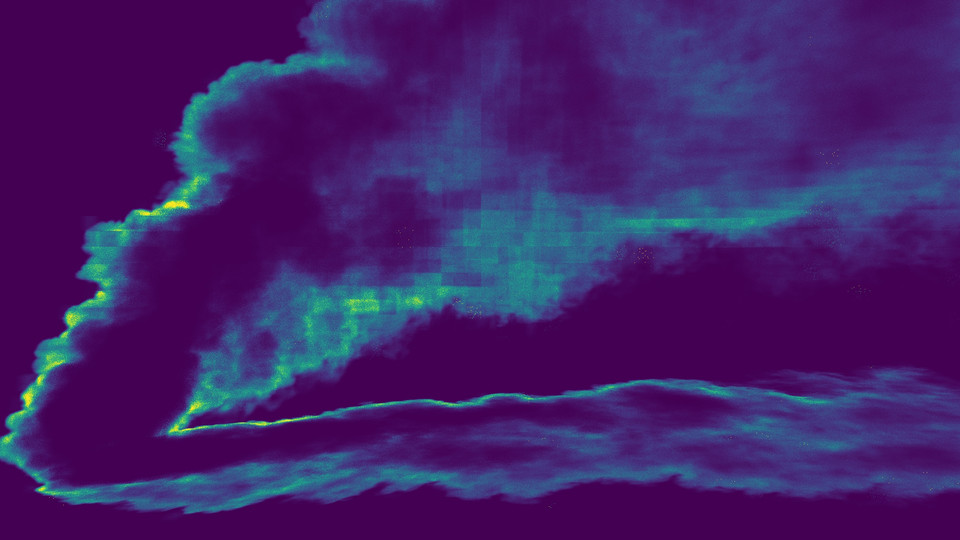}%
            \end{adjustbox}%
        \end{adjustbox}%
            &%
        \begin{adjustbox}{valign=t}%
            \begin{adjustbox}{totalheight=1.0\beautyHeight+2\bigFigureLengthB,tabular={c}}
                \includegraphics[height=\beautyHeight/3]{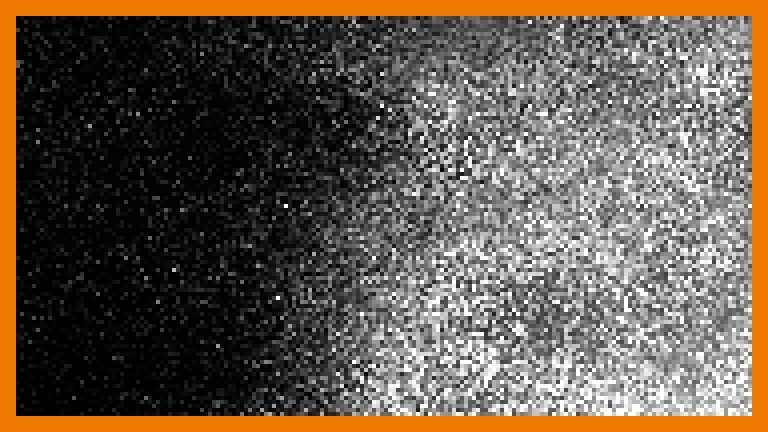}\vspace{\bigFigureLengthB}\\
                \includegraphics[height=\beautyHeight/3]{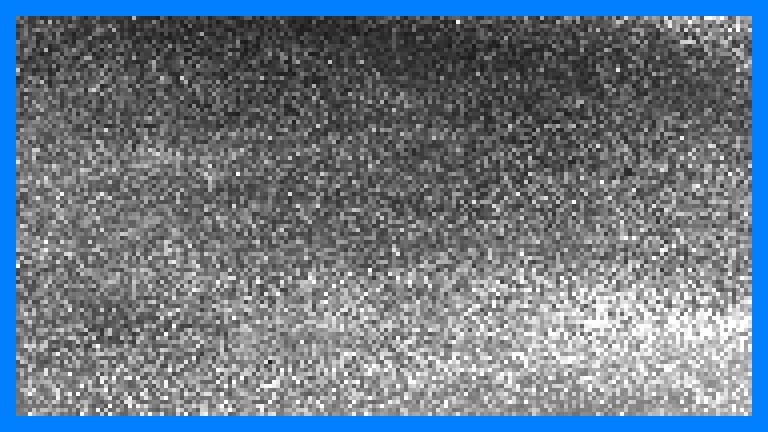}\vspace{\bigFigureLengthB}\\
                \includegraphics[height=\beautyHeight/3]{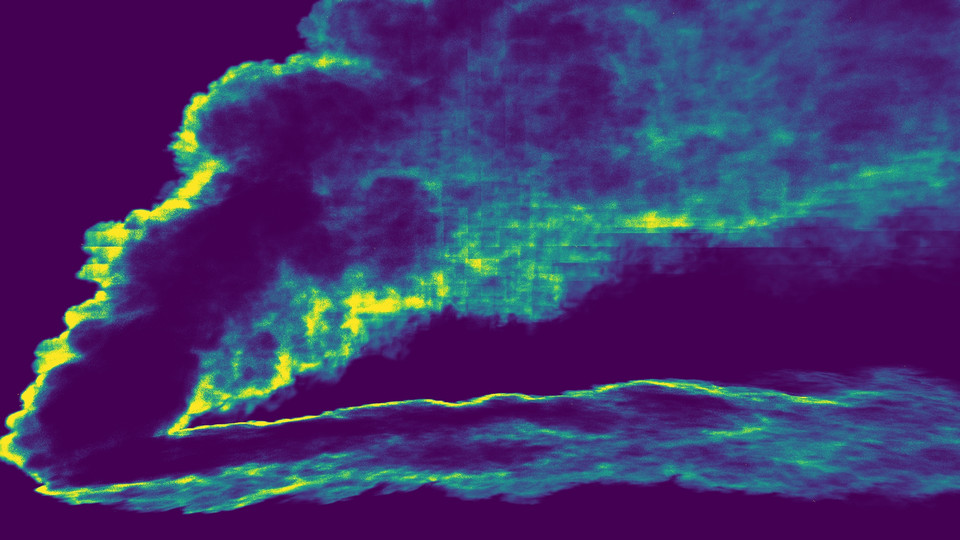}%
            \end{adjustbox}%
        \end{adjustbox}%
            &%
        \begin{adjustbox}{valign=t}%
            \begin{adjustbox}{totalheight=1.0\beautyHeight+2\bigFigureLengthB,tabular={c}}
                \includegraphics[height=\beautyHeight/3]{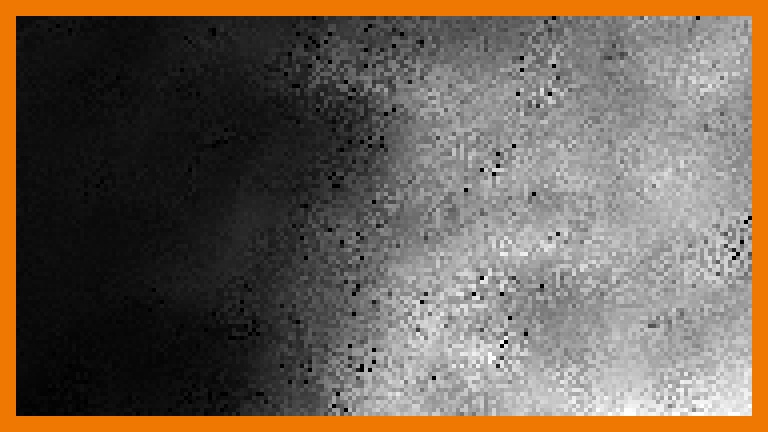}\vspace{\bigFigureLengthB}\\
                \includegraphics[height=\beautyHeight/3]{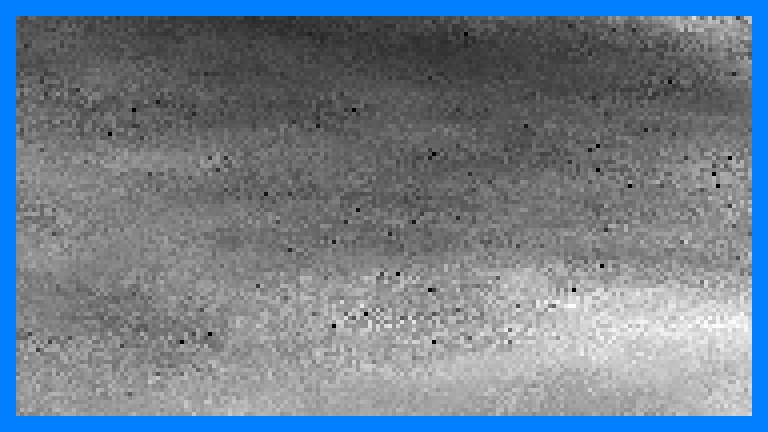}\vspace{\bigFigureLengthB}\\
                \includegraphics[height=\beautyHeight/3]{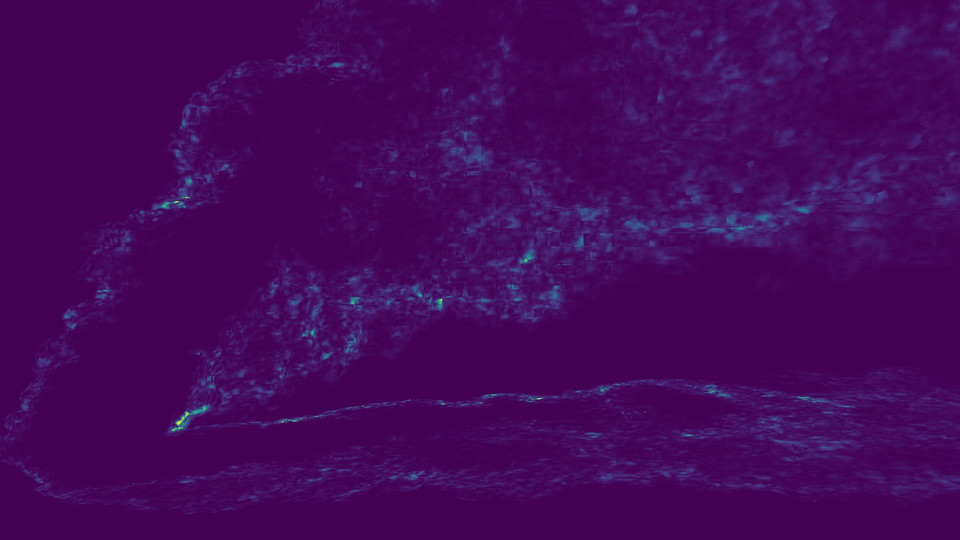}%
            \end{adjustbox}%
        \end{adjustbox}%
            &%
        \begin{adjustbox}{valign=t}%
            \begin{adjustbox}{totalheight=1.0\beautyHeight+2\bigFigureLengthB,tabular={c}}
                \includegraphics[height=\beautyHeight/3]{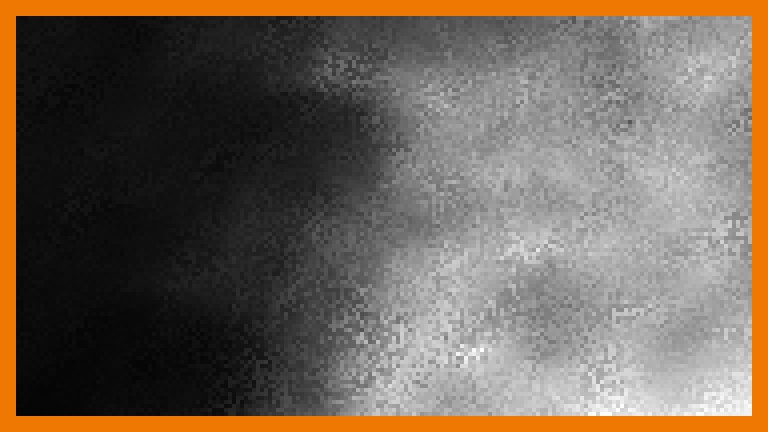}\vspace{\bigFigureLengthB}\\
                \includegraphics[height=\beautyHeight/3]{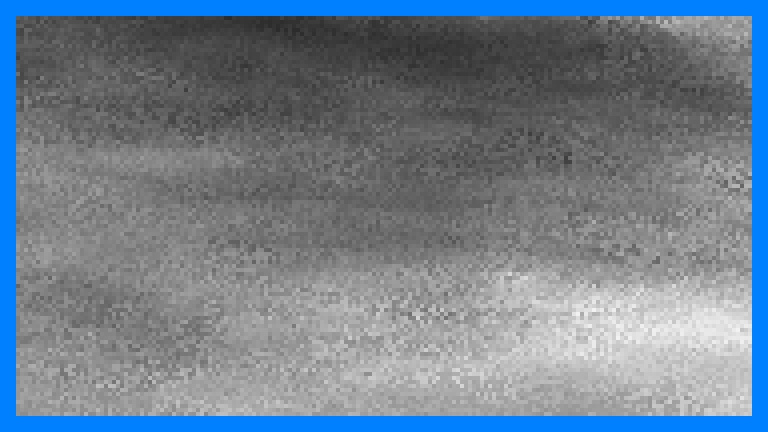}\vspace{\bigFigureLengthB}\\
                \includegraphics[height=\beautyHeight/3]{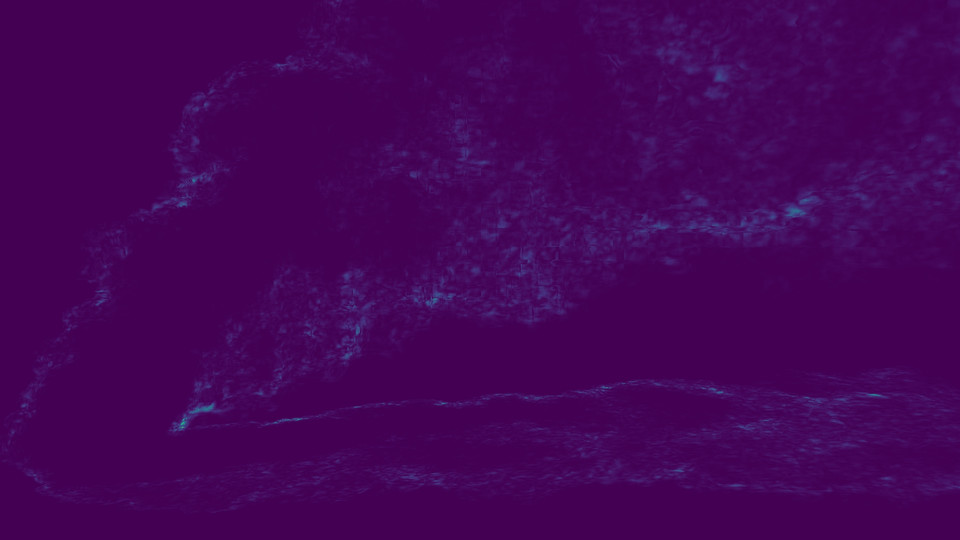}%
            \end{adjustbox}%
        \end{adjustbox}%
        \vspace{1mm}
    \\%
    &Variance (MSE):\hspace{\bigFigureLengthA}\strut%
    &
    &0.0277
    &0.0175
    &0.0287
    &3.87e-3
    &2.14e-3
    \\[1pt]
    &Cost:\hspace{\bigFigureLengthA}\strut%
    &
    &7.39
    &7.39
    &7.22
    &7.44
    &7.39
    \\[1pt]
    &Inv. efficiency:\hspace{\bigFigureLengthA}\strut%
    &
    &0.231
    &0.143
    &0.231
    &0.0301
    &0.0163
    \\[6pt]
	\begin{adjustbox}{valign=t}%
        \rotatebox{90}{%
            \makebox[\beautyHeight]{\textsc{Box}}%
        }%
    \end{adjustbox}\,\,%
    &%
    \begin{adjustbox}{valign=t}%
        \includegraphics[height=\beautyHeight+2\bigFigureLengthB]{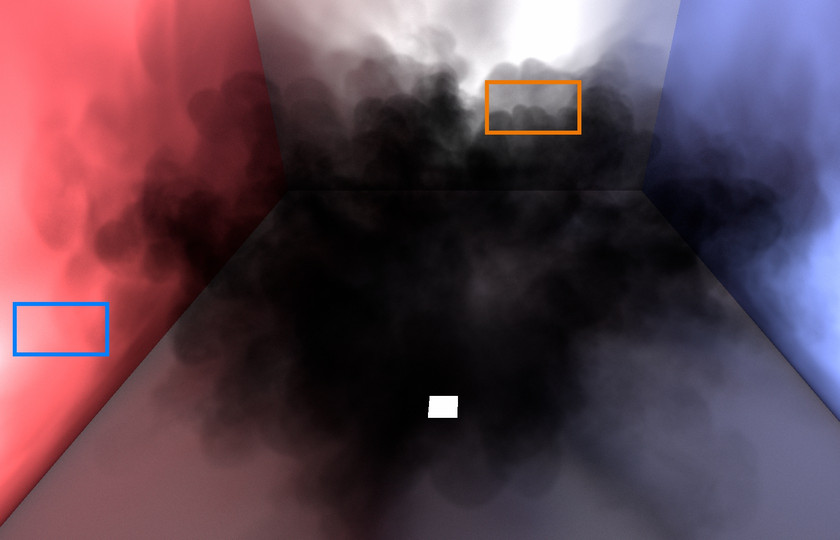}%
    \end{adjustbox}%
    \hspace{\bigFigureLengthA}\strut%
    &%
    \begin{adjustbox}{valign=t}%
        \rotatebox{90}{%
        	\makebox[\beautyHeight/3]{Inv.eff.}\hspace{\bigFigureLengthB}%
        	\makebox[\beautyHeight/3]{Inset 2}\hspace{\bigFigureLengthB}%
        	\makebox[\beautyHeight/3]{Inset 1}%
        }%
    \end{adjustbox}%
            &%
        \begin{adjustbox}{valign=t}%
            \begin{adjustbox}{totalheight=1.0\beautyHeight+2\bigFigureLengthB,tabular={c}}
                \includegraphics[height=\beautyHeight/3]{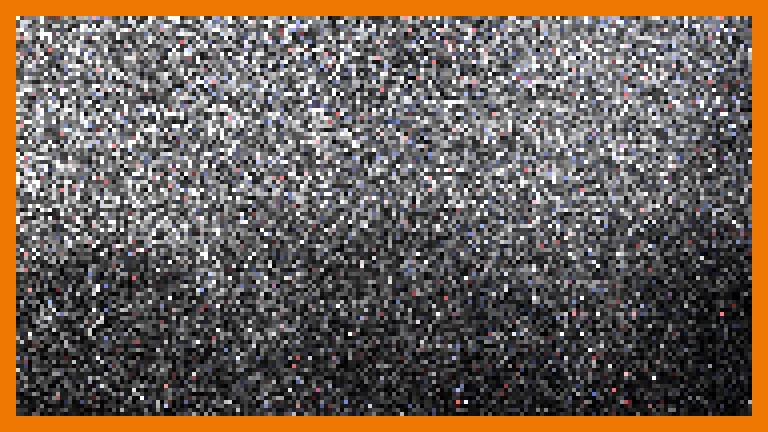}\vspace{\bigFigureLengthB}\\
                \includegraphics[height=\beautyHeight/3]{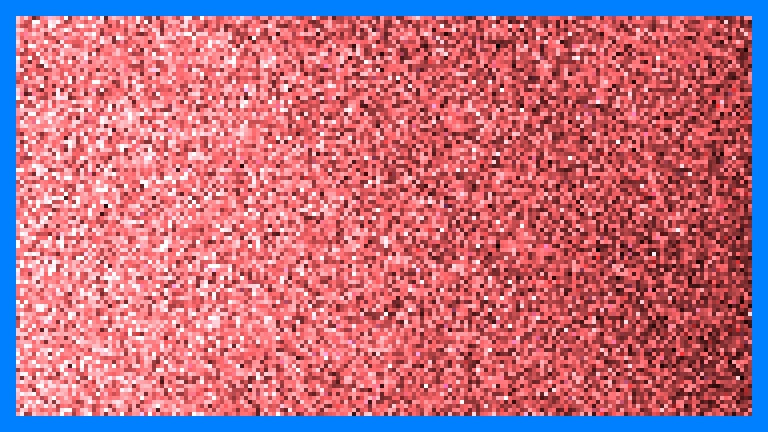}\vspace{\bigFigureLengthB}\\
                \includegraphics[height=\beautyHeight/3]{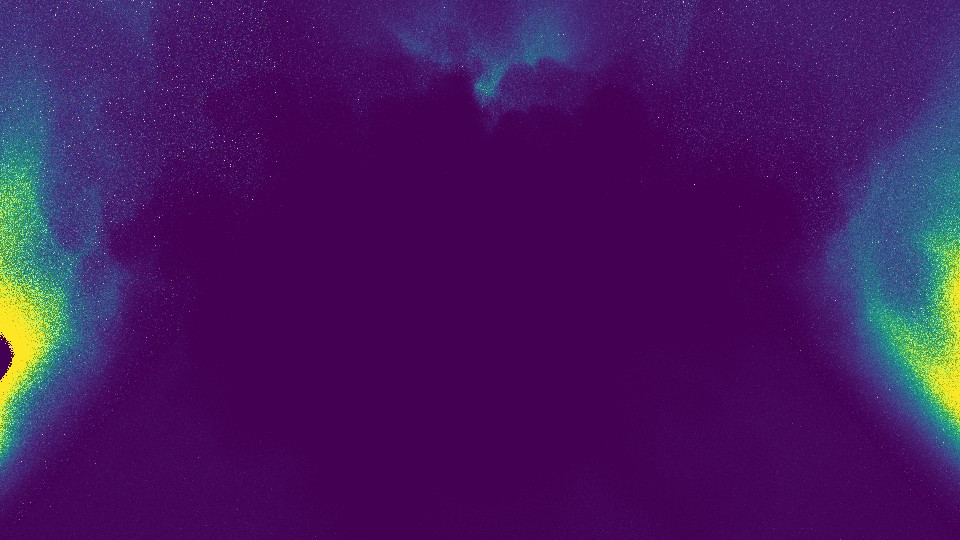}%
            \end{adjustbox}%
        \end{adjustbox}%
            &%
        \begin{adjustbox}{valign=t}%
            \begin{adjustbox}{totalheight=1.0\beautyHeight+2\bigFigureLengthB,tabular={c}}
                \includegraphics[height=\beautyHeight/3]{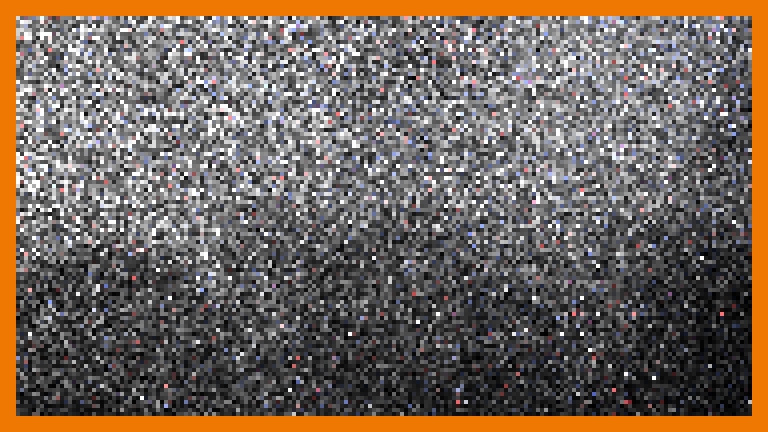}\vspace{\bigFigureLengthB}\\
                \includegraphics[height=\beautyHeight/3]{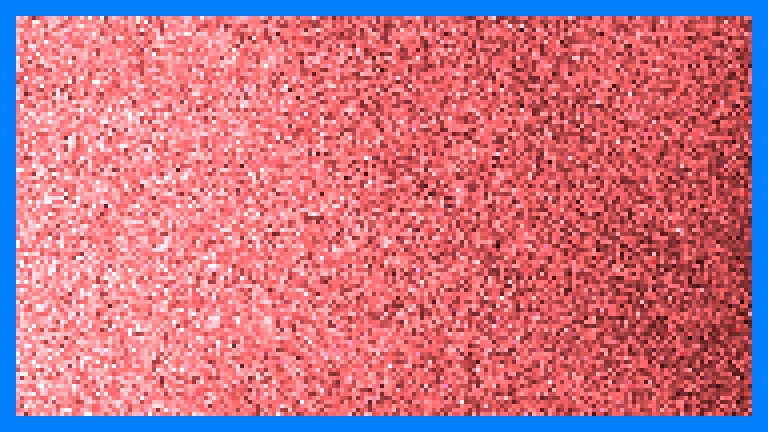}\vspace{\bigFigureLengthB}\\
                \includegraphics[height=\beautyHeight/3]{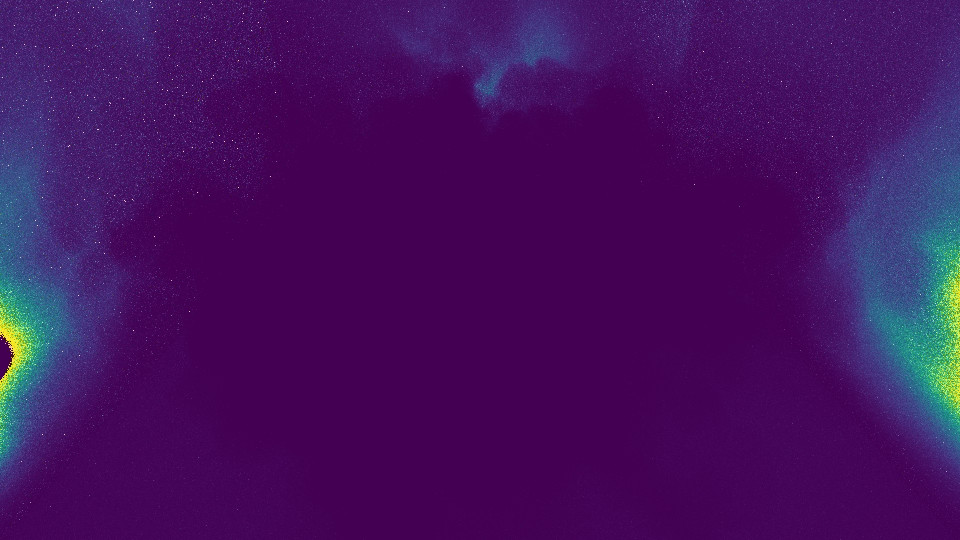}%
            \end{adjustbox}%
        \end{adjustbox}%
            &%
        \begin{adjustbox}{valign=t}%
            \begin{adjustbox}{totalheight=1.0\beautyHeight+2\bigFigureLengthB,tabular={c}}
                \includegraphics[height=\beautyHeight/3]{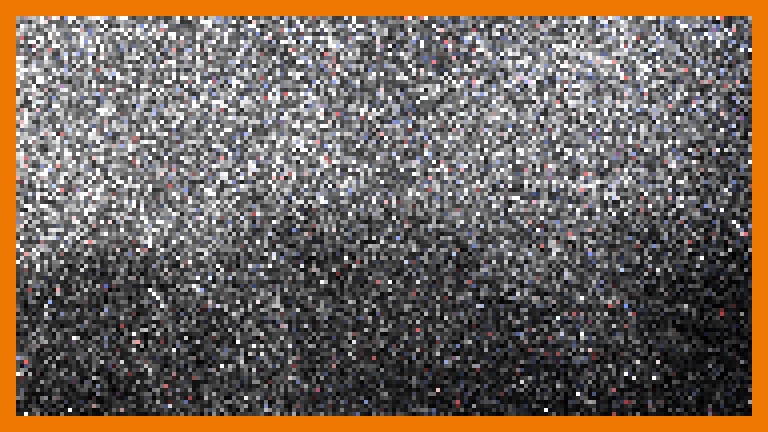}\vspace{\bigFigureLengthB}\\
                \includegraphics[height=\beautyHeight/3]{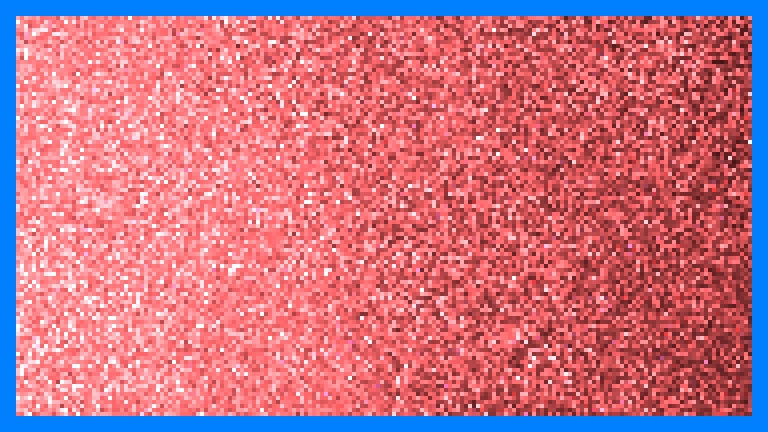}\vspace{\bigFigureLengthB}\\
                \includegraphics[height=\beautyHeight/3]{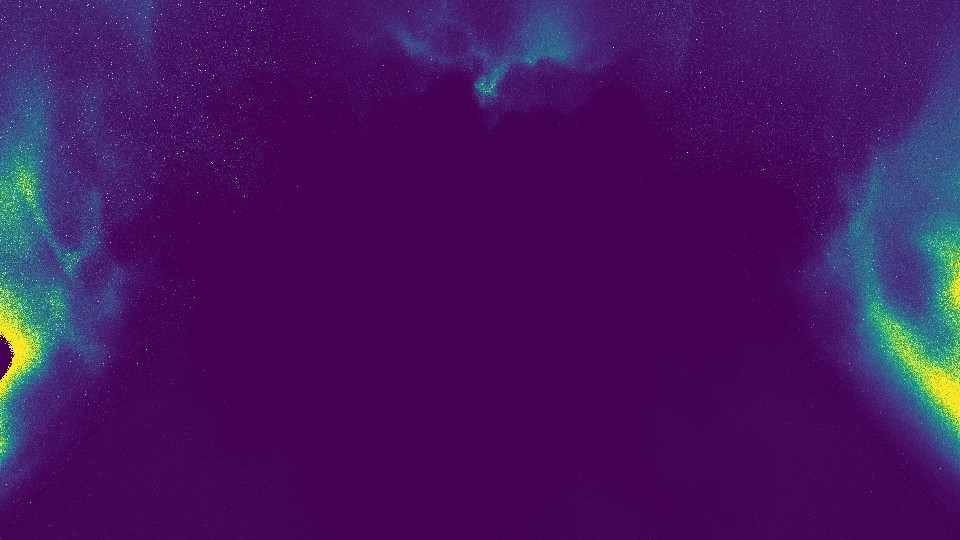}%
            \end{adjustbox}%
        \end{adjustbox}%
            &%
        \begin{adjustbox}{valign=t}%
            \begin{adjustbox}{totalheight=1.0\beautyHeight+2\bigFigureLengthB,tabular={c}}
                \includegraphics[height=\beautyHeight/3]{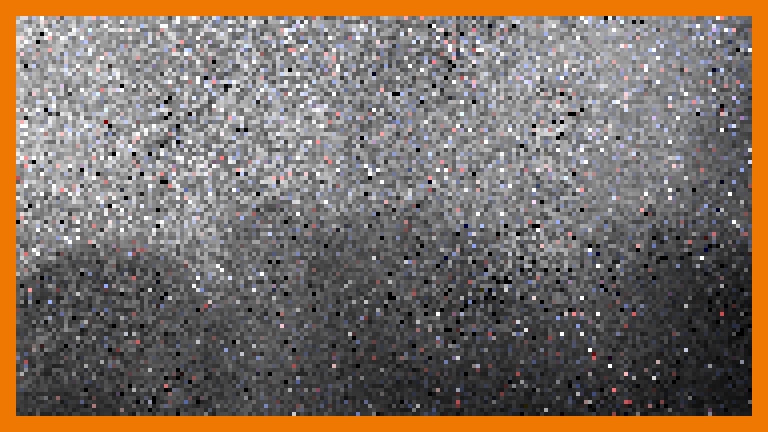}\vspace{\bigFigureLengthB}\\
                \includegraphics[height=\beautyHeight/3]{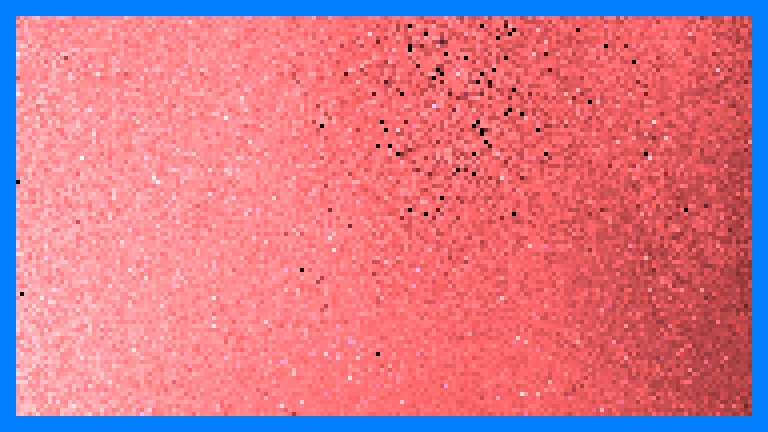}\vspace{\bigFigureLengthB}\\
                \includegraphics[height=\beautyHeight/3]{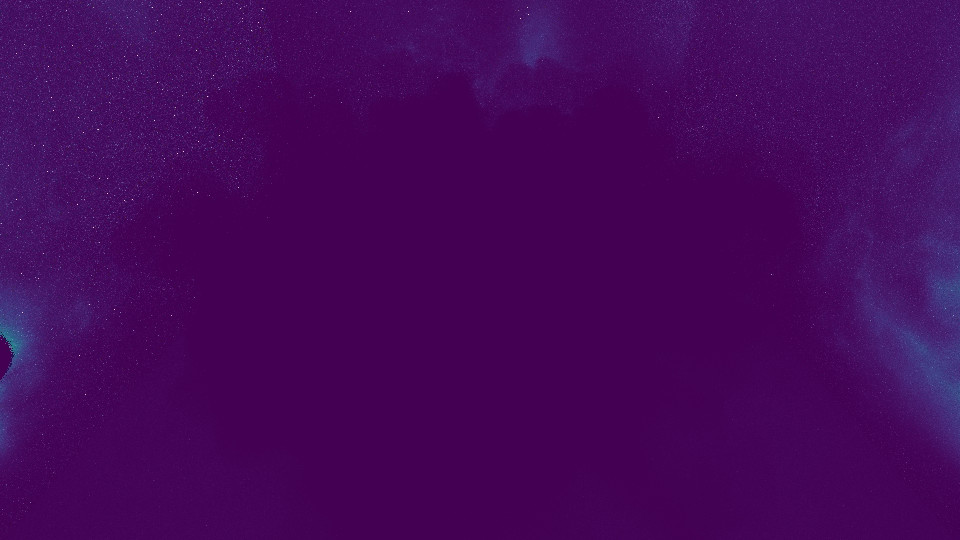}%
            \end{adjustbox}%
        \end{adjustbox}%
            &%
        \begin{adjustbox}{valign=t}%
            \begin{adjustbox}{totalheight=1.0\beautyHeight+2\bigFigureLengthB,tabular={c}}
                \includegraphics[height=\beautyHeight/3]{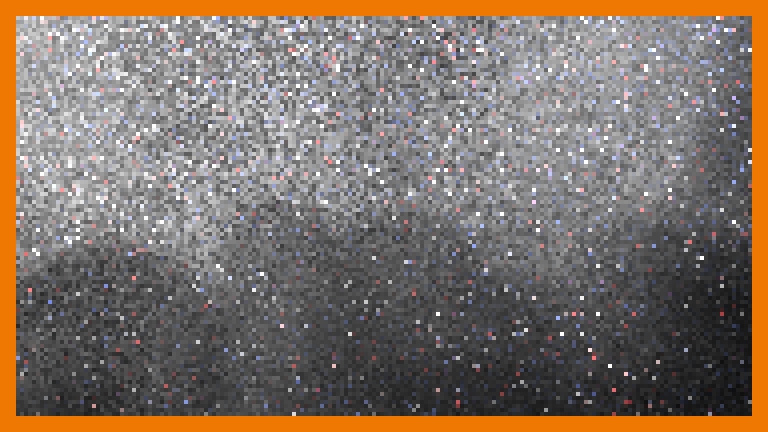}\vspace{\bigFigureLengthB}\\
                \includegraphics[height=\beautyHeight/3]{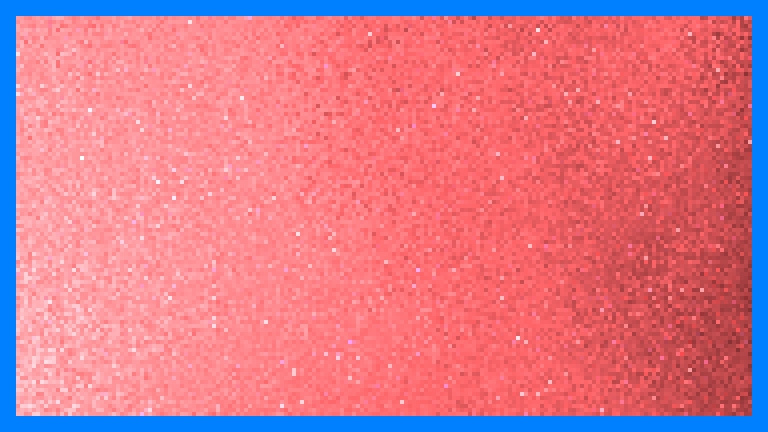}\vspace{\bigFigureLengthB}\\
                \includegraphics[height=\beautyHeight/3]{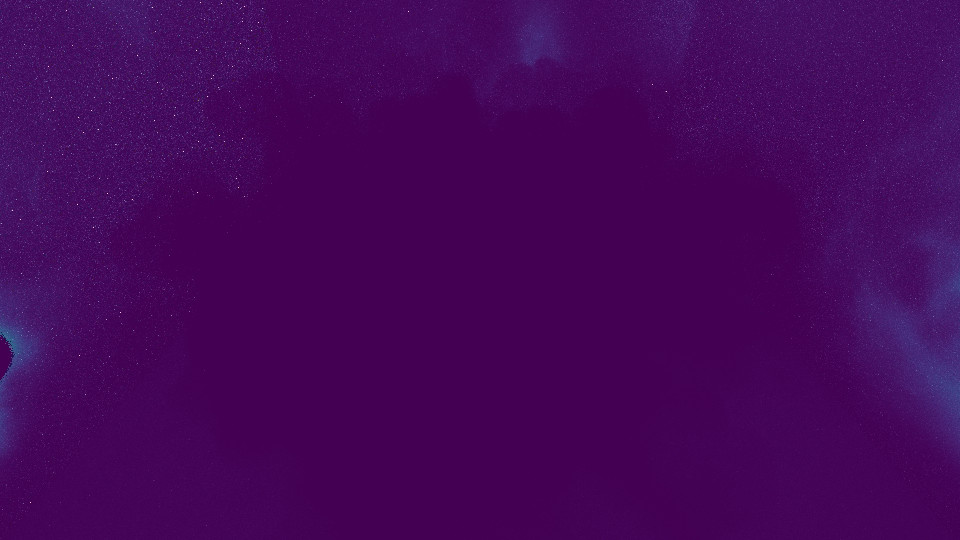}%
            \end{adjustbox}%
        \end{adjustbox}%
        \vspace{1mm}
    \\%
    &Variance (MSE):\hspace{\bigFigureLengthA}\strut%
    &
    &0.200
    &0.131
    &0.170
    &0.0518
    &0.0465
    \\[1pt]
    &Cost:\hspace{\bigFigureLengthA}\strut%
    &
    &46.2
    &46.2
    &45.2
    &46.5
    &46.2
    \\[1pt]
    &Inv. efficiency:\hspace{\bigFigureLengthA}\strut%
    &
    &8.96
    &5.79
    &7.35
    &2.27
    &2.02
    \\[6pt]
	\begin{adjustbox}{valign=t}%
        \rotatebox{90}{%
            \makebox[\beautyHeight]{\textsc{Cloud}}%
        }%
    \end{adjustbox}\,\,%
    &%
    \begin{adjustbox}{valign=t}%
        \includegraphics[height=\beautyHeight+2\bigFigureLengthB]{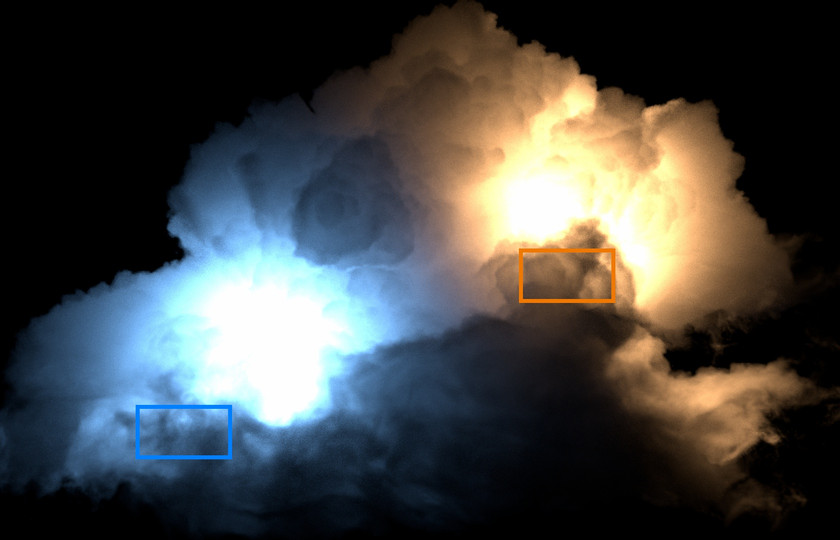}%
    \end{adjustbox}%
    \hspace{\bigFigureLengthA}\strut%
    &%
    \begin{adjustbox}{valign=t}%
        \rotatebox{90}{%
        	\makebox[\beautyHeight/3]{Inv.eff.}\hspace{\bigFigureLengthB}%
        	\makebox[\beautyHeight/3]{Inset 2}\hspace{\bigFigureLengthB}%
        	\makebox[\beautyHeight/3]{Inset 1}%
        }%
    \end{adjustbox}%
            &%
        \begin{adjustbox}{valign=t}%
            \begin{adjustbox}{totalheight=1.0\beautyHeight+2\bigFigureLengthB,tabular={c}}
                \includegraphics[height=\beautyHeight/3]{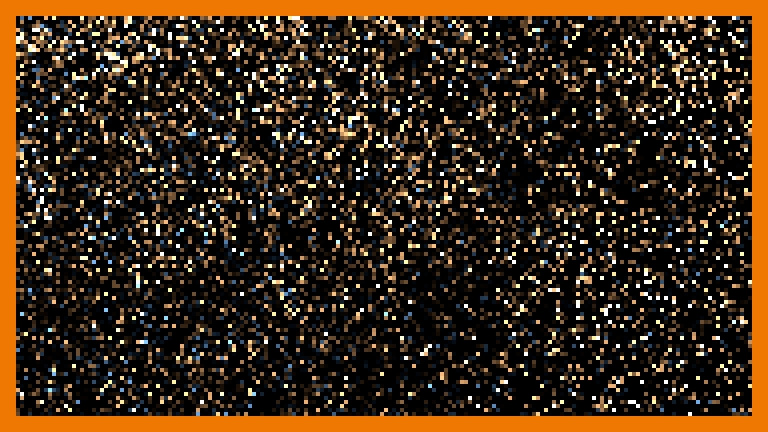}\vspace{\bigFigureLengthB}\\
                \includegraphics[height=\beautyHeight/3]{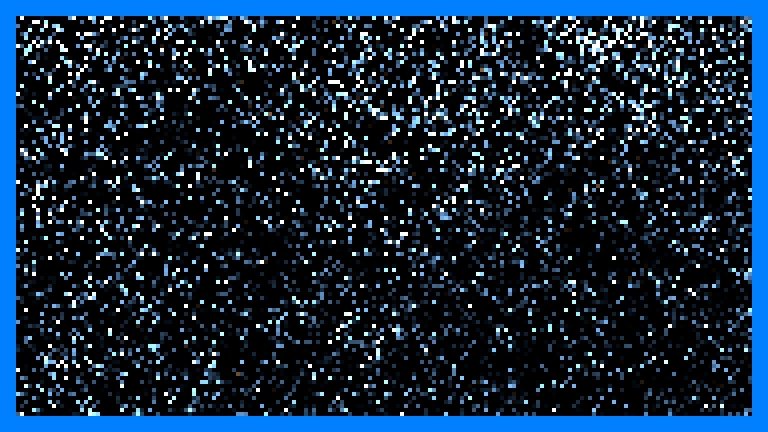}\vspace{\bigFigureLengthB}\\
                \includegraphics[height=\beautyHeight/3]{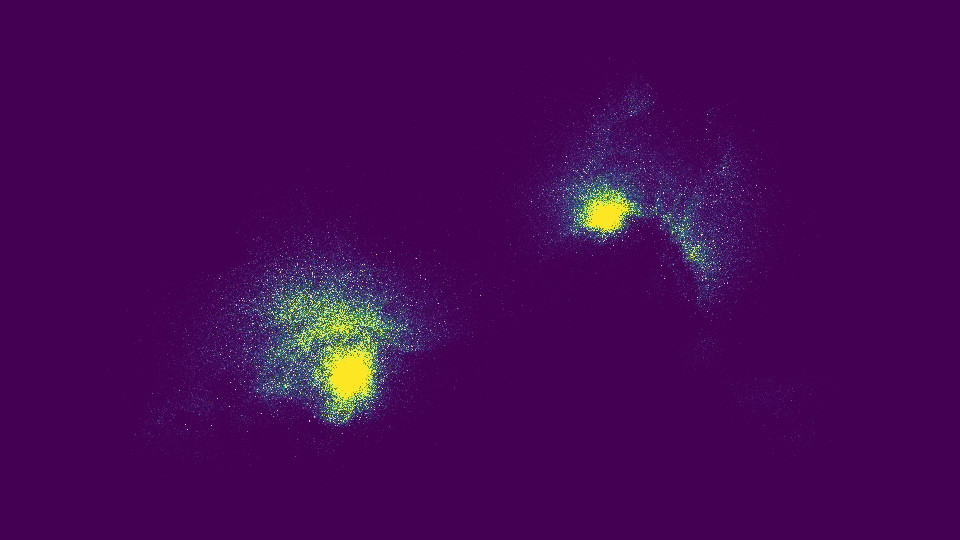}%
            \end{adjustbox}%
        \end{adjustbox}%
            &%
        \begin{adjustbox}{valign=t}%
            \begin{adjustbox}{totalheight=1.0\beautyHeight+2\bigFigureLengthB,tabular={c}}
                \includegraphics[height=\beautyHeight/3]{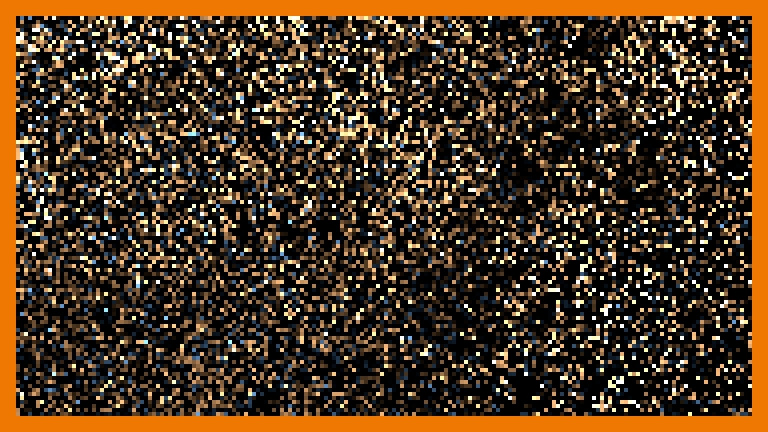}\vspace{\bigFigureLengthB}\\
                \includegraphics[height=\beautyHeight/3]{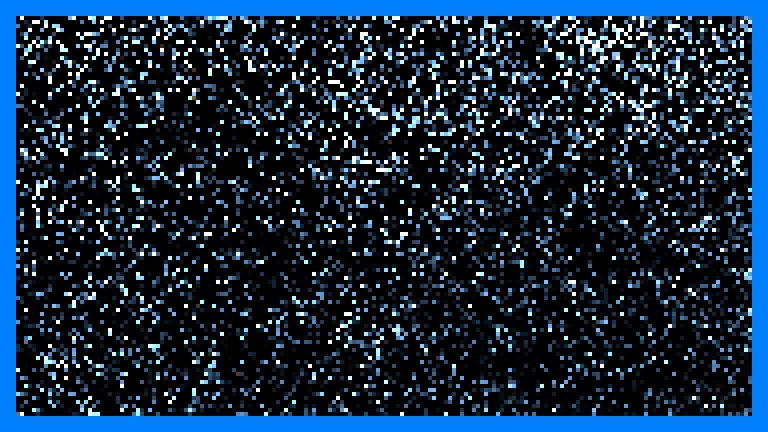}\vspace{\bigFigureLengthB}\\
                \includegraphics[height=\beautyHeight/3]{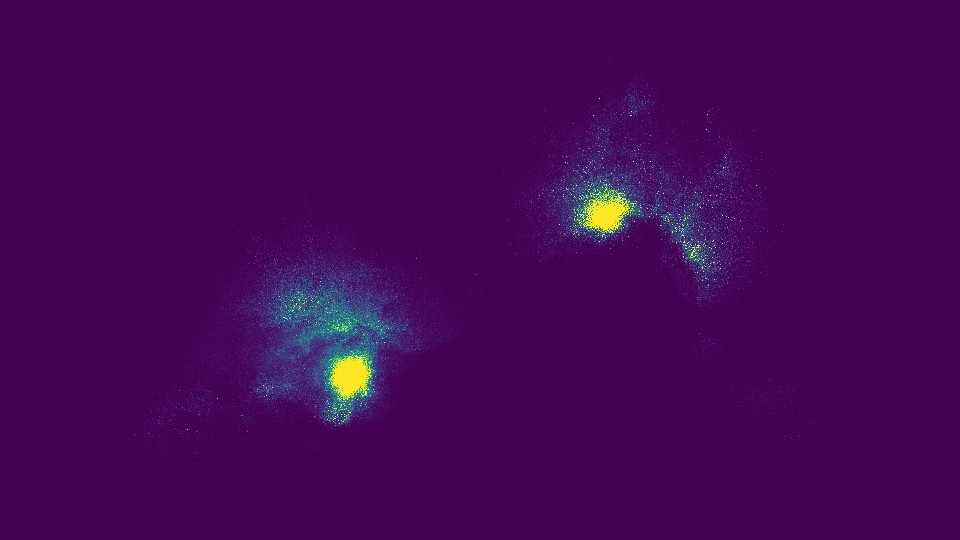}%
            \end{adjustbox}%
        \end{adjustbox}%
            &%
        \begin{adjustbox}{valign=t}%
            \begin{adjustbox}{totalheight=1.0\beautyHeight+2\bigFigureLengthB,tabular={c}}
                \includegraphics[height=\beautyHeight/3]{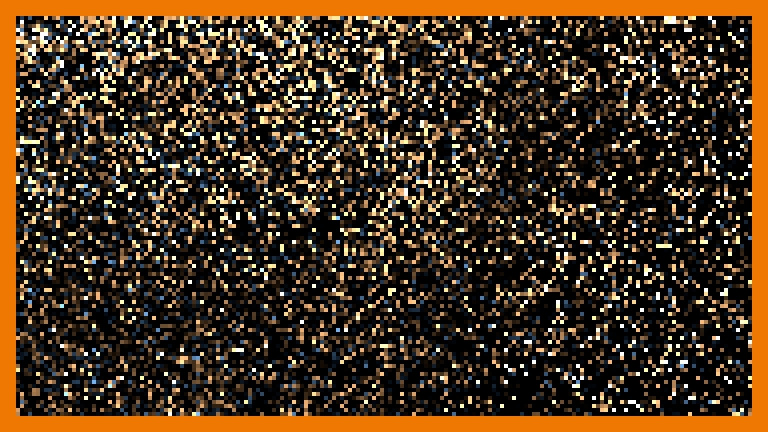}\vspace{\bigFigureLengthB}\\
                \includegraphics[height=\beautyHeight/3]{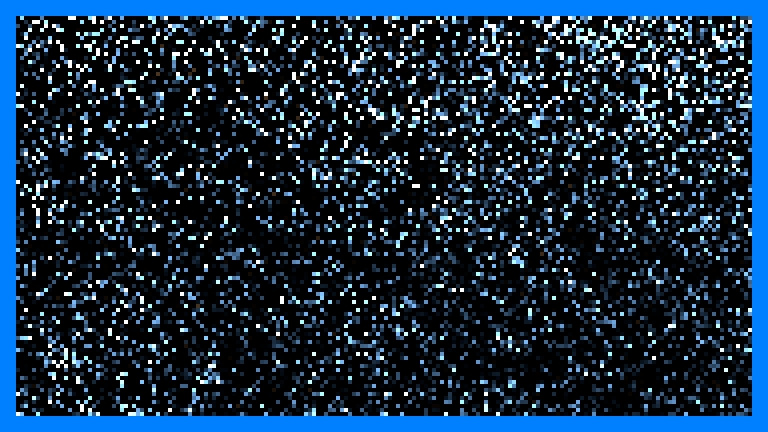}\vspace{\bigFigureLengthB}\\
                \includegraphics[height=\beautyHeight/3]{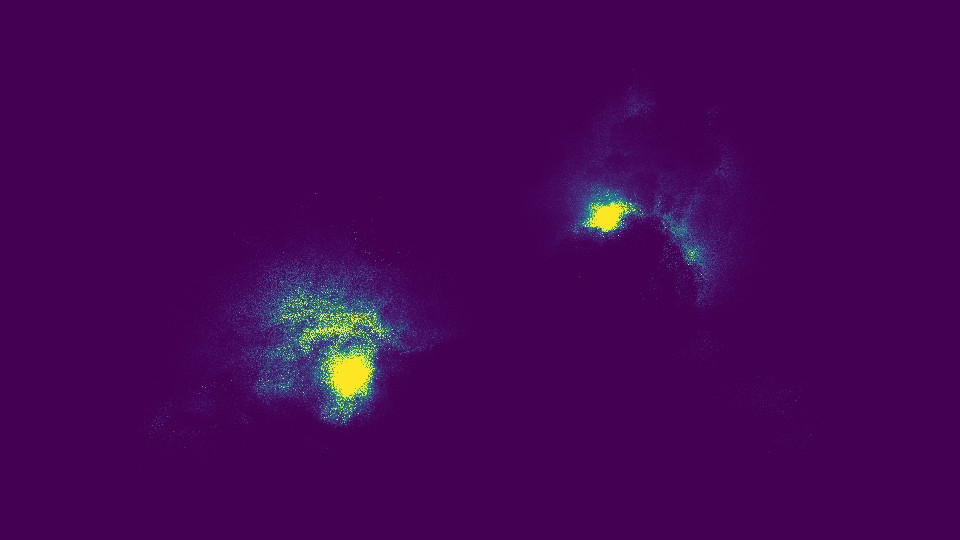}%
            \end{adjustbox}%
        \end{adjustbox}%
            &%
        \begin{adjustbox}{valign=t}%
            \begin{adjustbox}{totalheight=1.0\beautyHeight+2\bigFigureLengthB,tabular={c}}
                \includegraphics[height=\beautyHeight/3]{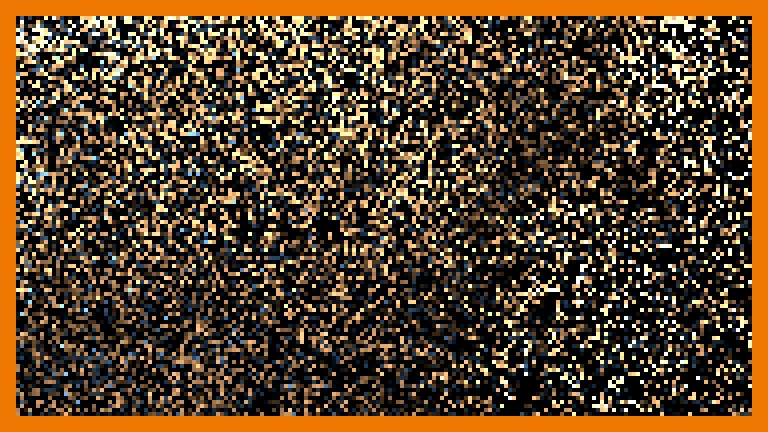}\vspace{\bigFigureLengthB}\\
                \includegraphics[height=\beautyHeight/3]{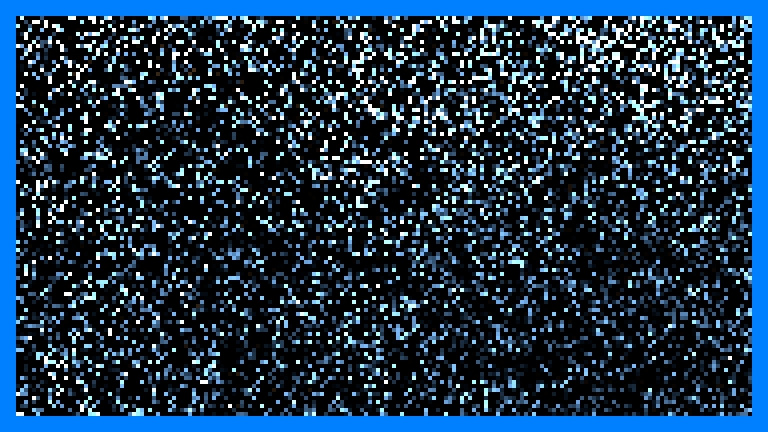}\vspace{\bigFigureLengthB}\\
                \includegraphics[height=\beautyHeight/3]{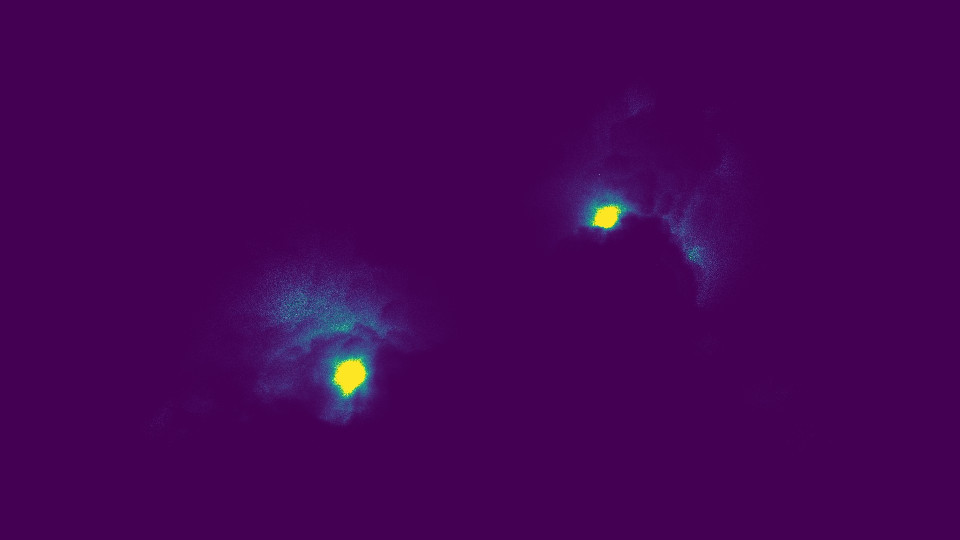}%
            \end{adjustbox}%
        \end{adjustbox}%
            &%
        \begin{adjustbox}{valign=t}%
            \begin{adjustbox}{totalheight=1.0\beautyHeight+2\bigFigureLengthB,tabular={c}}
                \includegraphics[height=\beautyHeight/3]{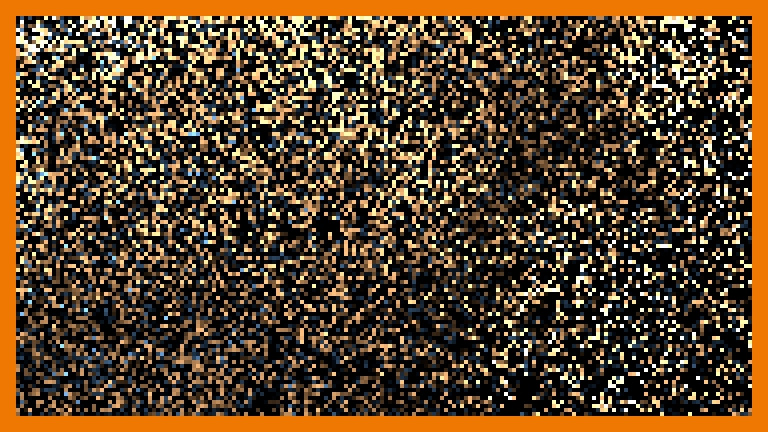}\vspace{\bigFigureLengthB}\\
                \includegraphics[height=\beautyHeight/3]{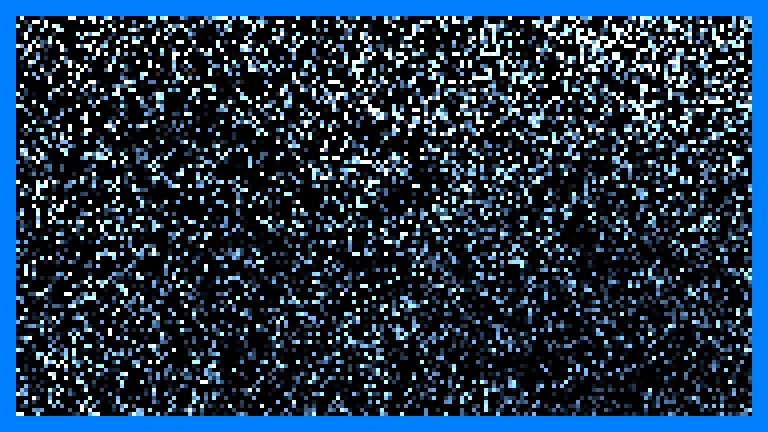}\vspace{\bigFigureLengthB}\\
                \includegraphics[height=\beautyHeight/3]{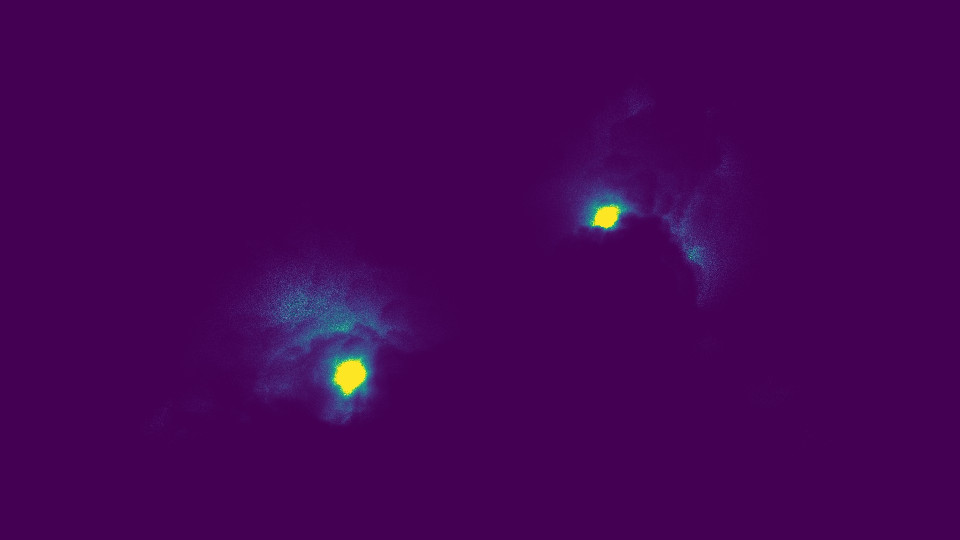}%
            \end{adjustbox}%
        \end{adjustbox}%
        \vspace{1mm}
    \\%
    &Variance (MSE):\hspace{\bigFigureLengthA}\strut%
    &
    &72.5
    &43.0
    &38.5
    &24.7
    &15.5
    \\[1pt]
    &Cost:\hspace{\bigFigureLengthA}\strut%
    &
    &22.4
    &22.4
    &22.4
    &22.5
    &22.4
    \\[1pt]
    &Inv. efficiency:\hspace{\bigFigureLengthA}\strut%
    &
    &3.06e+3
    &1.79e+3
    &1.61e+3
    &1.01e+3
    &645
    \\[6pt]
	\begin{adjustbox}{valign=t}%
        \rotatebox{90}{%
            \makebox[\beautyHeight]{\textsc{Glass with Smoke}}%
        }%
    \end{adjustbox}\,\,%
    &%
    \begin{adjustbox}{valign=t}%
        \includegraphics[height=\beautyHeight+2\bigFigureLengthB]{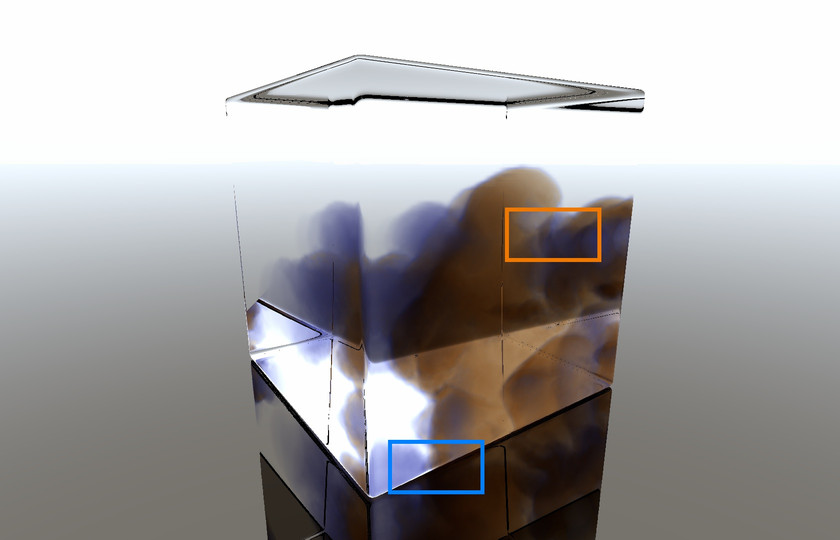}%
    \end{adjustbox}%
    \hspace{\bigFigureLengthA}\strut%
    &%
    \begin{adjustbox}{valign=t}%
        \rotatebox{90}{%
        	\makebox[\beautyHeight/3]{Inv.eff.}\hspace{\bigFigureLengthB}%
        	\makebox[\beautyHeight/3]{Inset 2}\hspace{\bigFigureLengthB}%
        	\makebox[\beautyHeight/3]{Inset 1}%
        }%
    \end{adjustbox}%
            &%
        \begin{adjustbox}{valign=t}%
            \begin{adjustbox}{totalheight=1.0\beautyHeight+2\bigFigureLengthB,tabular={c}}
                \includegraphics[height=\beautyHeight/3]{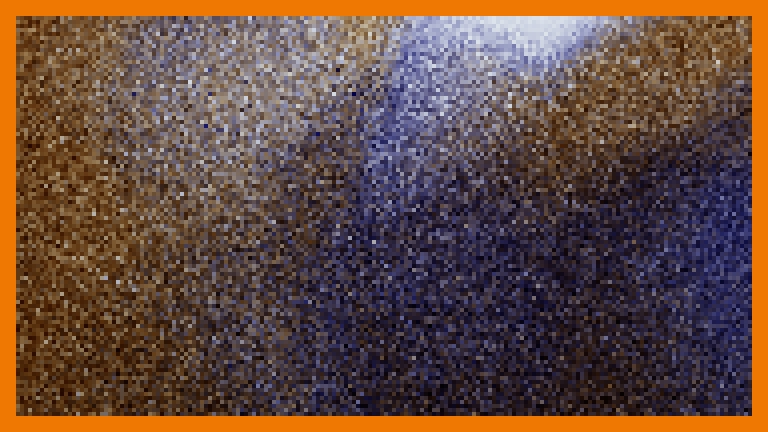}\vspace{\bigFigureLengthB}\\
                \includegraphics[height=\beautyHeight/3]{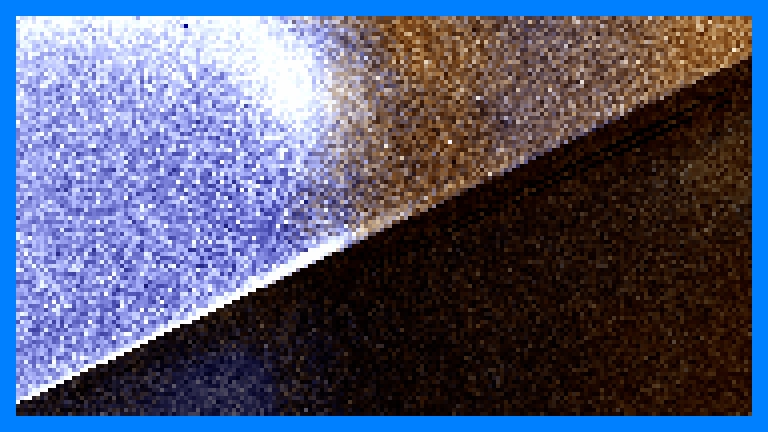}\vspace{\bigFigureLengthB}\\
                \includegraphics[height=\beautyHeight/3]{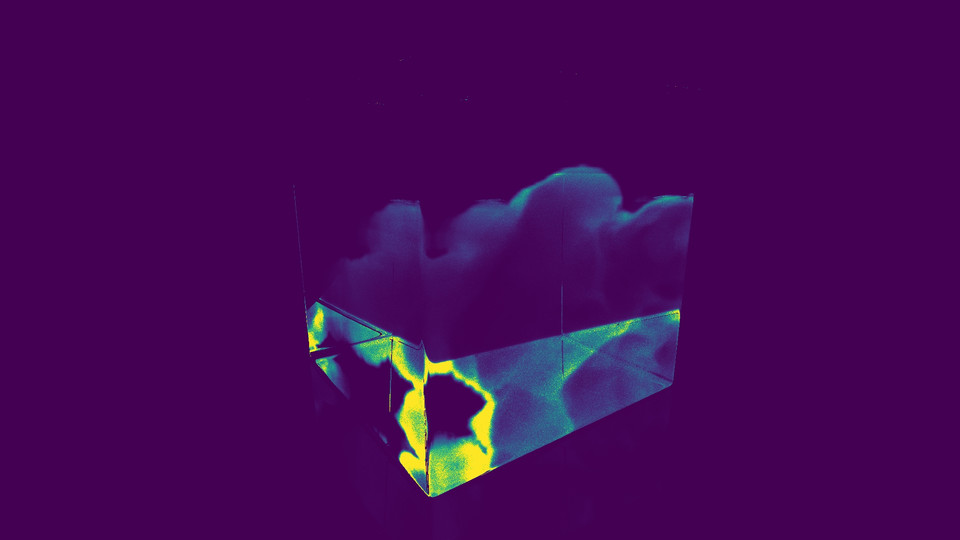}%
            \end{adjustbox}%
        \end{adjustbox}%
            &%
        \begin{adjustbox}{valign=t}%
            \begin{adjustbox}{totalheight=1.0\beautyHeight+2\bigFigureLengthB,tabular={c}}
                \includegraphics[height=\beautyHeight/3]{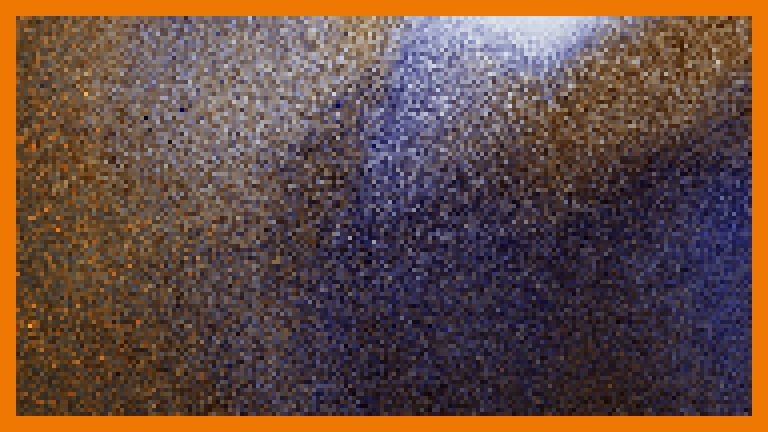}\vspace{\bigFigureLengthB}\\
                \includegraphics[height=\beautyHeight/3]{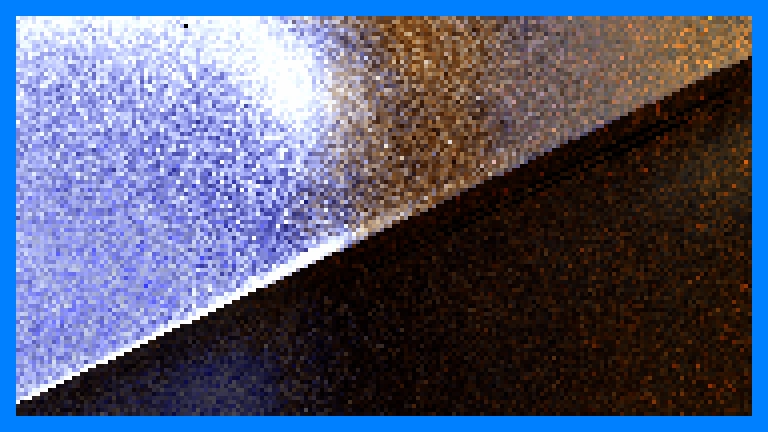}\vspace{\bigFigureLengthB}\\
                \includegraphics[height=\beautyHeight/3]{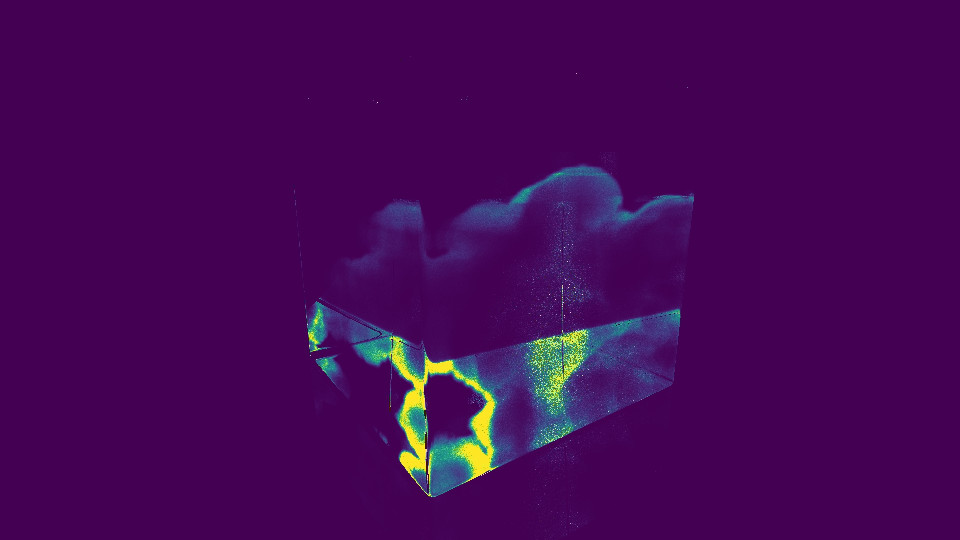}%
            \end{adjustbox}%
        \end{adjustbox}%
            &%
        \begin{adjustbox}{valign=t}%
            \begin{adjustbox}{totalheight=1.0\beautyHeight+2\bigFigureLengthB,tabular={c}}
                \includegraphics[height=\beautyHeight/3]{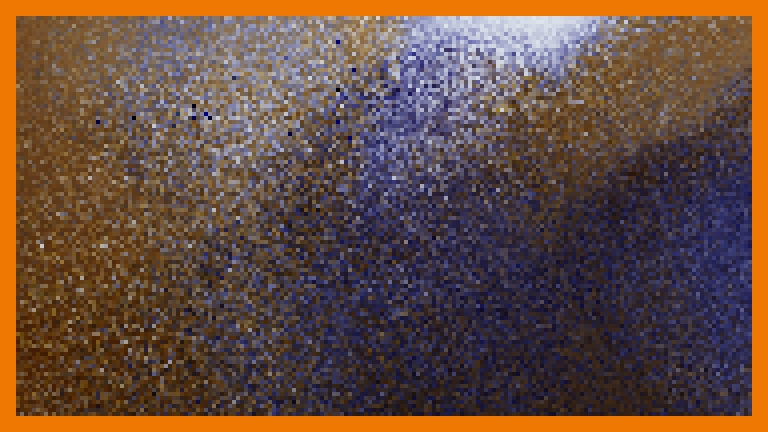}\vspace{\bigFigureLengthB}\\
                \includegraphics[height=\beautyHeight/3]{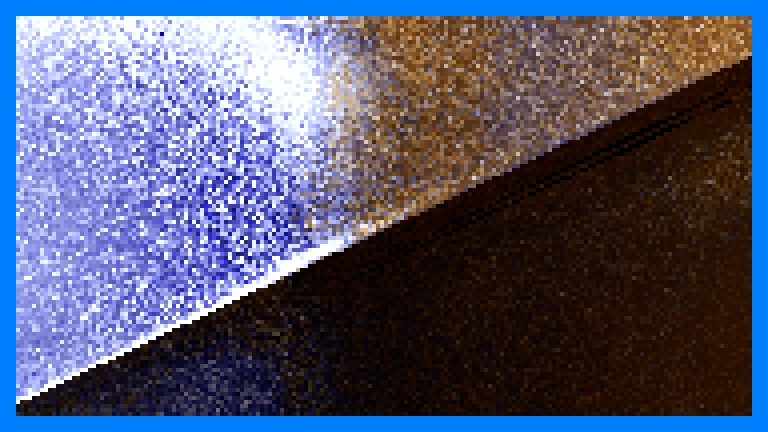}\vspace{\bigFigureLengthB}\\
                \includegraphics[height=\beautyHeight/3]{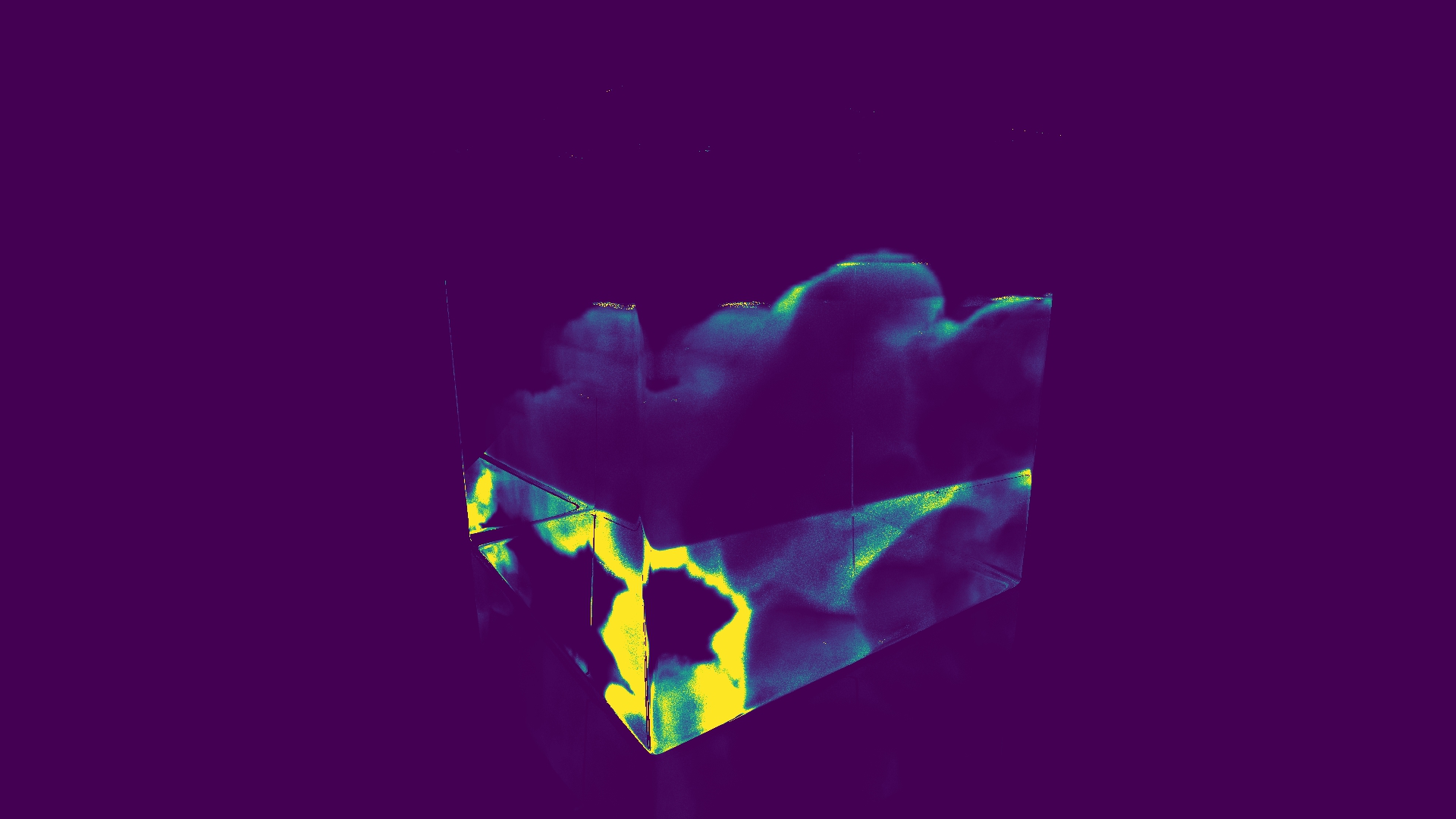}%
            \end{adjustbox}%
        \end{adjustbox}%
            &%
        \begin{adjustbox}{valign=t}%
            \begin{adjustbox}{totalheight=1.0\beautyHeight+2\bigFigureLengthB,tabular={c}}
                \includegraphics[height=\beautyHeight/3]{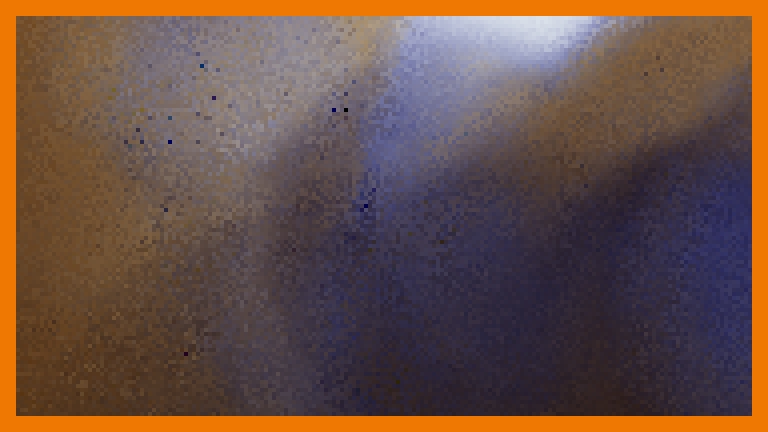}\vspace{\bigFigureLengthB}\\
                \includegraphics[height=\beautyHeight/3]{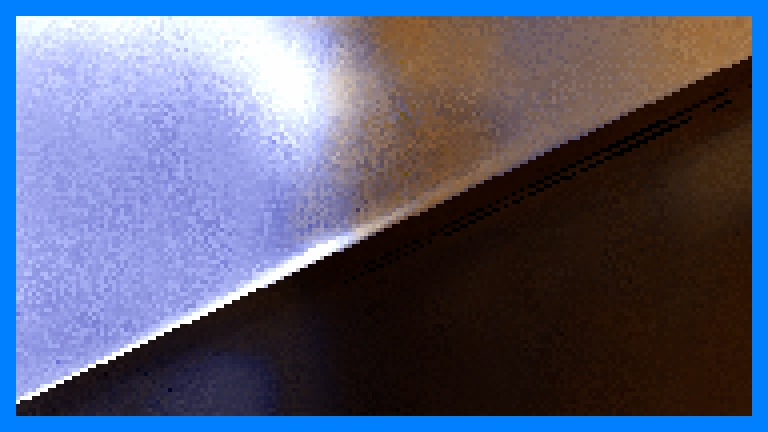}\vspace{\bigFigureLengthB}\\
                \includegraphics[height=\beautyHeight/3]{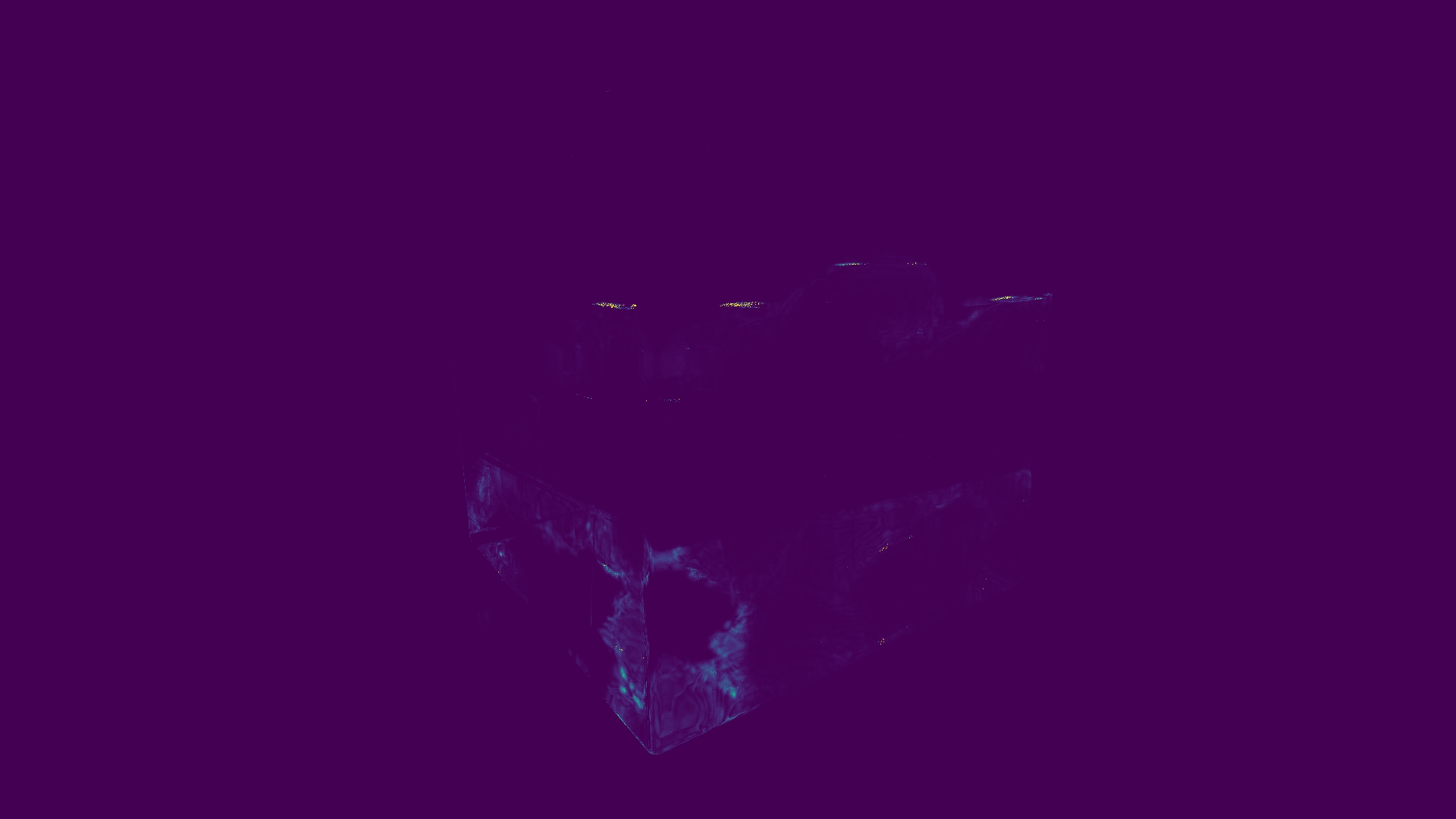}%
            \end{adjustbox}%
        \end{adjustbox}%
            &%
        \begin{adjustbox}{valign=t}%
            \begin{adjustbox}{totalheight=1.0\beautyHeight+2\bigFigureLengthB,tabular={c}}
                \includegraphics[height=\beautyHeight/3]{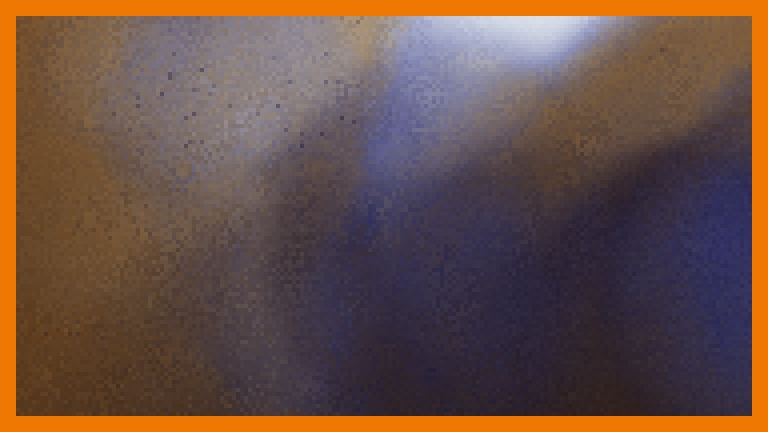}\vspace{\bigFigureLengthB}\\
                \includegraphics[height=\beautyHeight/3]{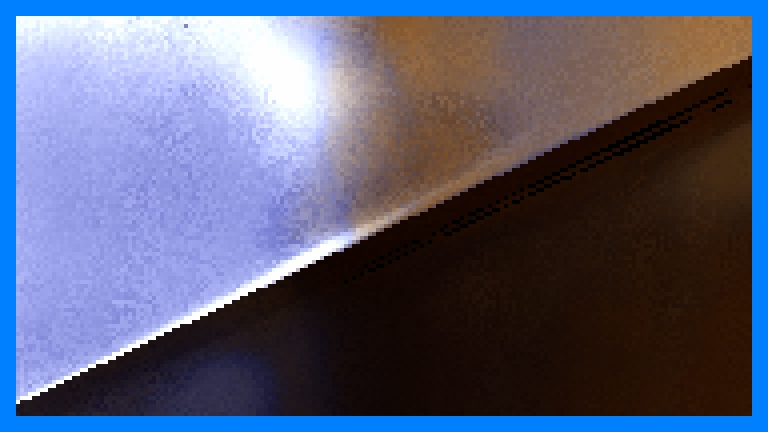}\vspace{\bigFigureLengthB}\\
                \includegraphics[height=\beautyHeight/3]{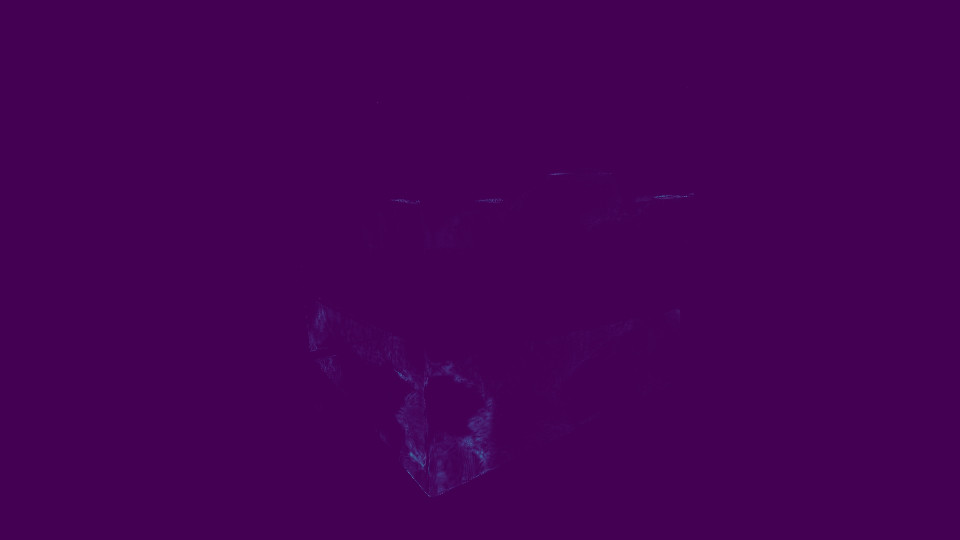}%
            \end{adjustbox}%
        \end{adjustbox}%
        \vspace{1mm}
    \\%
    &Variance (MSE):\hspace{\bigFigureLengthA}\strut%
    &
    &1.24e-3
    &1.12e-3
    &1.43e-3
    &1.02e-4
    &4.15e-5
    \\[1pt]
    &Cost:\hspace{\bigFigureLengthA}\strut%
    &
    &3.35
    &3.35
    &3.28
    &3.37
    &3.35
    \\[1pt]
    &Inv. efficiency:\hspace{\bigFigureLengthA}\strut%
    &
    &0.0167
    &0.0152
    &0.0183
    &1.14e-3
    &5.22e-4
    \\[6pt]
\end{tabular}
\caption{\bigFigureCaption}
\end{figure*}

    Our unbiased estimator obtains an MSE reduction between ~1.5 and ~13x across all scenes compared to previous state-of-the-art method for each scene. 
    Our biased estimator provides \emph{additional} improvement of ~1.1 to ~2x on top of that.
    Note that the MSE values include also other sources of noise (such as from global illumination in the Box scene, or single-scattering in the Cloud scene), which partly masks the improvements in transmittance estimation.

    \begin{figure}
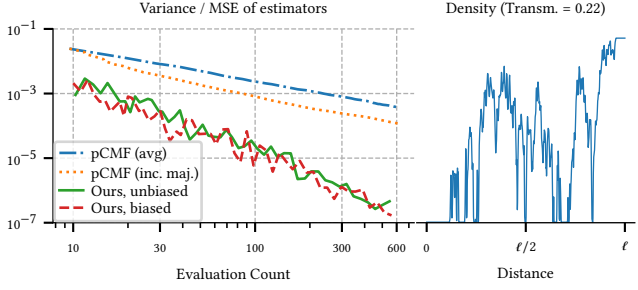

        \begin{center}
            \input{figures/plots/scene1longrange.pgf}%
            \input{figures/plots/scene1.pgf}
        \end{center}
        \caption{A graph of variance (respectively MSE) of our unbiased and biased estimators as well as that of Georgiev et al's p-series CMF as a function of sample count.
        For p-series CMF, we display two methods of increasing the expected sample count: the first (dashed blue line) is by averaging multiple evaluations, the second (dashed yellow line) is increasing the control optical thickness (in this case the majorant). 
        Both our estimators display a faster convergence rate.
        }
        \label{fig:graph-test}
    \end{figure}
    
    \begin{figure}
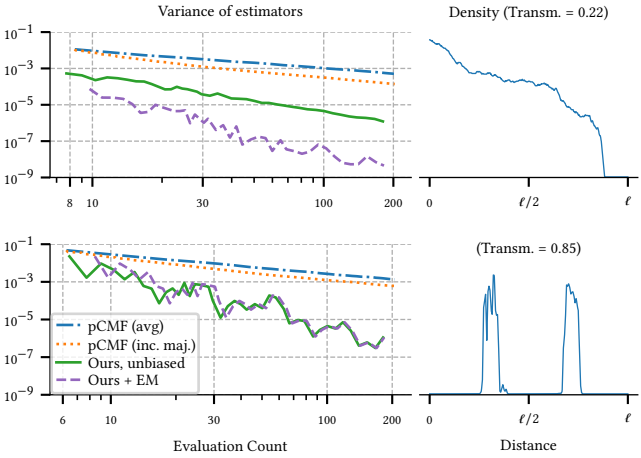

        \begin{center}
            \input{figures/plots/scene2shortrange.pgf}
            \input{figures/plots/scene2.pgf}
            \input{figures/plots/scene3shortrange.pgf} 
\begingroup%
\makeatletter%
\begin{pgfpicture}%
\pgfpathrectangle{\pgfpointorigin}{\pgfqpoint{1.155000in}{1.128600in}}%
\pgfusepath{use as bounding box, clip}%
\begin{pgfscope}%
\pgfsetbuttcap%
\pgfsetmiterjoin%
\definecolor{currentfill}{rgb}{1.000000,1.000000,1.000000}%
\pgfsetfillcolor{currentfill}%
\pgfsetlinewidth{0.000000pt}%
\definecolor{currentstroke}{rgb}{1.000000,1.000000,1.000000}%
\pgfsetstrokecolor{currentstroke}%
\pgfsetdash{}{0pt}%
\pgfpathmoveto{\pgfqpoint{0.000000in}{0.000000in}}%
\pgfpathlineto{\pgfqpoint{1.155000in}{0.000000in}}%
\pgfpathlineto{\pgfqpoint{1.155000in}{1.128600in}}%
\pgfpathlineto{\pgfqpoint{0.000000in}{1.128600in}}%
\pgfpathclose%
\pgfusepath{fill}%
\end{pgfscope}%
\begin{pgfscope}%
\pgfsetbuttcap%
\pgfsetmiterjoin%
\definecolor{currentfill}{rgb}{1.000000,1.000000,1.000000}%
\pgfsetfillcolor{currentfill}%
\pgfsetlinewidth{0.000000pt}%
\definecolor{currentstroke}{rgb}{0.000000,0.000000,0.000000}%
\pgfsetstrokecolor{currentstroke}%
\pgfsetstrokeopacity{0.000000}%
\pgfsetdash{}{0pt}%
\pgfpathmoveto{\pgfqpoint{0.006944in}{0.317130in}}%
\pgfpathlineto{\pgfqpoint{1.148056in}{0.317130in}}%
\pgfpathlineto{\pgfqpoint{1.148056in}{0.975822in}}%
\pgfpathlineto{\pgfqpoint{0.006944in}{0.975822in}}%
\pgfpathclose%
\pgfusepath{fill}%
\end{pgfscope}%
\begin{pgfscope}%
\pgfsetbuttcap%
\pgfsetroundjoin%
\definecolor{currentfill}{rgb}{0.000000,0.000000,0.000000}%
\pgfsetfillcolor{currentfill}%
\pgfsetlinewidth{0.803000pt}%
\definecolor{currentstroke}{rgb}{0.000000,0.000000,0.000000}%
\pgfsetstrokecolor{currentstroke}%
\pgfsetdash{}{0pt}%
\pgfsys@defobject{currentmarker}{\pgfqpoint{0.000000in}{-0.048611in}}{\pgfqpoint{0.000000in}{0.000000in}}{%
\pgfpathmoveto{\pgfqpoint{0.000000in}{0.000000in}}%
\pgfpathlineto{\pgfqpoint{0.000000in}{-0.048611in}}%
\pgfusepath{stroke,fill}%
}%
\begin{pgfscope}%
\pgfsys@transformshift{0.058813in}{0.317130in}%
\pgfsys@useobject{currentmarker}{}%
\end{pgfscope}%
\end{pgfscope}%
\begin{pgfscope}%
\definecolor{textcolor}{rgb}{0.000000,0.000000,0.000000}%
\pgfsetstrokecolor{textcolor}%
\pgfsetfillcolor{textcolor}%
\pgftext[x=0.058813in,y=0.219907in,,top]{\color{textcolor}\rmfamily\fontsize{5.000000}{6.000000}\selectfont \strut0\strut}%
\end{pgfscope}%
\begin{pgfscope}%
\pgfsetbuttcap%
\pgfsetroundjoin%
\definecolor{currentfill}{rgb}{0.000000,0.000000,0.000000}%
\pgfsetfillcolor{currentfill}%
\pgfsetlinewidth{0.803000pt}%
\definecolor{currentstroke}{rgb}{0.000000,0.000000,0.000000}%
\pgfsetstrokecolor{currentstroke}%
\pgfsetdash{}{0pt}%
\pgfsys@defobject{currentmarker}{\pgfqpoint{0.000000in}{-0.048611in}}{\pgfqpoint{0.000000in}{0.000000in}}{%
\pgfpathmoveto{\pgfqpoint{0.000000in}{0.000000in}}%
\pgfpathlineto{\pgfqpoint{0.000000in}{-0.048611in}}%
\pgfusepath{stroke,fill}%
}%
\begin{pgfscope}%
\pgfsys@transformshift{0.577500in}{0.317130in}%
\pgfsys@useobject{currentmarker}{}%
\end{pgfscope}%
\end{pgfscope}%
\begin{pgfscope}%
\definecolor{textcolor}{rgb}{0.000000,0.000000,0.000000}%
\pgfsetstrokecolor{textcolor}%
\pgfsetfillcolor{textcolor}%
\pgftext[x=0.577500in,y=0.219907in,,top]{\color{textcolor}\rmfamily\fontsize{5.000000}{6.000000}\selectfont \(\displaystyle \ell/2\)}%
\end{pgfscope}%
\begin{pgfscope}%
\pgfsetbuttcap%
\pgfsetroundjoin%
\definecolor{currentfill}{rgb}{0.000000,0.000000,0.000000}%
\pgfsetfillcolor{currentfill}%
\pgfsetlinewidth{0.803000pt}%
\definecolor{currentstroke}{rgb}{0.000000,0.000000,0.000000}%
\pgfsetstrokecolor{currentstroke}%
\pgfsetdash{}{0pt}%
\pgfsys@defobject{currentmarker}{\pgfqpoint{0.000000in}{-0.048611in}}{\pgfqpoint{0.000000in}{0.000000in}}{%
\pgfpathmoveto{\pgfqpoint{0.000000in}{0.000000in}}%
\pgfpathlineto{\pgfqpoint{0.000000in}{-0.048611in}}%
\pgfusepath{stroke,fill}%
}%
\begin{pgfscope}%
\pgfsys@transformshift{1.096187in}{0.317130in}%
\pgfsys@useobject{currentmarker}{}%
\end{pgfscope}%
\end{pgfscope}%
\begin{pgfscope}%
\definecolor{textcolor}{rgb}{0.000000,0.000000,0.000000}%
\pgfsetstrokecolor{textcolor}%
\pgfsetfillcolor{textcolor}%
\pgftext[x=1.096187in,y=0.219907in,,top]{\color{textcolor}\rmfamily\fontsize{5.000000}{6.000000}\selectfont \(\displaystyle \ell\)}%
\end{pgfscope}%
\begin{pgfscope}%
\definecolor{textcolor}{rgb}{0.000000,0.000000,0.000000}%
\pgfsetstrokecolor{textcolor}%
\pgfsetfillcolor{textcolor}%
\pgftext[x=0.577500in,y=0.081019in,,top]{\color{textcolor}\rmfamily\fontsize{6.000000}{7.200000}\selectfont Distance}%
\end{pgfscope}%
\begin{pgfscope}%
\pgfpathrectangle{\pgfqpoint{0.006944in}{0.317130in}}{\pgfqpoint{1.141111in}{0.658693in}}%
\pgfusepath{clip}%
\pgfsetrectcap%
\pgfsetroundjoin%
\pgfsetlinewidth{0.501875pt}%
\definecolor{currentstroke}{rgb}{0.121569,0.466667,0.705882}%
\pgfsetstrokecolor{currentstroke}%
\pgfsetdash{}{0pt}%
\pgfpathmoveto{\pgfqpoint{0.058813in}{0.321933in}}%
\pgfpathlineto{\pgfqpoint{0.319455in}{0.322672in}}%
\pgfpathlineto{\pgfqpoint{0.324647in}{0.326886in}}%
\pgfpathlineto{\pgfqpoint{0.328800in}{0.335382in}}%
\pgfpathlineto{\pgfqpoint{0.330877in}{0.351805in}}%
\pgfpathlineto{\pgfqpoint{0.335031in}{0.395306in}}%
\pgfpathlineto{\pgfqpoint{0.341261in}{0.774317in}}%
\pgfpathlineto{\pgfqpoint{0.350607in}{0.837484in}}%
\pgfpathlineto{\pgfqpoint{0.351645in}{0.843445in}}%
\pgfpathlineto{\pgfqpoint{0.353722in}{0.766811in}}%
\pgfpathlineto{\pgfqpoint{0.357876in}{0.602319in}}%
\pgfpathlineto{\pgfqpoint{0.364106in}{0.857707in}}%
\pgfpathlineto{\pgfqpoint{0.369298in}{0.885978in}}%
\pgfpathlineto{\pgfqpoint{0.374490in}{0.864921in}}%
\pgfpathlineto{\pgfqpoint{0.381759in}{0.754902in}}%
\pgfpathlineto{\pgfqpoint{0.385913in}{0.738713in}}%
\pgfpathlineto{\pgfqpoint{0.392143in}{0.944685in}}%
\pgfpathlineto{\pgfqpoint{0.396297in}{0.943129in}}%
\pgfpathlineto{\pgfqpoint{0.397335in}{0.934932in}}%
\pgfpathlineto{\pgfqpoint{0.403566in}{0.758964in}}%
\pgfpathlineto{\pgfqpoint{0.408758in}{0.733817in}}%
\pgfpathlineto{\pgfqpoint{0.413950in}{0.737705in}}%
\pgfpathlineto{\pgfqpoint{0.420181in}{0.553671in}}%
\pgfpathlineto{\pgfqpoint{0.426411in}{0.336234in}}%
\pgfpathlineto{\pgfqpoint{0.430565in}{0.330674in}}%
\pgfpathlineto{\pgfqpoint{0.432642in}{0.337241in}}%
\pgfpathlineto{\pgfqpoint{0.440949in}{0.366771in}}%
\pgfpathlineto{\pgfqpoint{0.441987in}{0.369983in}}%
\pgfpathlineto{\pgfqpoint{0.445102in}{0.356535in}}%
\pgfpathlineto{\pgfqpoint{0.449256in}{0.341349in}}%
\pgfpathlineto{\pgfqpoint{0.453410in}{0.335501in}}%
\pgfpathlineto{\pgfqpoint{0.458602in}{0.346106in}}%
\pgfpathlineto{\pgfqpoint{0.459640in}{0.345113in}}%
\pgfpathlineto{\pgfqpoint{0.465871in}{0.324957in}}%
\pgfpathlineto{\pgfqpoint{0.472101in}{0.321933in}}%
\pgfpathlineto{\pgfqpoint{0.726512in}{0.322778in}}%
\pgfpathlineto{\pgfqpoint{0.732743in}{0.324013in}}%
\pgfpathlineto{\pgfqpoint{0.747280in}{0.325116in}}%
\pgfpathlineto{\pgfqpoint{0.751434in}{0.333504in}}%
\pgfpathlineto{\pgfqpoint{0.752472in}{0.342859in}}%
\pgfpathlineto{\pgfqpoint{0.758703in}{0.561302in}}%
\pgfpathlineto{\pgfqpoint{0.764933in}{0.630962in}}%
\pgfpathlineto{\pgfqpoint{0.770125in}{0.734541in}}%
\pgfpathlineto{\pgfqpoint{0.775318in}{0.892558in}}%
\pgfpathlineto{\pgfqpoint{0.780510in}{0.885603in}}%
\pgfpathlineto{\pgfqpoint{0.785702in}{0.898238in}}%
\pgfpathlineto{\pgfqpoint{0.786740in}{0.897183in}}%
\pgfpathlineto{\pgfqpoint{0.791932in}{0.879978in}}%
\pgfpathlineto{\pgfqpoint{0.798163in}{0.844845in}}%
\pgfpathlineto{\pgfqpoint{0.804393in}{0.855051in}}%
\pgfpathlineto{\pgfqpoint{0.808547in}{0.856607in}}%
\pgfpathlineto{\pgfqpoint{0.814777in}{0.831395in}}%
\pgfpathlineto{\pgfqpoint{0.819969in}{0.798769in}}%
\pgfpathlineto{\pgfqpoint{0.825161in}{0.820799in}}%
\pgfpathlineto{\pgfqpoint{0.826200in}{0.817988in}}%
\pgfpathlineto{\pgfqpoint{0.833469in}{0.772325in}}%
\pgfpathlineto{\pgfqpoint{0.836584in}{0.758912in}}%
\pgfpathlineto{\pgfqpoint{0.840737in}{0.620149in}}%
\pgfpathlineto{\pgfqpoint{0.849045in}{0.363410in}}%
\pgfpathlineto{\pgfqpoint{0.854237in}{0.327565in}}%
\pgfpathlineto{\pgfqpoint{0.860467in}{0.321933in}}%
\pgfpathlineto{\pgfqpoint{1.096187in}{0.321933in}}%
\pgfpathlineto{\pgfqpoint{1.096187in}{0.321933in}}%
\pgfusepath{stroke}%
\end{pgfscope}%
\begin{pgfscope}%
\pgfsetrectcap%
\pgfsetmiterjoin%
\pgfsetlinewidth{0.501875pt}%
\definecolor{currentstroke}{rgb}{0.000000,0.000000,0.000000}%
\pgfsetstrokecolor{currentstroke}%
\pgfsetdash{}{0pt}%
\pgfpathmoveto{\pgfqpoint{0.006944in}{0.317130in}}%
\pgfpathlineto{\pgfqpoint{1.148056in}{0.317130in}}%
\pgfusepath{stroke}%
\end{pgfscope}%
\begin{pgfscope}%
\definecolor{textcolor}{rgb}{0.000000,0.000000,0.000000}%
\pgfsetstrokecolor{textcolor}%
\pgfsetfillcolor{textcolor}%
\pgftext[x=0.577500in,y=1.059156in,,base]{\color{textcolor}\rmfamily\fontsize{6.000000}{7.200000}\selectfont \!\!\!\!\!\!\!\!\!\!(Transm. = 0.85)\!\!\!\!\!\!\!\!\!\!}%
\end{pgfscope}%
\end{pgfpicture}%
\makeatother%
\endgroup%
        \end{center}
        \caption{\label{fig:endpoint-matching}Endpoint matching (top, dashed purple) may improve the convergence rate over the base method (green) when the interval ends are at very different densities compared to the general density variation (top right). Endpoint matching is not beneficial when the ends are at similar densities (bottom).}
        \label{fig:endpoint-graph}
    \end{figure}

    \begin{figure*}
        \centering
        \includegraphics[width=0.98\linewidth]{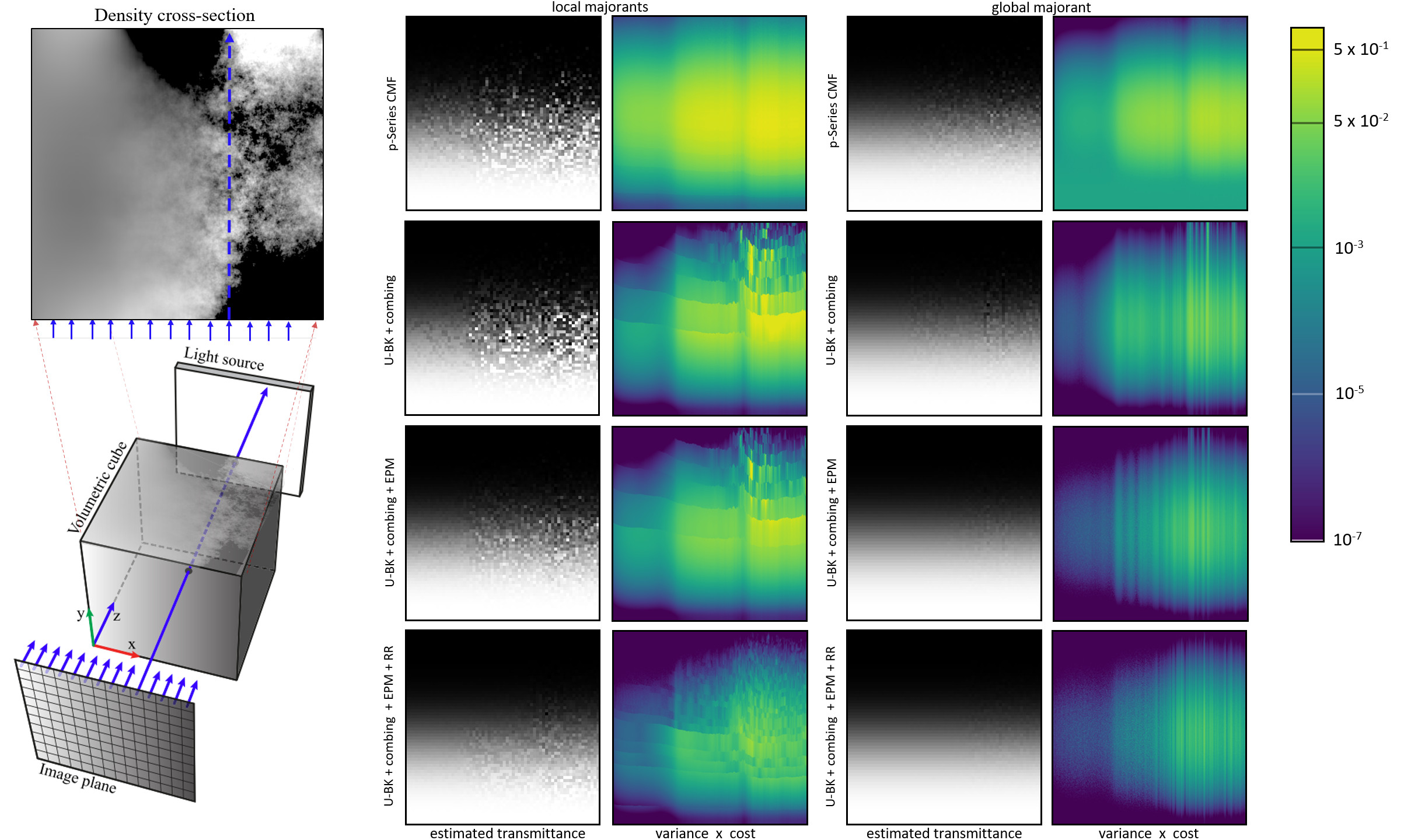}
        \caption{This setup compares our techniques to the power-series estimator (p-series CMF) from Georgiev et al. simulating transmittance through a 3d slab with varying density. The (X,Z) cross-section of the density field, shown in the upper-left corner, features a higher and higher fractal dimension going from left to right. The average density is varied across the vertical axis Y, so as to have near-zero density at the bottom of the slab and a maximum optical thickness of 10 towards the top. The slab is illuminated by a uniform directional light source on the back, so that each pixel in the image plane records the amount of light transmitted through a single ray through the slab. 
        The two leftmost columns show equal sample count results using tight per-pixel majorants, whereas the rightmost columns show equal sample count results using a single global majorant $\maj = 25$. 
        Note the use of a logarithmic color scale: our final estimators provide a
        2 to 5 orders of magnitude efficiency increase over previous state-of-the-art.  }
        \label{fig:slab-tests}
    \end{figure*}

    \begin{figure*}
        \centering
        \includegraphics[width=0.98\linewidth]{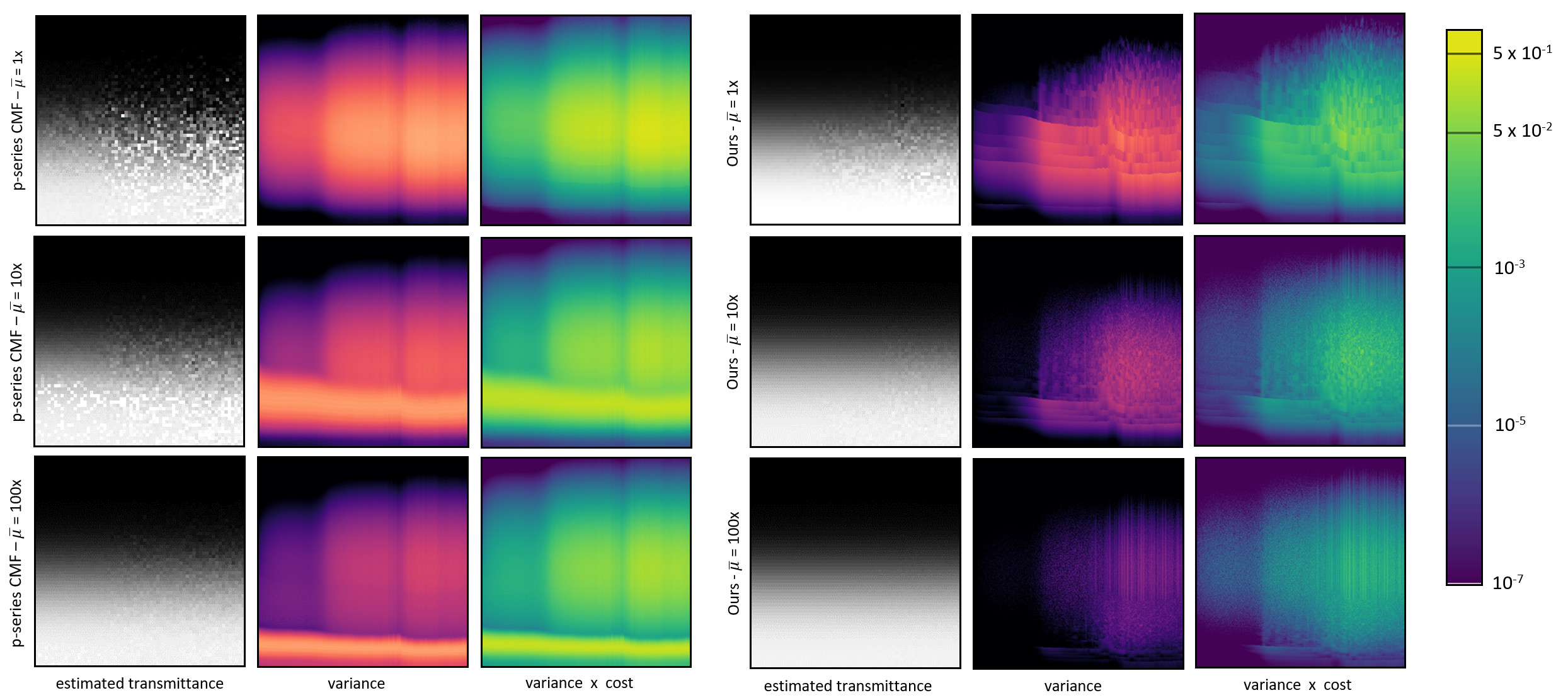}
        \caption{
        This figure uses the same setup of Figure \ref{fig:slab-tests} to compare the variance and inverse efficiency of the p-series CMF estimator (left 3 columns) to that of our unbiased estimator (right 3 columns) when the majorant is raised respectively by 1, 10, and 100 times compared to the tight per-pixel majorant, increasing the number of density evaluations. Notice how at low optical thickness values the original p-series CMF estimator can even suffer from raising the majorant above a certain point, as the error from the low order terms increases exponentially due to the use of a very bad pivot, without ever being fully recovered. Our estimator is able to use all the available density evaluations to reduce variance and improve efficiency.
        }
        \label{fig:slab-tests-k}
    \end{figure*}
    
    \paragraph{Scaling to higher quality} Figure \ref{fig:graph-test} shows a simple test comparing the variance of our estimators to Georgiev et al's p-series CMF as a function of the number of density evaluations on a single example density that exhibits high frequencies and fractal behavior.
    For p-series CMF, we examined two methods of increasing the expected number of samples: by averaging multiple evaluations, and by multiplying the optical thickness (in this case the majorant) by a constant greater than 1. For our methods we increase the control optical thickness similarly, but it is only used for calculating the tuple size. 
    
    Increasing the majorant helps the p-series CMF estimator in this particular instance, but this is not always the case, as we will see later on. Our method always benefits from increasing the tuple size due to the improved pivot and lower-variance correction samples, but we see bumps in the convergence curves due to the non-uniform frequency response of our equidistant sampling combs. Despite the bumps, we always found equidistant sampling to perform better than breaking the frequency response with e.g. stratified sampling or a low-discrepancy pattern.
    
    Our methods clearly show a higher rate of convergence which continually increases their lead by orders of magnitude when targeting noise-free transmittance estimates, with the biased variant featuring slightly lower MSE at the cost of a small amount of bias.
    
    \paragraph{Endpoint matching} Figure \ref{fig:endpoint-graph} shows another test where we analyze the behavior of our estimator with and without endpoint matching on two different densities. The top of the figure shows a case where the density is very different at the two endpoints, which creates a strong discontinuity in the periodic extension of the function. Our endpoint-matching control variate removes this discontinuity, greatly reducing the variance and improving the convergence rate. The effect is particularly large when the discontinuity is high compared to the other variation in the density function (as in this example). The bottom plot shows a counter example where the control variate does not yield any improvement. The plot shows that the overhead of performing the extra lookups at the endpoints is relatively low; we see only mild reduction in efficiency, especially when targeting high-quality transmittance estimates.
    
    \paragraph{Pure transmittance estimation}
    Figure~\ref{fig:slab-tests} analyzes the impact of gradually enabling some of our proposed techniques; the results from the p-series CMF estimator are used as a baseline.
    All methods are adjusted to utilize roughly the same number of lookups.
    We estimate transmittance through a uniformly lit volumetric slab with a variable density field. The slab is viewed from the +Z direction; its (X,Z) cross-section features a 2D fractal density field with increasing fractal dimension going from left to right, modulated along the vertical Y axis so as to have near-zero density at the bottom, and a maximum optical thickness of 10 at the top.
    Each pixel in the rendered insets represents transmittance along a single ray through the slab, evaluated using one of the tested estimators.
    We performed two tests: the first test (two leftmost columns) employs tight per-pixel majorants, whereas the second test (two rightmost columns) employs a single global majorant for the entire volume.
    The four rows compare:
    \begin{itemize}
    \item the p-series CMF estimator;
    \item our U-BK estimator with $\g = 2$ using a sampled pivot and combing;
    \item our U-BK estimator with $\g = 2$ using a sampled pivot, combing and our endpoint matching control variate;
    \item our final U-BK estimator using the sampled pivot, combing, the endpoint matching control variate and our aggressive roulette scheme detailed in \autoref{section:aggressive-roulette}.
    \end{itemize}
    All variants of our estimators employ the automatic tuple size deduction algorithms described in \autoref{sec:tuple-size-deduction} in order to match the expected sample count of the p-series CMF estimator.
    Odd columns show the result of a single evaluation, whereas even column show a plot of inverse efficiency.

    We make the following observations:
    \begin{itemize}
    \item Using tight per-pixel majorants causes the p-series CMF estimator to take discrete jumps in the base number of terms evaluated, due to it activating RR after reaching 99\% mass only - this appears as blocky variations in variance/efficiency.
    \item With enough samples (e.g. with the global majorant), our equidistant sampling combs coupled with the symmetrization provided by U-statistics already provide a significant efficiency improvement.
    \item Enabling the endpoint matching control variate in some areas allows a relatively large variance reduction, but the largest improvement is obtained by combining the previous techniques with our aggressive roulette, that allows using even large tuples by sampling fewer orders.
    \item In regions with a low-frequency density function, we obtain up to 5 orders of magnitude improvements in efficiency. With higher frequencies our final estimator achieves 2 to 3 orders of magnitude lower variance.
    \end{itemize}

    Since our estimator gains efficiency with larger and larger tuples, we also compared the evaluation of the p-series CMF estimator with varying majorants $\maj$ (respectively 1, 10 and 100 times larger than the tight per-pixel majorant) against single evaluations of our estimators with a tuple size $M = DetermineTupleSize( \maj )$; see \autoref{fig:slab-tests-k}.
    This comparison reveals that using larger majorants with the p-series CMF estimator can be very detrimental at low optical thicknesses; the majorant effectively acts as a worse and worse pivot. This leads to an exponential increase of the error of the low order terms that is never fully recovered, as the Russian roulette continuation probability after the CMF threshold of 99\% approaches zero. In the second and third row, increasing the majorant appears to squeeze the large variance (low efficiency) bump at the center of the first row (using the tight majorant) towards the bottom of the slab, where the transmittance $T$ approaches $1$.

\section{Discussion}

    In the Results section, we have seen how our novel unbiased ray-marching estimator provides a major efficiency improvement across all our tests compared to previous state-of-the-art, and how the biased ray-marching solution reaches even lower MSE at equal cost.

    In the following we discuss a different perspective on our U-statistics estimator as well as alternative strategies to equidistant combing and connections to the more general theme of sample stratification.

    \paragraph{Complex factorization of the truncated Taylor polynomial}

    Another path to obtaining our U-statistics estimator is to apply the complex factorization of the truncated power-series polynomial:
    \begin{equation}
    \sum_{k=1}^N \frac{1}{k! Q(k)} \prod_i^k X_i = c_0 \prod_{i=1}^N (X_i - c_i)
    \end{equation}
    and apply the generic estimator of products of unbiased estimators recently suggested by \citet[Eq.(6)]{lee2019unbiased}.
    The resulting permuted estimator matches exactly our U-statistics estimator, and despite the presence of complex coefficients, the imaginary part cancels out.
    What is most interesting, though, is that unlike the generic estimators of \citet{lee2019unbiased}, our algorithms can exploit the structure of the truncated Taylor series to evaluate all combinations in $O(NZ)$ time, whereas the direct evaluation of \citep[Eq.(6)]{lee2019unbiased} is $\#P$-hard.
    
    \paragraph{Connections of combing to stratified sampling}\label{sec:stratified-sampling}
    Combing can be seen as a form of stratified sampling applied to each \emph{individual} estimate of the integral of the null density. It is important to note that the separate integral estimates are uncorrelated. Using a single stratified set of random numbers across all orders is not possible, as that would result in correlated integral estimates whose products would result in biased estimates of the powers of $-\tau$.
    \citet{georgiev2019integral} had previously suggested another form of stratification, \emph{across multiple evaluations} of the transmittance integral.
    This form of stratification is orthogonal and can be combined with our approach: it is sufficient to stratify the Cranley-Patterson rotations $(x_0, ..., x_N)$ across different evaluations of the estimator, for example using Latin hypercube sampling or some other randomized QMC sequence. 

    \paragraph{Alternative strategies for reducing $Y$-variance}
    \label{sec:lds-tuples}
     
    A regular comb using an equidistant sampling tuple works well under the assumption that the density has bounded slope: in this case it can potentially reduce the integration error to $O(1/M)$ or less.
    
    For highly discontinuous densities, or densities with very high fractal dimension, this might no longer be the case.
    An alternative in these extreme cases could be using CP-rotated low discrepancy blue-noise combs that react less to the spectrum of the integrand. An example of such a comb can be easily obtained using
    $u_{j} = \len \cdot \phi \cdot j$, 
    where $\phi$ is the well known golden ratio.
    
    In practice, however, we have found equidistant sampling to always outperform any other low-discrepancy set we have tried.
    This might be related to the observations of \citet{ramamoorthi2012visibility}.

    \subsection{Future Work}\label{sec:future-work}

    While in this work we have focused on matching the sample budgets of previously known methods, as briefly mentioned in Section \ref{sec:tuple-size-deduction} a natural and needed extension of this work would be a scheme for adaptive allocation of tuple sizes in the presence of additional sources of noise: the superlinear convergence properties of our estimators might in fact allow to highly benefit from taking more samples in important regions of path space, while taking fewer in less important ones.
    Another potentially related point that deserves attention is a more thorough investigation of the bias/variance tradeoff of our biased and unbiased estimators.
    
    There may be scenarios where negative transmittance estimates are undesired or the sample budget is fixed independent of a majorant optical depth.  What estimator performs best in these cases remains an open question.  Finally, power series estimation of zero-order probabilities for random media with non-exponential transmission laws where $\ext(x)$ is a random variable is an interesting open area~\cite{jarabo18,bitterli2018radiative,deon2018reciprocal}, and some steps in this direction using the Master equation for binary mixtures has already been made~\cite{longo2002direct}.

\section{Conclusion}
\label{sec:conclusion}
    We presented a novel in-depth variance analysis (\autoref{sec:variance}) of existing unbiased transmittance estimators, revealing weaknesses and areas for improvement.
    We then proposed a series of techniques (\autoref{sec:power-series}) exploiting these insights, specifically:
    \begin{itemize}
        \item We have presented a novel power-series estimator utilizing all samples efficiently using U-statistics, a recipe for evaluating the estimator in quadratic time, and a numerically robust, incremental elementary symmetric means algorithm. 
        
        \item We have demonstrated how to further reduce variance by using sampled mean pivots instead of majorant derived ones; a development enabled by the U-statistics.
        
        \item We described a combed estimator for evaluating optical depth using $M$ rotated equidistant samples and proposed an affine CV to preserve its superlinear convergence rate.
        
        \item We have proposed to alter the BK roulette and make it vastly more aggressive, enabling us to use larger combs and attain even higher overall efficiency.
    \end{itemize}

    Since the zeroth order term of our final power-series estimator is analogous to the classical ray marching solution (with the addition of our endpoint matching control variate), we refer to the novel estimator as \emph{unbiased ray marching}.
    We have shown that unbiased ray marching is universally faster than any of the previously known unbiased estimators, and often offers several orders of magnitude lower variance at equal sample count.
    Moreover, we have shown that stopping the power-series evaluation at the zeroth-order and effectively getting back to simple ray marching results in a very low-bias estimator that attains lower MSE than any known unbiased estimator, even at relatively low sample counts.
    This latter result might have interesting consequences for real-time rendering and other applications where unbiasedness is not crucial.


\bibliographystyle{ACM-Reference-Format}
\bibliography{Tr2021}

\appendix

\section{Optimality of the Mean Pivot}
\label{appendix:OptimalityOfTheMeanPivot}

Earlier, we discussed how non-symmetric power series estimators benefit from a majorant pivot and how U-statistics changes this behaviour. 
With this important difference, we perform a similar analysis for the pivot as in earlier work (e.g.~\cite{georgiev2019integral}): We analyze the sum of the absolute values of the different order contributions in the Taylor series of $e^{\E[X]}$ with different pivots $p$. The sum of the absolute values of the contributions from all orders with pivot $p$ is
\begin{equation}
    e^p \left(1 + \left|\E[X] - p\right| + \frac{\left|\E[X] - p\right|^2}{2!} + \cdots \right) = e^{p + \left|\E[X] - p\right|} .
\end{equation}
This says that in some sense, it is optimal to use any pivot less than the expectation, $p \le \E[x]$, as any such pivot minimizes the above expression. In terms of positive densities, this says that the control density should be at least as high as the mean density. However, the control density does not need to be greater than all of the density samples---there is no need to use a majorant.

However, the picture changes drastically when we take Russian roulette into account: As noted earlier, moving the pivot closer to $-\optd$ implies faster convergence for the Taylor series. This means that we need to evaluate fewer orders of the power series for good estimates, which means that we can employ more aggressive Russian roulette.

Interestingly, with a good pivot, even aggressive Russian roulette will not make efficiency worse:  for $N$ total evaluations of an estimator $X$, returning the roulette-compensated variable with probability $p$ and otherwise zero, the actual number of evaluations is $N p$. The inverse efficiency of the estimator is thus proportional to 
\begin{equation}
\begin{split}
    p \Var\left[\frac{R X}{p}\right] = p \left(\frac{\E[R^2] \E[X^2]}{p^2} - \frac{\E[R]^2 \E[X]^2}{p^2}\right) \\
                         = \E[X^2] - p \E[X]^2 = \Var[X] + (1 - p) \E[X]^2 ,
\end{split}
\end{equation}
where $R$ is the random binary choice variable. This says that the efficiency of the roulette is maximized when $\E[X] = 0$, that is, when we use the theoretical mean pivot. The efficiency of the rouletted estimator decreases as the pivot moves farther from the real expectation.

Therefore, using an approximate mean pivot allows the use of a more aggressive roulette.

The resulting lower mean estimation order from more aggressive roulette means that we now need a smaller number of independent samples, and we can use our density evaluation budget to make those samples higher-quality by performing variance reduction techniques such as stratification or numerical integration rules, as discussed in Section 4.

\section{Elementary symmetric means}
\label{appendix:ElementarySymmetricMeans}

In this appendix we derive the elementary symmetric means formulas that lead to Algorithm~\ref{Algorithm:ElementarySymmetricMeansNew}.

Elementary symmetric sums $e_k = e_k(x_1, \cdots, x_n)$ are defined as
\begin{equation}
    e_k = \sum\limits_{1 \le i_1 < \cdots < i_k \le n} x_{i_1} \cdots x_{i_k} ,
\end{equation}
with $e_0 = 1$. To distinguish between different numbers of parameters, in this appendix we denote the elementary sums of $x_1 \cdots x_n$ by
\begin{equation}
    e_k^{n} = e_k(x_1, \cdots, x_n) ,
\end{equation}
and elementary symmetric means by
\begin{equation}
    m_k^n = m_k(x_1, \cdots, x_n) = \frac{e_k(x_1, \cdots, x_n)}{ {n \choose k} } .
\end{equation}

A simple derivation leads to a formula for iteratively constructing elementary symmetric sums:
\begin{equation}
\begin{split}
e_k^{n+1} & = \sum\limits_{1 \le i_1 < \cdots < i_k \le n+1} x_{i_1} \cdots x_{i_k} \\
          & = \sum\limits_{1 \le i_1 < \cdots < i_k \le n} x_{i_1} \cdots x_{i_k} + \sum\limits_{1 \le i_1 < \cdots < i_k = n+1} x_{i_1} \cdots x_{i_k} \\
          & = e_k^n + \left(\sum\limits_{1 \le i_1 < \cdots < i_{k-1} \le n} x_{i_1} \cdots x_{i_{k-1}}\right) x_{n+1} \\
          & = e_k^n + e_{k-1}^n x_{n+1} . \\
\end{split}
\end{equation}

Then, by substituting the definition of elementary symmetric means, we reach the recurrence formula
\begin{equation}
\begin{split}
m_k^{n+1} & = m_k^n + \frac{k}{n+1} \left(m_{k-1}^n x_{n+1} - m_k^n\right) .
\end{split}
\end{equation}
Observing the directions of the dependencies in this formula leads to Algorithm~\ref{Algorithm:ElementarySymmetricMeansNew}.

\section{Efficiency derivations}\label{appendix:efficiency:deriv}

In this appendix we review known analytic results for the variance and cost of transmittance estimators as well as present some new derivations for power-series estimators.

\subsection{Costs}

\paragraph{Tracking estimators} The expected number $\E[N]$ of optical-depth estimates for a tracking estimator follows from the mean of the Poisson distribution, which is simply the rate, $\E[N] = \lambda_\len = \optdmajres$.  Therefore, for residual ratio tracking with query size $\M$,
\begin{equation}
    \Cost[\widehat{T}_{rrt}] = \M \, \optdmajres.
\end{equation}
Delta-tracking with $n = 1$ is an exception in that the estimator can perform early termination as soon as the first real (non-thinned) estimate is performed.  The cost for $n = 1$ delta-tracking is therefore (assuming $\control = 0$) \cite{georgiev2019integral}
\begin{equation}
    \Cost[\widehat{T}_{dt}] = \M \optdmaj \frac{\left(1-e^{-\optd }\right) }{\optd }, \quad (n = 1)
\end{equation}
and otherwise tracking must completely traverse the interval $n$ times, $\Cost[\widehat{T}_{J}] = \M \optdmaj n$.

\paragraph{Bhanot and Kennedy roulette}

Using the continuation probabilities of the generalized BK estimator (\ref{eq:bhanot}) we find the probability $Q_{\text{BK}}(N)$ of evaluating term $N$ to be
\begin{equation}
    Q_{\text{BK}}(N) = \frac{\g}{K+1} \frac{\g}{K+2} \dots \frac{\g}{K+N} = \frac{\g^{N-K}}{N! / K!}, \quad N > K = \myfloor{c},
\end{equation}
and $1$ otherwise.  The expected number of evaluated orders is thus
\begin{equation}\label{eq:cost:BK}
   \E[N_{BK}] = K + \sum_{N=K+1}^\infty \frac{\g^{N-K}}{N! / K!} = K + \frac{K!}{\g^K} \left(e^\g - \sum\limits_{N=0}^K \frac{\g^N}{N!}\right) .
\end{equation}

\subsection{Variances}

\paragraph{Delta tracking}\label{sec:variance:DT}

The exact variance of delta-tracking is known \cite{glasser1962minimum} and agrees with a derivation for the special case of $n = 1$ and a uniform-medium~\cite{georgiev2019integral}
\begin{equation}
    \Var[\widehat{T}_{dt}] = e^{-\optd }-e^{-2 \optd }.
\end{equation}
This result is exact for any input, and generalizes for Johnson's $n > 1$ estimator to \cite{glasser1962minimum}
\begin{equation}
    \Var[\widehat{T}_{J}] = e^{-2 \optd } \left(e^{\frac{\optd }{n}}-1\right).
\end{equation}
We investigate the efficiency of Johnson's estimator in the supplementary material.

\paragraph{Residual ratio tracking}

The exact variance for residual ratio tracking is also known.  The rate of the Poisson process follows from the difference of known optical depths $\lambda = \optdmaj - \optdctrl$, which are the integrals of the upper $\maj(x)$ and lower $\extctrl(x)$ control variates.  The variance is then~\citep[Eq.(4.17)]{papaspiliopoulos2011monte}
\begin{align}
    \Variance[\widehat{T}_{rrt}] &= e^{-2 \, \optdmaj+\lambda+\frac{V}{\lambda }} - e^{-2 \optd}, \label{eq:var:RRT}
    \\ V & = \frac{1}{(b-a)} \int_a^b \left((b-a)(\maj - \ext(x)\right)^2 dx. \label{eq:V}
\end{align}

\subsubsection{Truncated estimators}
\paragraph{Roulette variance}\label{sec:var:BK:constdens}

With uniform density, the negative residual optical depth $Y$ is estimated with zero variance and the full variance of the generalized BK estimator is (see supplemental material)
  \begin{equation*}
    \Var[\widehat{T}_{BK}] = e^{-2 \optdctrl } \sum_{j=0}^\infty \frac{\g^j \left(1-\frac{\g}{j+K+1}\right)}{(K+1)_j} \left( \sum_{n=0}^K \frac{\estoptdres^n}{n!} + \sum_{i=1}^j \frac{\estoptdres^{K+i}}{\g^i K!} \right)^2 - T^2.
  \end{equation*}
  
  \paragraph{Variance at the optimal pivot}
    
    Consider the BK estimator with pivot $p = -\optdctrl = -\optd$, where truncation is fixed to deterministic order $N$. The estimator will then have a biased expectation $T_{b} \approx e^{-\optd}$.  In the supplemental material we show that this estimator has variance
    \begin{equation}
        \Var[\widehat{T}_{BK}] = e^{-2 \optd} \sum_{k=0}^N \frac{\E[Y^2]^k}{(k!)^2} - T_b^2 \approx e^{-2 \optd} \sum_{k=1}^N \frac{\E[Y^2]^k}{(k!)^2}.
    \end{equation}
    With U-statistics the variance becomes
    \begin{equation}
        \Var[\widehat{T}_{UBK}] = e^{-2 \optd} \sum_{k=0}^N \frac{\E[Y^2]^k}{\binom{N}{k}(k!)^2} - T_b^2 \approx e^{-2 \optd} \sum_{k=1}^N \frac{\E[Y^2]^k}{\binom{N}{k} (k!)^2}.
    \end{equation}
    The binomial denominators reduce the variance relative to the non-symmetrized estimator.  For small $\E[Y^2]$ (low $Y$-variance), the linear term sees a variance reduction of $1/N$ relative to the non-symmetrized version, with diminishing gains for the higher order terms.  So at the optimal pivot, the variance reduction between UBK and BK approaches $1 / N$ as $Y$-variance goes to $0$.
    
    Roulette significantly complicates the variance derivation for the UBK estimator, but at the optimal pivot we found
    \begin{equation*}
        \E[T^2] = e^{-2 \optd} \sum _{k=0}^{\infty } \frac{\left(\g^k \left(1-\frac{\g}{1+K+k}\right)\right) }{(1+K)_k} \left( \sum _{n=0}^K
   \frac{\E[Y^2]^n}{(n!)^2 \binom{k+K}{n}} + \sum _{i=1}^k \frac{\g^{-2 i} \E[Y^2]^{K+i}}{(K!)^2 \binom{K+k}{K+i}} \right)
    \end{equation*}
    from which the variance follows ($\Var[T] = \E[T^2] - \E[T]^2$).  This can be used to rigorously analyze the tradeoffs between decreasing $K$ in favour of reducing $Y$-variance (by increasing $\M$).
\section{Endpoint Matching Formulas}\label{appendix:epm}

    Integration with endpoint matching and equidistant combs can be further simplified. Since the order of the samples doesn't matter, we can write the integral estimate as
    \begin{equation}
    X_i = -\frac{\ell}{M} \sum_{i=0}^{M-1} \mu^{\star}\left(\frac{\ell}{M} (u + i)\right)
    \end{equation}
    where $u$ is a uniform random number in $[0, 1)$. This directly simplifies into
    \begin{equation}
    X_i = \underbrace{-\frac{\ell}{M} \sum_{i=0}^{M-1} \mu\left(\frac{\ell}{M} (u + i)\right)}_{\text{original estimate}}
        - \underbrace{\frac{\ell}{M}\left(\frac{1}{2} - u\right)\left(\mu(\ell) - \mu(0)\right)}_{\text{endpoint matching}}.
    \end{equation}
    The left-hand-side term is the integral estimate without endpoint matching, and the right-hand-side is the zero-expectation term from endpoint matching that often improves the convergence rate.
    
    This reshuffling can cause the resulting integrand to assume negative values.  If non-negativity is a constraint, an alternate option is to symmetrize the estimator over the interval by using the mean of mirrored lookups~\cite{buslenko1966monte} (p.106).  This is equivalent to blending the interval of scattering material with its reversed copy.

\end{document}